\documentclass[aps,prd,twocolumn]{revtex4-2}

\usepackage{orcidlink}

\usepackage{graphicx}
\usepackage[caption=false]{subfig}

\usepackage{amsmath}
\usepackage{amssymb}

\usepackage{xcolor}
\usepackage{hyperref}
\usepackage{soul}

\usepackage{tabularx}
\usepackage{float}

\newcommand{\Msolar}{\text{M}_{\odot}}
\newcommand{\umin}{_{\text{min}}}
\newcommand{\umax}{_{\text{max}}}
\newcommand{\uref}{_{\text{ref}}}
\newcommand{\ustart}{_{\text{start}}}
\newcommand{\ufinish}{_{\text{finish}}}
\newcommand{\uGW}{_{\text{GW}}}
\newcommand{\upw}{_{\text{PW}}}
\newcommand{\urss}{_{\text{rss}}}
\newcommand{\ugte}{_{\text{GTE}}}

\newcommand{\cF}{\mathcal{F}}
\newcommand{\vecl}{\vec{\lambda}}

\newcommand{\mf}[2]{\mathfrak{f}_{#1, #2}}

\begin{document}

\title{Gravitational Wave Searches for Post-Merger Remnants of GW170817 and GW190425}

\author{Benjamin Grace\,\orcidlink{0009-0009-9349-9317}}
\email{Benjamin.Grace@anu.edu.au}
\author{Karl Wette\,\orcidlink{0000-0002-4394-7179}}
\email{Karl.Wette@anu.edu.au}
\author{Susan M. Scott\,\orcidlink{0000-0002-9875-7700}}

\affiliation{OzGrav-ANU, Centre for Gravitational Astrophysics, Australian National University, Canberra ACT 2601, Australia}

\begin{abstract}
    We present the results of two searches for gravitational waves from the post-merger remnants of the binary neutron star coalescence events GW170817 and GW190425. The searches are fully coherent over 1800~s of data from the 2nd (for GW170817) and 3rd (for GW190425) observing runs of the LIGO and Virgo observatories. The searches compute the matched filter $\cF$-statistic, and use a piecewise model of the rapidly changing frequency evolution appropriate for young neutron stars. No detection is claimed. The peak root-sum-squared strain upper limit at 50\% detection probability ($h\urss^{50\%})$ of both searches occurs at 1700~Hz and is estimated at $1.64 \times 10^{-22}~\text{Hz}^{-1/2}$ for GW170817, and $1.0 \times 10^{-22}~\text{Hz}^{-1/2}$ for GW190425. This is the first gravitational wave search for a neutron star remnant of GW190425.
\end{abstract}

\maketitle

\section{Introduction}

Of the over 90 gravitational wave detections made to date, the majority have been binary black hole mergers; only two are confirmed to be from the coalescence of a binary system of neutron stars~\cite{Collaboration2021}. Binary neutron star (BNS) mergers are signals of particular interest as, for example, they are expected to produce electromagnetic counterparts which may be detectable on Earth and provide a precise sky position of the source. The BNS signals detected to date have contributed to an understanding of the neutron star equation of state~\cite{Abbott2018d, Bhat2019} and electromagnetic observations have shown that BNS mergers are responsible for the majority of nucleosynthesis for numerous neutron-rich elements~\cite{Abbott2017a, Tanvir2017, Wu2019}.

Despite electromagnetic observations, the nature of the object left after a BNS coalescence is uncertain. The remnant object after coalescence may immediately collapse into a black hole, form a momentarily stable neutron star before collapsing into a black hole, or form a stable long-lived neutron star~\cite{Ravi2014}. Electromagnetic observations of the remnant object may give us insight into the evolution of a BNS remnant. It is not guaranteed, however, that electromagnetic observations will always be possible after the detection of a BNS signal. If the remnant object does not immediately collapse into a black hole, it is expected to be radiating gravitational waves. Gravitational waves are therefore an invaluable tool for studying these systems.

The first gravitational wave detection of a BNS merger was GW170817~\cite{Abbott2017}. This event was significant in that the a gamma ray burst (GRB) was detected 1.7~s after the collision. An extensive electromagnetic follow up campaign allowed for the precise sky position of the source to be known, constraining its position to the galaxy NGC-4993, at the cosmologically close distance of 40~Mpc to Earth~\cite{Abbott2017a}. The exact nature of the remnant is unknown. The general consensus is that the remnant is now a black hole, but between coalescence and today there was an unknown period of time when a neutron star might have been present. This is supported by the sharp cut off seen in x-ray observations, as we would expect to observe extended emission in this spectrum if a long-lived neutron star was formed~\cite{Abbott2017a}. There are claims that a stable neutron star may have formed after the coalescence~\cite{VanPutten2019,AghaeiAbchouyeh2023}.

The second BNS detection was GW190425~\cite{Abbott190425}. This event had no detectable electromagnetic counterpart and was located at a much greater distance than GW170817 at $159_{-72}^{+69}$~Mpc. This event was only detected by the LIGO Livingston detector, which led to poor sky localisation of the source. Due to its greater distance than GW170817, and the absence of a counterpart, GW190425 has not been extensively studied in electromagnetic and gravitational waves for a possible remnant. GW170817 and GW190425 remain the only two BNS detections to date.

Gravitational wave searches have been conducted to search for a remnant from GW170817~\cite{Abbott2017b,TheLIGOScientificCollaboration2019, Abbott2019, Miller:2019jtp}, however no detection has been claimed. A claimed detection of a signal originating from a neutron star remnant following GW170817 was made~\cite{VanPutten2019,VanPutten2019b}, but is considered implausible due to energy constraints, as discussed in~\cite{OlivEtAl2019-MtcSEBSNDtGrvExEmG}. The authors of~\cite{VanPutten2019,VanPutten2019b} later changed the interpretation of their gravitational wave signal claim to originate from an accretion disk surrounding a remnant black hole, which is not ruled out by the energy arguments presented in~\cite{OlivEtAl2019-MtcSEBSNDtGrvExEmG}. No other claims of detection have been made. No searches have been carried out for GW190425.

Neutron stars formed as remnants from BNS coalescences are expected to be born spinning extremely rapidly and spinning down over short periods of time (seconds to hours). A signal of such duration is often referred to as a \emph{long transient}. In contrast, standard continuous wave search techniques are specifically built to search for much longer-lived spinning neutron stars ($\gg$~years). Due to the short period over which a remnant neutron star spins down, and the rate at which its frequency is changing, continuous wave search techniques are not immediately applicable. The adaptation of continuous wave search techniques, however, is a promising path forward to detecting long-lived transient signals. The modification of search techniques to be appropriate for different targets has been carried out previously, such as in~\cite{PhysRevD.77.082001} where a stochastic search technique was modified to be used for periodic signals. This method was further adapted to be used for supernovae searches~\cite{10.1111/j.1365-2966.2011.18585.x}, long-duration transients~\cite{PhysRevD.93.104059} and a follow-up of GW170817~\cite{PhysRevD.100.124041}. Unified frameworks of gravitational wave data analysis strongly motivate the adaptation of established techniques for different emission sources~\cite{PhysRevD.87.122003}. Here, we use a piecewise model as a modification to standard continuous wave search techniques to make them suitable for young neutron stars. The piecewise model is fully described in~\cite{Grace2023}.

In this work we present the results of two searches for a remnant neutron star following the BNS coalescences GW170817 and GW190425. In Sec.~\ref{sec:Background} we introduce the framework of a typical continuous wave search. We then summarise previous searches carried out for binary neutron star remnants. In Sec.~\ref{sec:Method} we introduce the piecewise model and the setup used for the searches. In Sec.~\ref{sec:Data} we describe the data used. In Sec.~\ref{sec:Application} we detail how the searches have been implemented. In Sec.~\ref{sec:Results} we discuss the results of the searches and the vetoing processes used. In Sec.~\ref{sec:Sensitivity} we estimate the strain upper limits of the searches and present results for GW170817 and GW190425. Finally, in Sec.~\ref{sec:Conclusion} we summarise our results and discuss future applications of the piecewise model search method.

\section{Background} \label{sec:Background}

\subsection{Continuous Wave Search Methods} \label{sec:Cont_Search_Methods}

For a rotating solid body, the strain amplitude of radiated gravitational waves is~\cite{Jaranowski1998}
\begin{align}
    h_{0} &= \frac{16\pi^{2}G}{c^{4}}\frac{\epsilon I_{zz} f^{2}}{D}, \label{eq:strain}
\end{align}
where $I_{zz}$ is the moment of inertia of the body about its axis of rotation, $\epsilon$ is its ellipticity, $f$ the frequency of the emitted gravitational waves (assumed here to be twice that of the body's rotational frequency), $D$ is the distance to the source from the detector, $G$ is the gravitational constant, and $c$ is the speed of light.

Continuous gravitational wave search methods typically use a matched filtering process, where a signal model is compared to data in order to calculate a detection statistic. The gravitational-wave signal $s$ at a detector at time $t$ depends on the characteristic strain of the incoming gravitational wave [Eq.~\eqref{eq:strain}], and the antenna pattern of the detector, encoded in four functions $h_{i}, i=1,\dots,4$~\cite{Prix2007, Jaranowski1998}. The signal $s$ is then given by
\begin{align}
    s(t, \mathcal{A}, \Vec{\lambda}) &= \sum_{i = 1}^{4}\mathcal{A}_{i}h_{i}(t, \Vec{\lambda}).
\end{align}
The parameters $\mathcal{A}_{i}$ are known as the canonical amplitudes of the gravitational wave. They are functions of the parameters $\phi_{0}, \psi, h_{0}$ and $\cos\iota$~\cite{Prix}. The parameters $\phi_{0}$ and $\psi$ are, respectively, the initial phase of the gravitational wave at reference time $t = t\uref$ and its polarisation angle. The final parameter $\iota$ is the angle between the rotational axis of the source and its sky position vector $\vec{n}$. Finally, the vector of phase parameters $\vec{\lambda}$ is composed of the sky position, frequency, and derivatives of frequency in time of the gravitational wave.

The likelihood ratio $\Lambda$ is the ratio of the probability that a signal $s(t, \mathcal{A})$ is present within noisy data against the probability that only noise is present. It is defined by~\cite{Jaranowski1998}
\begin{align}
    \ln\Lambda (\mathcal{A}, \vec\lambda) &= (x | s) - \frac{1}{2}( s| s ), \label{eq:loglikeli}
\end{align}
where $x$ is a continuous representation of the detector data. The scalar product $( \cdot | \cdot )$ is defined as
\begin{align}
    (x | y ) &= \frac{2}{\mathcal{S}_{h}} \int_{t\uref}^{t\uref + T}x(t)y(t)dt,
\end{align}
where $T$ is the length of data used, and $\mathcal{S}_{h}$ is the single-sided spectral density of the detector noise (here assumed, for simplicity, to be constant in time).

The $\cF$-statistic is the maximisation of Eq.~\eqref{eq:loglikeli} over the canonical amplitudes $\mathcal{A}_{i}$, and is commonly used in continuous gravitational wave searches~\cite{Jaranowski1998}. We observe that Eq.~\eqref{eq:loglikeli} is linear in the parameters $\mathcal{A}_{i}$. This allows for the $\cF$-statistic to be analytically maximised in these parameters:
\begin{align}
    2\mathcal{F} &= \max_{\mathcal{A}}\big\{ \ln\Lambda (\mathcal{A}, \vec{\lambda}) \big\}.
\end{align}
We then wish to find the optimal phase parameters $\vec\lambda$ which maximise $2\cF$.

We compute $2\cF$ for a set of phase parameters $\{\vec{\lambda}\}$. This set is the template bank, and each $\vec{\lambda}$ is a template. If a signal is present in the data with parameters $\vecl_{S}$, it is unlikely that they will coincide exactly with any given $\vecl$ in the template bank. Any recovered signal from some $\vecl$ within the template bank will therefore have some \emph{mismatch} between it and the signal parameters $\vecl_{S}$. The mismatch between $\vecl$ and $\vecl_{S}$ is defined using the signal-to-noise ratio $\rho(\mathcal{A}, \vecl_{S}, \vecl)$, and is given as~\cite{Prix2007}
\begin{align}
    \mu &= \frac{\rho^{2}(\mathcal{A}, \vecl_{S}, \vecl_{S}) - \rho^{2}(\mathcal{A}, \vecl_{S}, \vecl)}{\rho^{2}(\mathcal{A}, \vecl_{S}, \vecl_{S})}. \label{eq:MismatchSNR}
\end{align}
If a particular template $\vecl$ lies close to the signal parameters $\vecl_{S}$ such that $\Delta\vecl = \vecl_{S} - \vecl$ is small, a second order Taylor expansion of Eq.~\eqref{eq:MismatchSNR} leads us to the parameter space metric $\mathbf{g}$~\cite{Owen1996-STmGrvWInsBnCTmS, Astone2002, Prix2007}:
\begin{align}
    \mu &\approx \Delta\vecl^{T} \frac{-1}{2\rho^{2}(\mathcal{A}, \vecl_{S}, \vecl_{S})} \frac{\partial\rho^{2}(\mathcal{A}, \vecl_{S}, \vecl)}{\partial \vecl} \bigg|_{\vecl = \vecl_{S}} \Delta\vecl \\
            &= \Delta\vecl^{T} \mathbf{g} \Delta\vecl, \label{eq:MismatchAsDistance}
\end{align}
where $\cdot^{T}$ represents the matrix transpose. We can see that Eq.~\eqref{eq:MismatchAsDistance} defines $\mu$ as the magnitude of the vector $\Delta\vecl$, as measured by $\mathbf{g}$, and one can therefore treat the mismatch as a geometric distance between $\vecl$ and $\vecl_{S}$ within the parameter space.

A convenient approximation to $\mathbf{g}$ is the phase metric, $\mathbf{g}_{\phi}$. The phase metric relies only on the parameters $\vecl$, and is defined as~\cite{Brady1998, Prix2007}
\begin{equation}
    \begin{split}
        [\mathbf{g}_{\phi}]_{ij} &= \big\langle \partial_{i} \phi(t, \vec{\lambda})\partial_{j} \phi(t, \vec{\lambda})\big\rangle \\ &\quad - \big\langle\partial_{i} \phi(t, \vec{\lambda})\big\rangle \big\langle\partial_{j} \phi(t, \vec{\lambda})\big\rangle.
    \end{split}
    \label{eq:phase_metric}
\end{equation}
The $\partial_{i}$ are partial derivatives with respect to the phase parameters which make up $\vecl$. The $\langle\cdot\rangle$ are time averages defined as
\begin{align}
    \langle x \rangle &= \frac{1}{T}\int_{t\uref}^{t\uref + T}x(t)dt.
\end{align}
The function $\phi(t, \vecl)$ describes the phase evolution of the gravitational wave signal. The phase is typically given as~\cite{Jaranowski1998}
\begin{align}
    \phi(t, \vecl) &= 2\pi \sum_{l = 0}^{l\umax} f^{(l)}\frac{(t - t\uref)^{l + 1}}{(l + 1)!}+ 2\pi \frac{\vec{r} \cdot \vec{n}}{c}f\umax,
\end{align}
where $\vec{r}$ is the position vector of the detector with respect to the Solar System Barycentre (SSB), $\vec{n}$ is the unit vector pointing from the SSB to the source, $f\umax$ is the maximum frequency of the gravitational wave frequency over the search band, and the superscript $(l)$ represents a time derivative. The phase model can also be written as a time integral of a gravitational wave frequency model, $f\uGW(t, \vecl)$:
\begin{align}
    \phi(t, \vecl) &= 2\pi\int_{t\ustart}^{t_{\text{start}} + t}f\uGW(t', \vecl)dt'. \label{eq:phase_integral}
\end{align}

The frequency model $f\uGW$ is typically chosen to be a first or second-order Taylor expansion \cite[e.g.][]{Abbott2022a, Owen2022, Abbott2022b}. Taylor expansions are good signal models for the close to monochromatic signals that continuous wave search methods typically target. Additionally, Taylor expansion signal models are linear in their parameters $\vecl$ by construction. Signal models with this property minimise the number of templates needed to cover the parameter space, and hence the computational cost of carrying out a search. Linear signal models achieve this by facilitating the implementation of lattice-based algorithms which minimise the size of the template bank while still fully covering the parameter space~\cite{Wette2014, Prix2007-TmpSrGrvWEfLCFPrS}.

In this section we have given an overview of the techniques used to search for continuous gravitational waves in the case of a single detector. These methods can be generalised to cases where multiple detectors are used; specifically, the definition of the $\cF$-statistic provided here is extended to scenarios involving more than one detector as well as different noise contributions in~\cite{Cutler2005}. Searches which use the multi-detector $\cF$-statistic have improved sensitivity as they provide more data for which signal power is accumulated. The single detector $\cF$-statistic is useful in vetoing signal candidates, as seen in Section~\ref{sec:Detector2FVetoing}. In the searches presented here, we use both multi-detector and single detector $\cF$-statistics.

\subsection{Binary Neutron Star Remnants}

Four scenarios are typically considered for the evolution of the remnant of a BNS coalescence event. The remnant may: i) immediately collapse into a black hole; ii) form a hypermassive neutron star before collapsing into a black hole; iii) form a supramassive neutron star before collapsing into a black hole; or iv) form a stable long-lived neutron star~\cite{Ravi2014}. Each of these scenarios is expected to be accompanied by a different gravitational wave signal.

Hypermassive and supramassive neutron stars are born with masses above what a non-rotating neutron star is able to support without collapsing into a black hole. The threshold mass to form a stable non-rotating neutron star is unknown, due to the uncertainty surrounding the neutron star equation of state. It is generally agreed, however, that this threshold mass exceeds $2~\Msolar$, as electromagnetic observations have found pulsars with masses exceeding this limit~\cite{Romani2022, Antoniadis2013}. Some equations of state predict maximum neutron star masses as large as $2.8~\Msolar$~\cite{Bauswein2015, Takami2015, Ai2020}. Hypermassive neutron stars are expected to support their weight through thermal pressure and differential rotation, and have expected lifetimes on the order of milliseconds~\cite{Baumgarte2000, Hotokezaka2013, Shapiro2000}. Supramassive neutron stars are less massive than hypermassive neutron stars, support their weight through rotation alone, and are expected to live anywhere from seconds to hours~\cite{Ravi2014}, to possibly up to two years~\cite{Suvorov2022}.

Remnant neutron stars are expected to be born with large ellipticities and spinning at high frequencies~\cite{Ai2020}, making them strong candidates for gravitational wave follow-up searches. Hypermassive neutron stars may have extreme deformations, with possible ellipticities as large as 0.87~\cite{Shibata2005}. The estimated ellipticities of supramassive neutron stars are affected by their assumed lifespan, and range from $10^{-7}$ to $10^{-4}$~\cite{Ai2020}. Remnant neutron stars which are stable have estimated initial ellipticities ranging from $10^{-7}$--$10^{-2}$ provided they possess a large ($10^{15}$~G) magnetic field~\cite{Colaiuda2008, Haskell2007a, Ciolfi2010}; other estimates of maximum ellipticities range between $10^{-8}$--$10^{-6}$~\cite{Ciolfi2010, Haskell2006, Ushomirsky2000, DeLillo2022}.

\subsection{Prior Searches for BNS Remnants}

The gravitational wave event GW170817 was the first detection of a BNS coalescence. This event coincided with the detection of a gamma ray burst (GRB). Electromagnetic observations following the initial gravitational wave detection suggest that a short-lived neutron star was present immediately after the merger~\cite{Gill2019}. Initial gravitational wave follow-up of GW170817 focused on short- to long-transient length signals of $<$1~s to $<$500~s~\cite{Abbott2017b}. The searches carried out in~\cite{Abbott2017b} made use of the STAMP~\cite{Thrane2011} and coherent WaveBurst (cWB)~\cite{Klimenko2008} algorithms. These search methods make use of unmodelled techniques, which are typically used to search for signals with unknown waveforms or large parameter spaces. Across the 1--4~kHz frequency band which~\cite{Abbott2017b} covered, peak search upper limits for 50\% search confidence of $2.1 \times 10^{-22}~\text{Hz}^{-1/2}$ and $5.9 \times 10^{-22}~\text{Hz}^{-1/2}$ were achieved for signals of 1~s and 500~s duration respectively.

Subsequent searches looked for longer-lived signals for a GW170817 remnant, ranging from 3 hours to 8.5 days~\cite{TheLIGOScientificCollaboration2019}. Four independent algorithms were used: STAMP~\cite{Thrane2011}; unmodelled Hidden Markov model tracking in conjunction with the Viterbi algorithm~\cite{Viterbi1967,Suvorova2016,Sun2019}; and adaptations of the modelled Hough transform method~\cite{Houg1959-McAnlBbChPct} known as FrequencyHough~\cite{Astone:2014esa, Miller:2018rbg} and Adaptive Transient Hough~\cite{Oliver2019}.
These algorithms searched for a potential gravitational wave signal covering frequency bands from 300--4000~Hz; no detection was claimed. The searches were sensitive to a BNS remnant signal at a distance of 1~Mpc, and therefore not capable of detecting a post-merger signal from GW170817.

A search carried out in~\cite{Abbott2019} used \texttt{BayesWAVE}~\cite{Helou2015}, a Bayesian inference analysis  to achieve upper limits on strain amplitude. No claim of detection was made.  A search using a machine learning method was performed on one week of data in~\cite{Miller:2019jtp}. Sensitivity upper limit estimates were consistent with those achieved in~\cite{TheLIGOScientificCollaboration2019}.

At the time of GW190425, only two detectors were online. This resulted in poor sky localisation of the source, and despite electromagnetic follow-up efforts, no counterpart was found~\cite{Coughlin2019, Hosseinzadeh2019}. The total system mass for GW190425 sits 5 standard deviations above the expected galactic distribution for neutron stars~\cite{Abbott190425}. The remnant mass was estimated at $3.4~\Msolar$. If a neutron star formed after the coalescence, at such a high mass it is likely to have promptly collapsed into a black hole. A search for a gravitational wave signal following GW190425 may then shed light on the remnant object of this event. The detection of a gravitational wave signal from this event would indicate that massive neutron stars are possible and give further insight into the neutron star equation of state. Alternatively, it may confirm that neutron stars of these masses are unstable and again inform the equation of state.

\section{Piecewise Model} \label{sec:Method}

In this work we present the results of an $\cF$-statistic search using a piecewise model for remnants from the BNS merger events GW170817 and GW190425. Typical continuous wave searches use a Taylor expansion gravitational wave frequency model in place of $f\uGW$ in Eq.~\eqref{eq:phase_integral}. We substitute the Taylor expansion with a piecewise model, fully described in~\cite{Grace2023}, and summarised in this section.

Young neutron stars, such as those born from BNS coalescences, are expected to be spinning at high frequencies and spinning down over short periods of time. Taylor expansions do not model these kinds of signals well. In extreme cases, they have a finite interval of convergence, beyond which they diverge from the expected frequency evolution of young neutron stars~\cite{Grace2023}. The piecewise model overcomes this issue by starting a new piecewise segment whenever the frequency starts to diverge from the expected spin-down of a young neutron star. In this way, a piecewise model has the flexibility to fit rapidly changing gravitational wave signals.

The piecewise model, $f\upw$, of the gravitational wave frequency on each piecewise segment $i$ is
\begin{align}
    f_{i}(t) &= \sum_{s=0}^{S - 1} \mathfrak{f}_{i,s} B^{0}_{i, s}(t) + \mathfrak{f}_{i + 1, s}B^{1}_{i, s}(t). \label{eq:PWModel}
\end{align}
Here, $S$ is the number of frequency parameters used in the piecewise model. The dimensionality of the parameter space scales as $S(N + 1)$, where $N$ is the number of piecewise segments. The $B_{i, s}^{0/1}$ are the basis functions of the piecewise model. The subscripts $i$ and $s$ and the superscripts $0/1$ inform us to which of the phase parameters $\mf{i}{s}$ the basis functions are attached. The basis functions are chosen such that
\begin{align}
    \frac{d^{s} f_{i}(t)}{dt^{s}}\bigg|_{t = p_{i}} &= \mathfrak{f}_{i, s} , \label{eq:FirstPWCondition} \\
    \frac{d^{s} f_{i}(t)}{dt^{s}}\bigg|_{t = p_{i + 1}} &= \mathfrak{f}_{i + 1, s}. \label{eq:SecondPWCondition}
\end{align}
The points in time $p_{i}$ are the \emph{knots} of the piecewise function, where we switch between piecewise segments. The frequency parameters $\mf{i}{s}$ make up our parameter space. The subscript $i$ refers to the knot to which the parameter is attached, and the subscript $s$ is the derivative order, in time, of the frequency parameter.

\begin{figure}
    \centering
    \includegraphics[width=\columnwidth]{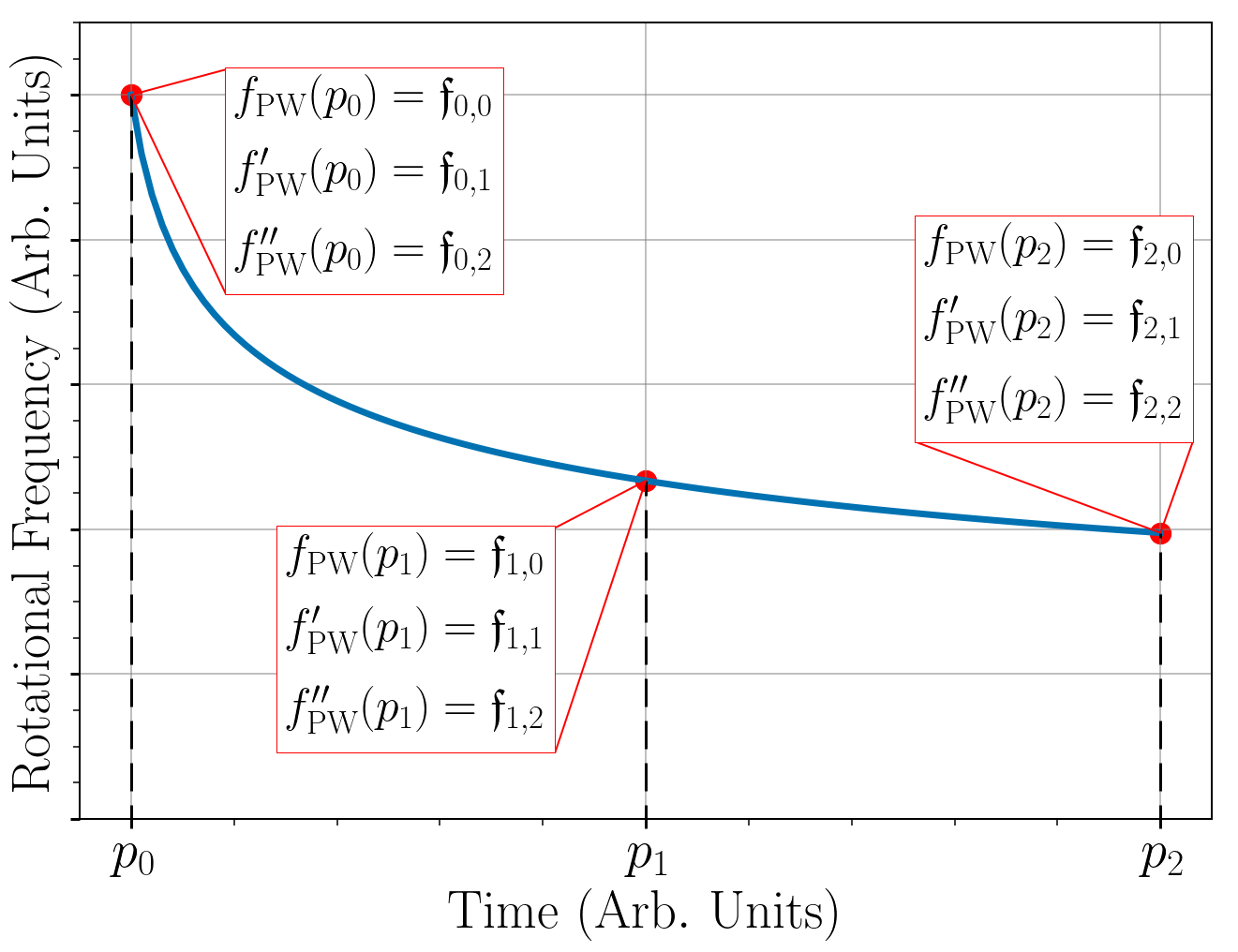}
    \caption{Representation of how the piecewise model, plotted in blue, takes on the values of the parameters $\mf{i}{s}$ at each piecewise knot. At each of the red dots, the values of the piecewise model, and its time derivatives at those points, are shown.}
    \label{fig:PW_Visualisation}
\end{figure}

Equations~\eqref{eq:FirstPWCondition} and~\eqref{eq:SecondPWCondition} state that Eq.~\eqref{eq:PWModel} must take the values of the parameters $\mf{i}{s}$ at their associated knots. Similarly, the derivatives of Eq.~\eqref{eq:PWModel} at each knot must equal the $\mf{i}{s}$ of the same derivative order. This is represented visually in Fig.~\ref{fig:PW_Visualisation}. 
Observe that Eq.~\eqref{eq:PWModel} is linear in its parameters $\mf{i}{s}$. Thus, the piecewise model can be used for a $\cF$-statistic search by substituting $f\upw$ into Eq.~\eqref{eq:phase_integral} as $f\uGW$ and applying the method described in Sec.~\ref{sec:Cont_Search_Methods}.

The piecewise model uses the \emph{general torque equation} (GTE) to define the boundaries of the parameter space. The GTE is commonly used to model the frequency evolution of a neutron star over time, and arises from considering a rotating solid body losing energy via electromagnetic or gravitational radiation~\cite{Ostriker1969}. As we have assumed that gravitational wave emission from a neutron star is at twice the star's rotational frequency, the GTE can also be used to model the frequency evolution of the radiated gravitational waves.  The GTE is given as
\begin{align}
    \frac{df\uGW}{dt} &= -k f\uGW(t)^{n}. \label{eq:GTE}
\end{align}
The constant $k$ contains information on the physical properties of the neutron star. The range of possible values for $k$ is uncertain, as it depends on unknown properties of the remnant neutron star, such as its moment of inertia, ellipticity, and magnetic dipole moment~\cite{Ostriker1969}. The exponent $n$ is known as the braking index. The braking index indicates the energy mechanism through which the neutron star is losing energy. Values of interest are $n = 3$ for energy loss via electromagnetic energy, and $n = 5$ for energy loss through gravitational radiation via a mass quadrupole. Another possibility is $n = 7$, for gravitational wave energy loss via a current quadrupole. Let $f\ugte(f_{0}, n, k, t)$ represent the solution to Eq.~\eqref{eq:GTE}, where $f_{0}$ is the frequency of the gravitational wave at reference time $t = 0$. The boundaries of the parameter space for the piecewise model are:
\begin{align}
    \begin{split}
        \mathfrak{f}_{0, 0} &\geq  f\umin , \\
        \mathfrak{f}_{0, 0} &\leq f\umax ,
    \end{split} \label{eq:f00_Condition} \\
    \begin{split}
        \mathfrak{f}_{i, 0} &\geq f\ugte(\mathfrak{f}_{i - 1, 0}, n\umax, k\umax, p_{i} - p_{i - 1}) , \\
        \mathfrak{f}_{i, 0} &\leq f\ugte(\mathfrak{f}_{i - 1, 0}, n\umin, k\umin, p_{i} - p_{i - 1}) ,
    \end{split} \label{eq:fi0_Condition} \\
    \begin{split}
        \mathfrak{f}_{i, 1} &\geq f\ugte'(\mathfrak{f}_{i, 0}, n\umax, k\umax, 0) , \\
        \mathfrak{f}_{i, 1} &\leq f\ugte'(\mathfrak{f}_{i, 0}, n\umin, k\umin, 0) ,
    \end{split} \label{eq:fi1_Condition} \\
    \begin{split}
        \mathfrak{f}_{i, 2} &\geq f\ugte''(\mathfrak{f}_{i, 0}, n\umin, k\umin, 0) , \\
        \mathfrak{f}_{i, 2} &\leq f\ugte''(\mathfrak{f}_{i, 0}, n\umax, k\umax, 0) ,
    \end{split} \label{eq:fi2_Condition}
\end{align}
where $f_{\text{min/max}}$, $n_{\text{min/max}}$ and $k_{\text{min/max}}$ are the minimum and maximum values of the initial frequency, braking index and $k$ value used for the search.

The boundary conditions given in Eqs.~\eqref{eq:f00_Condition}--\eqref{eq:fi2_Condition} allow for the value of the braking index and $k$ value to evolve between knots. This gives the piecewise model templates additional flexibility to model unknown signals or signals which may not follow the GTE exactly. This optimistically reflects how these physical properties change between young neutron stars and their long-lived counterparts.

\section{Gravitational Wave Data} \label{sec:Data}

\begin{figure*}
    \centering
    \includegraphics[width=\textwidth]{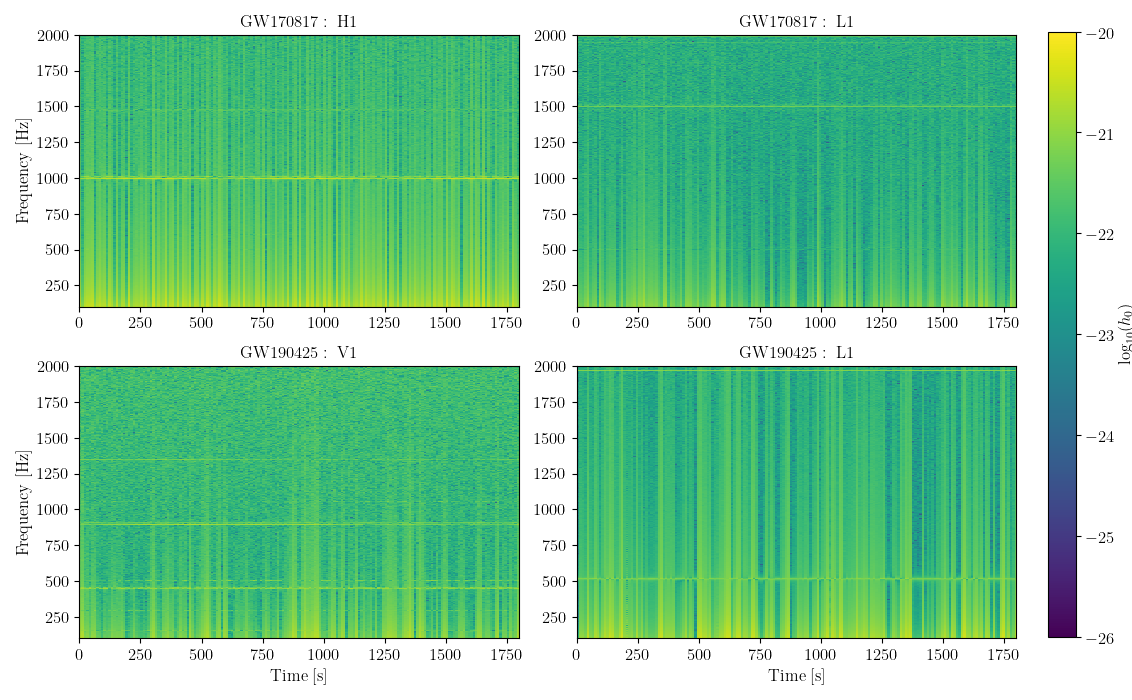}
    \caption{Spectrograms of the data used for both searches. Top: the 1800~s of data following  GW170817 from the (left) H1 and (right) L1 detectors. Bottom: the data following GW190425 from the (left) V1 and (right) L1 detectors. The noise levels in all detectors can be seen to be lower at high frequencies. Strong noise lines can also be seen in the spectrograms. These show evidence of \textit{spectral leakage}, where the noise from these lines is not well resolved into a signal frequency bin and has contaminated neighbouring bins.}
    \label{fig:Spectrograms}
\end{figure*}

Data for the GW170817 search comes from the LIGO Hanford (H1) and LIGO Livingston (L1) detectors~\cite{Aasi2015}. Data for the GW190425 search comes from the L1 and Virgo (V1) detectors~\cite{Aasi2015, Acernese2014}. The data used is publicly available through the Gravitational Wave Open Science Centre (GWOSC)~\cite{Abbott2023}.

Data were acquired at a sample rate of 16384~Hz, and calibrated following~\cite{CahiEtAl2017-ClbUncAdLFScObR,SunEtAl2020-ChrSyErAdLClb}. All data passes the CAT1 data quality vetoes~\cite{Davis2021}. A glitch occurred in L1 data in the 1800~s following GW170817, and was subtracted from the data~\cite{GlitchSub}. The data is filtered with a $10^{\text{th}}$-order Butterworth high-pass filter with a knee frequency of 7~Hz. The filtered time series is then divided into 10~s segments which are Fourier transformed to create Short Fourier Transforms (SFTs) using the program \textsc{lalpulsar\_MakeSFTs}~\cite{lalsuite}. No windowing is applied to the SFTs.

Figure~\ref{fig:Spectrograms} shows the spectrograms of the data used for each of the searches presented here. From this figure we can observe that the noise levels in each detector appear to be lower at high frequencies. Strong noise lines can also be seen to contaminate the data at specific frequencies. These lines show evidence of \textit{spectral leakage}, where instrumental noise between frequency bins contaminates neighbouring bins. This is expected to be due to no windowing being applied to the SFTs used for this search.

\section{Post-Merger Remnant Searches} \label{sec:Application}

\subsection{Setup} \label{sec:Setup}

\begin{table}
    \begin{tabularx}{\columnwidth}{l@{\extracolsep{\fill}}l@{\extracolsep{\fill}}r}
        \hline \hline
        Parameter & Symbol & Value \\
        \hline
        Number of spin-downs & $S$ &  2 \\
        Minimum braking index & $n\umin$ & 2 \\
        Maximum braking index & $n\umax$ & 5 \\
        Minimum initial frequency & $f\umin$ & 100~Hz \\
        Maximum initial frequency & $f\umax$ & 2000~Hz \\
        Principal moment of inertia & $I_{zz}$ & $10^{38}$ $\text{kg m}^{2}$ \\
        Maximum ellipticity & $\epsilon$ & $10^{-4}$ \\
        Minimum $k$ value & $k\umin$ & $1.72 \times 10^{-20} \text{ s}^{3}$ \\
        Maximum $k$ value & $k\umax$ & $1.72 \times 10^{-19} \text{ s}^{3}$ \\
        Maximum mismatch & $\mu\umax$ & 0.2 \\
        Short Fourier Transform timebase & $T_{\text{SFT}}$ & 10~s  \\
        Knots & $p_{0}, p_{1}$ & 0, 1800~s  \\
        \hline \hline
    \end{tabularx}
    \caption{Values of the parameters used for the searches carried out in this work.}
    \label{tab:SearchParams}
\end{table}

\begin{table}
    \begin{tabularx}{\columnwidth}{l@{\extracolsep{\fill}}r@{\extracolsep{\fill}}r}
        \hline \hline
        & GW170817 & GW190425 \\
        \hline
        $t\ustart$ & 1187008882  &  1240215503 \\
        Right ascension & 13h 9m 46s & 16h 6m 42s \\
        Declination & $-23.38^{\circ}$ & $22.98^{\circ}$ \\
        \hline \hline
    \end{tabularx}
    \caption{Search parameters specific to each source. The parameter $t\ustart$ is the starting GPS time used for each search. The sources' sky positions are given by their right ascension and declination.}
    \label{tab:SearchSourceParams}
\end{table}

The common parameters of the searches are summarised in Table~\ref{tab:SearchParams}; the setup was studied in detail in~\cite{Grace2023}. Search parameters specific to each source are given in Table~\ref{tab:SearchSourceParams}.
The searches take place over a frequency band of 100--2000~Hz on the 1800~s of data following the GW170817 and GW190425 events.  We choose 1800~s of data for computational cost constraints and because a signal duration of this length following GW170817 has not previously been explored. The value $k\umax$ is determined using the values of $I_{zz}$ and $\epsilon$ substituted into Eq.~(34) of~\cite{Grace2023}. We use fiducial values for $I_{zz}$ and optimistic accepted values of $\epsilon$. The value $k\umin$ is chosen to be 10\% of $k\umax$. The value of $S = 2$ has been chosen as a value of $S = 3$ increases the computational cost of the search with no expected improvements to search sensitivity upper limits. A single piecewise segment is used since for the 1800~s of data multiple segments are not required, however, suggestions for multiple segment searches are made in ~\cite{Grace2023}. The value of $\mu\umax = 0.2$ is chosen as a commonly used value for continuous wave searches.

The sky position of GW170817 is taken as that of its host galaxy, NGC 4993, known precisely from its electromagnetic counterpart~\cite{Abbott2017a}. Despite two detectors being online at the time of GW190425, only the L1 detector was able to observe the event due to its greater sensitivity. Its sky localisation was therefore poor, and no electromagnetic counterpart was detected. For this search, we take the sky position of GW190425 to be its most likely location according to the sky map available from GWOSC~\cite{Abbott2023}. This choice is justified, as follows.

Unlike searches for binary black hole/neutron star systems, where the sky position is determined by triangulation of time delays between detectors, continuous wave search methods rely on the sky-position-dependent Doppler modulation of the signal phase due to the Earth's spin and orbit. When the time-span of the data being analysed is long ($\gg$~days), the motion of the Earth causes an appreciable Doppler modulation of the signal phase, and therefore it is important to know the sky position of the source to high accuracy. In this search we use 1800~s of data, however, which means that the phase of a potential signal will only have minimal Doppler modulation. Moreover, over such a short time-span, the sinusoidal Doppler phase modulation, when Taylor expanded in time, is well-approximated by linear and quadratic terms, and is therefore degenerate with the terms which arise from the frequency $f$ and spin-down $f^{(1)}$ phase evolution, respectively:
\begin{equation*}
\phi_{\mathrm{sky}} \sim \sin t + \cos t \sim 1 + t + t^2 \sim \phi_0 + \phi_{f} + \phi_{f^{(1)}} .
\end{equation*}
Any error between the chosen sky position used and the true sky position of GW190425 should therefore not affect the detectability of any signal, except that such a signal may be recovered with errors in the frequency and spin-down parameters.

\subsection{Search Jobs}

\begin{figure}
    \subfloat[]{\includegraphics[width=\columnwidth]{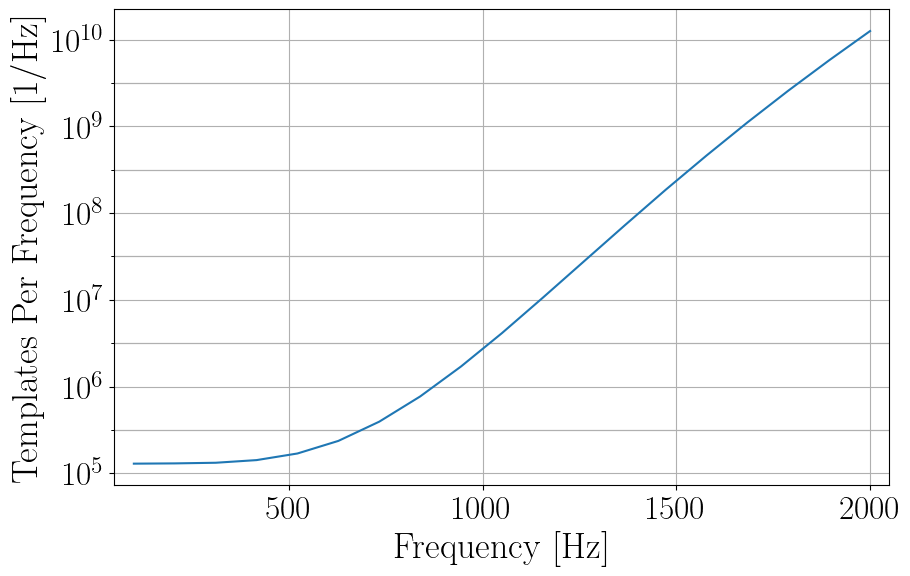}\label{fig:Temp_Density_First}} \\
    \subfloat[]{\includegraphics[width=\columnwidth]{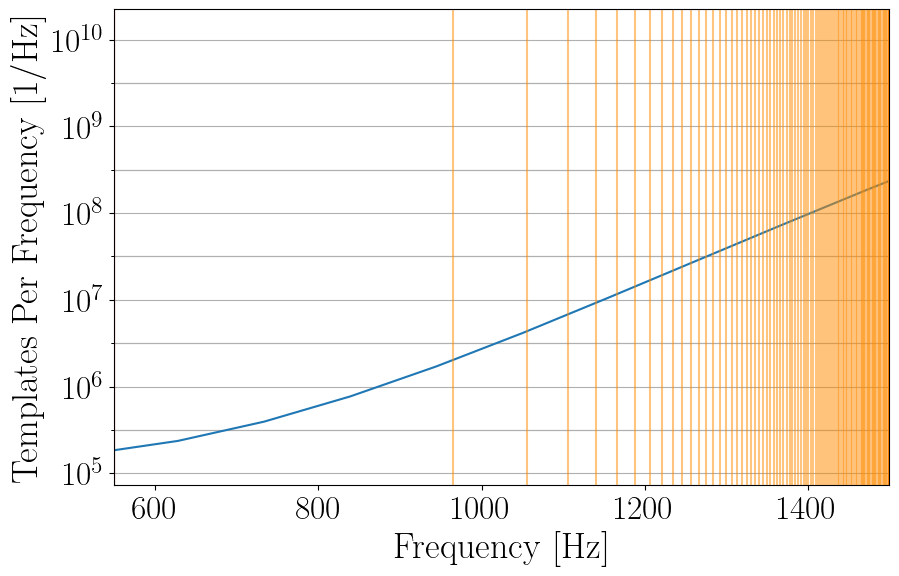}\label{fig:SearchBands}}
    \caption{\protect\subref{fig:Temp_Density_First} Number of templates per unit frequency across the search band 100--2000Hz. \protect\subref{fig:GW190425_line_veto} Individual search bands used for each search job for the 550--1500~Hz frequency range shown in orange, with the blue line representing the same as that in \protect\subref{fig:Temp_Density_First}.}
    \label{fig:Temp_Density}
\end{figure}

To carry out the search, the frequency range of 100--2000~Hz is split into four sub-bands; 100--550~Hz, 550--1500~Hz, 1500--1990~Hz and 1990--2000~Hz. Each of these sub-bands is then further divided into smaller frequency ranges to make up individual search jobs. Using these sub-bands is necessary in order to request the appropriate computational resources and for minimising the time to carry out the search. The density of templates throughout the parameter space changes with $f\umax$, as shown in Fig.~\ref{fig:Temp_Density}. Consequently, the search setup also changes with frequency. At frequencies below 550~Hz, the parameter space is extremely narrow. This leads to a significant portion of the parameter space, at its boundaries, being unaccounted for by templates. To remedy this, template \textit{padding} is applied at the boundaries of the parameter space~\cite{Grace2023}. Conversely, the density of templates progressively increases in the frequency bands 550--1500~Hz, 1500--1990~Hz and 1990--2000~Hz.

The density of templates per unit frequency is determined by finding small frequency ranges across the 100--2000~Hz band which give template banks of sizes of approximately $10^{6}$. These small frequency ranges were chosen to have their upper frequency limit at 100~Hz intervals beginning at 100~Hz. The lower frequency limit was then found by manual selection until a value which gave a template bank size of $10^{6}$ was found. The density of templates was then determined by dividing the size of each of these small template banks by the width of its associated frequency band. Linear interpolation between these results then achieved an estimate of the number of templates per unit frequency across the full 100--2000~Hz range~\cite{Grace2023}. Figure~\ref{fig:Temp_Density} presents this template density.

The rapid increase in template density with frequency seen in Figure~\ref{fig:Temp_Density} is due to the GTE. Each time derivative of the GTE introduces a factor of $f_{0}^{n - 1}$. Each derivative of the GTE is then proportional to $f_{0}$, and hence $f\umax$, by; GTE $\propto f\umax$, GTE$'$ $\propto f\umax^{n}$ and GTE$''$ $\propto f\umax^{2n - 1}$. As these form the boundaries of the parameter space, Eqs.\eqref{eq:f00_Condition}--\eqref{eq:fi2_Condition}, the size of the parameter space then scales strongly with $f\umax$. This increase in parameter space size, even for a fixed frequency band, results in the strong frequency scaling of template density seen in Fig.~\ref{fig:Temp_Density}.

Using Fig.~\ref{fig:Temp_Density_First}, each of the frequency ranges 100--550~Hz, 550--1500~Hz, 1500--1990~Hz and 1990--2000~Hz is divided into smaller ranges for each search job, each with an equal number of templates. To get the frequency range of each search job, the curve given in Fig.~\ref{fig:Temp_Density} is integrated from the start of the frequency sub-band (e.g.\ 550~Hz for the 550--1500~Hz band) up to a frequency which gives the desired number of templates for the given job. This sets the frequency range of the first search job. By then repeating this process from the final frequency of the previous job, a frequency range for each job is calculated, with each job containing approximately the same number of templates. An example of how the frequency range has been divided for the 550--1500~Hz band is shown in Fig.~\ref{fig:SearchBands}. The number of templates and estimated computational cost of each job in each frequency sub-band is shown in Table~\ref{tab:SearchJobSpecifics}.

\begin{table}
    \begin{tabularx}{\columnwidth}{r@{\extracolsep{\fill}}r@{\extracolsep{\fill}}r@{\extracolsep{\fill}}r}
        \hline \hline
        Sub-Band~(Hz) & Jobs & Templates & Cost~(hours) \\
        \hline
        100--550 &   5 & $1.26 \times 10^{7}$ & 0.14\\
        550--1500 & 100 & $2.90 \times 10^{8}$ & 3 \\
        1500--1990 & 500 & $4.32 \times 10^{9}$ & 36 \\
        1990--2000 &  50 & $6.14 \times 10^{9}$ & 66 \\
        \hline \hline
    \end{tabularx}
    \caption{Estimated number of templates and computational cost of each search job in the different frequency sub-bands.}
    \label{tab:SearchJobSpecifics}
\end{table}

The total number of templates is calculated by integrating the curve given in Fig.~\ref{fig:Temp_Density}. Once known, an acceptable computational cost for each job was chosen. The search was performed on the OzSTAR supercomputing cluster, for which it was known that approximately $2.5 \times 10^{4}$ templates per second could be computed~\cite{Grace2023}. Using this number, the computational cost of a search job is estimated from its frequency range and template bank size.

One might consider why in Table~\ref{tab:SearchJobSpecifics} we have split the 655 search jobs unevenly across four frequency sub-bands rather than use 655 search jobs all of equal computational cost. The search jobs have been split this way due to the higher density of templates at large frequencies, as demonstrated in Fig.~\ref{fig:Temp_Density_First}. To have 655 search jobs each of equal computational cost, one search job would cover a frequency band of 100--1242~Hz, calculated using data from Fig.~\ref{fig:Temp_Density_First}. Such a wide band is likely to contain instrumental lines which will saturate the loudest recorded $2\cF$s. This problem is alleviated by purposefully splitting jobs into smaller frequency bands. Furthermore, at high frequencies to split jobs equally by computational cost, the search bands will be forced to become extremely narrow. This narrowing would force that padding be required, which would unnecessarily increase the computational cost of carrying out these jobs. The total number of CPU hours used to carry out the search can then be further reduced by using fewer jobs at high frequencies which cover wider frequency bands. Furthermore, search jobs cannot be divided to have equal frequency as this would make the computational cost of jobs at high frequencies impractical.

In Table~\ref{tab:SearchJobSpecifics} the number of templates, and corresponding computational cost, per individual job differs between each frequency sub-band. The frequency band 100--550~Hz must be separated from the other sub-bands due to template bank padding being required at frequencies below 550~Hz~\cite{Grace2023}. As the density of templates is lower in this region, the search jobs in this band are considerably cheaper in computational cost than those in the other frequency bands. The 100--550~Hz band is still split across 5 search jobs, despite their lower computational cost. This is to reduce the impact of instrumental lines saturating the loudest recorded $2\cF$s. This is discussed in Section~\ref{sec:Recording_Results}. The remaining sub-bands are separated from one another due to the changing shape of the parameter space. If they were treated as a single sub-band, such that each search job from this sub-band had equal computational cost, the frequency range of each job would decrease as higher frequencies are reached. For the high frequency jobs, their frequency bands would narrow sufficiently such that they would require padding. This would then increase the computational cost of the search. Furthermore, the additional padding templates used by each search job would overlap in frequency, leading to double counting of templates at similar frequencies. This would again unnecessarily increase the computational cost of the search.

\subsection{Recording of Results} \label{sec:Recording_Results}

For each job, the largest $1,000$ multi-detector $\cF$-statistics were recorded, alongside their associated templates, and the corresponding individual detector $\cF$-statistics. With 655 search jobs, a total of 655,000 templates and their corresponding $\cF$-statistics were recorded.

At low frequencies, the frequency bands of each search job are wide, covering many tens of Hz. Frequency bands of these widths are likely to encompass numerous instrumental artefacts, or lines, arising from noise sources in the detectors. If a strong instrumental line occurs within a search job band, templates with frequencies coinciding with the line will likely have very large $\cF$-statistics. Given that a list of only $1000$ $2\cF$s are stored from each job, it is possible that the templates coincident with the instrumental line will saturate this list; no templates at other frequencies in the search band will then have a $2\cF$ value large enough to be recorded in the top 1000 templates, and any gravitational wave signal may be missed.

To remedy this behaviour, additional search jobs were run to cover the frequency ranges where no $2\cF$s were recorded. To cover these frequency ranges, 7 additional searches jobs for GW170817 and 42 search jobs for GW190425 were used. For brevity we will continue to state that 655 search jobs and 655,000 $2\cF$s were recorded despite these additional jobs. By including results from these additional jobs we achieve $2\cF$ values across the entire frequency range. These results are shown in Fig.~\ref{fig:Fstat_Line_Vetos}.

\section{Candidate Post-Processing} \label{sec:Results}

\subsection{Line Veto} \label{sec:line_vetoing}

\begin{figure}
    \subfloat[]{\includegraphics[width=\columnwidth]{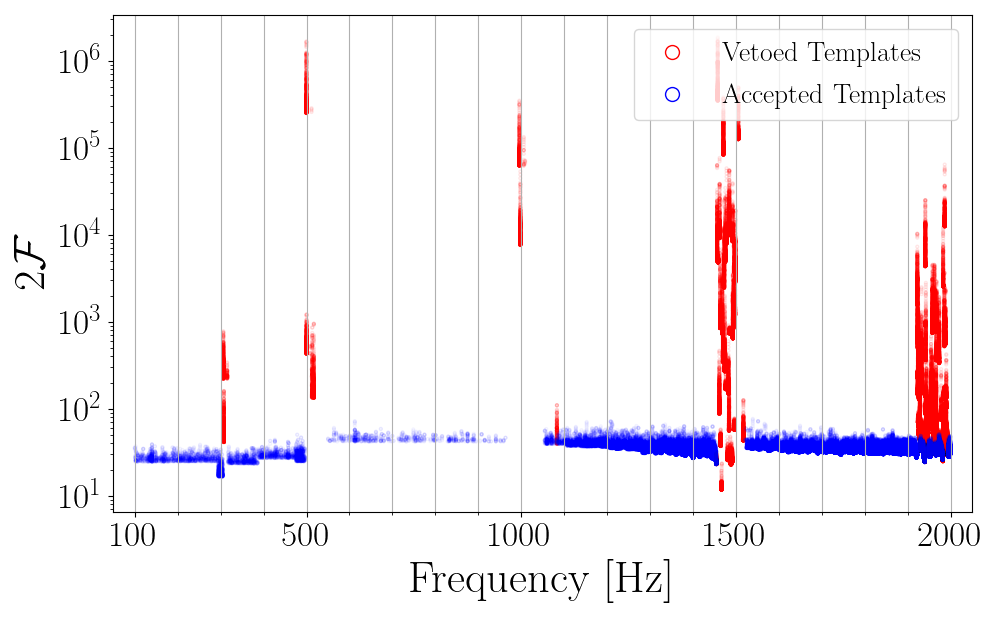}\label{fig:GW170817_line_veto}} \\
    \subfloat[]{\includegraphics[width=\columnwidth]{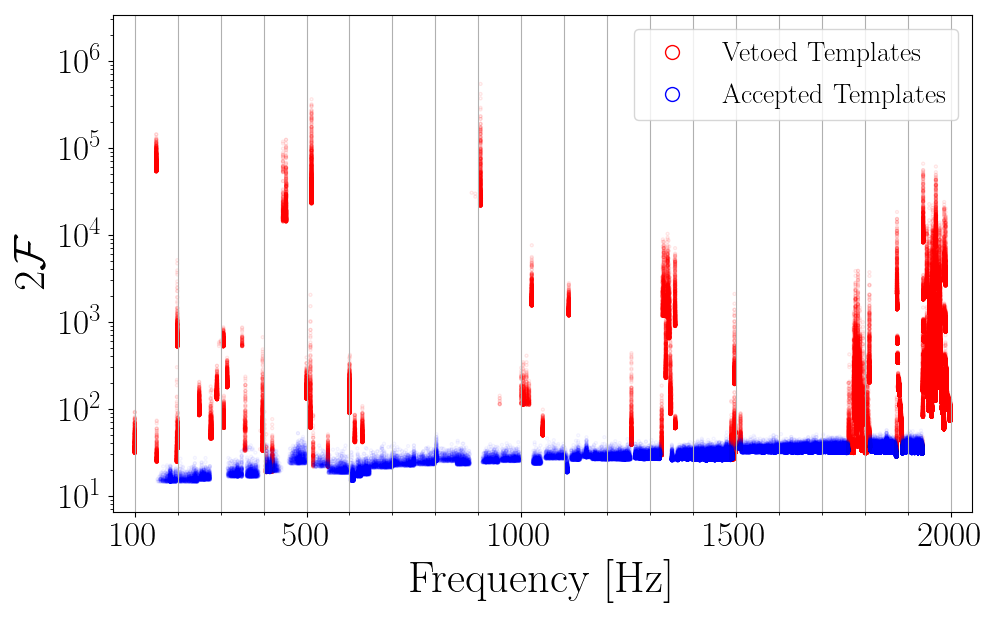}\label{fig:GW190425_line_veto}}
    \caption{Top 1,000 $\cF$-statistics recorded for each search job. The $2\cF$s are plotted against the frequency parameters $\mf{0}{0}$ for its associated template. Figure~\protect\subref{fig:GW170817_line_veto} and Figure~\protect\subref{fig:GW190425_line_veto} show the results for the GW170817 and GW190425 searches respectively.}
    \label{fig:Fstat_Line_Vetos}
\end{figure}

Figure~\ref{fig:Fstat_Line_Vetos} presents the top 1,000 candidates from each search. Through visual inspection of Fig.~\ref{fig:Fstat_Line_Vetos}, groups of large-$2\cF$ candidates are identified; the central frequencies of these groups are then cross-examined with known line lists from the second (O2) and third (O3) observing runs for GW170817 and GW190425, respectively~\cite{Covas2018, GoetzO3Lines}. Groups of large-$2\cF$ candidates which coincide with known lines are then vetoed. In Fig.~\ref{fig:Fstat_Line_Vetos} all vetoed candidates are shown in red, and the remaining potential candidates in blue. The identified lines used for vetoing are listed in Tables~\ref{tab:GW170817Lines} and~\ref{tab:GW190425Lines} in the Appendix.

\subsection{Detector \texorpdfstring{$2\cF$}{2F} Veto} \label{sec:Detector2FVetoing}

Another candidate veto commonly used in continuous wave searches is a detector $2\cF$ veto~\cite{LIGOVirg2013-EnsAlSrPrGrvWLSD,LIGOVirg2013-DrSrCntGrvWGlC, Leaci2015}. For this veto, the multi-detector $\cF$-statistic must be greater in value than the individual detector $2\cF$s for a candidate to pass. It is applied to ensure that a signal is present in both detectors. If a candidate has a large $2\cF$ in only one detector, it is likely to originate from a local source of noise.

All candidates presented in Fig.~\ref{fig:Fstat_Line_Vetos} pass the detector $2\cF$ veto, however. This illustrates a limitation of the detector $2\cF$ veto when applied to the searches presented in this work.
First, the 1800~s of data used in these searches is much shorter than typical continuous wave searches (which span weeks to years of data). Over such long time spans, a continuous wave signal will be Doppler modulated by the motion of the Earth. Instrumental lines, however, are not affected by the movement of the Earth, and can therefore be more readily identified. The 1800~s used in these searches, however, is not long enough to observe any appreciable Doppler modulation, and thus weak instrumental lines are less well distinguished from possible signals.

\begin{figure*}
    \subfloat[100--1000~Hz]{\includegraphics[width=0.33\textwidth]{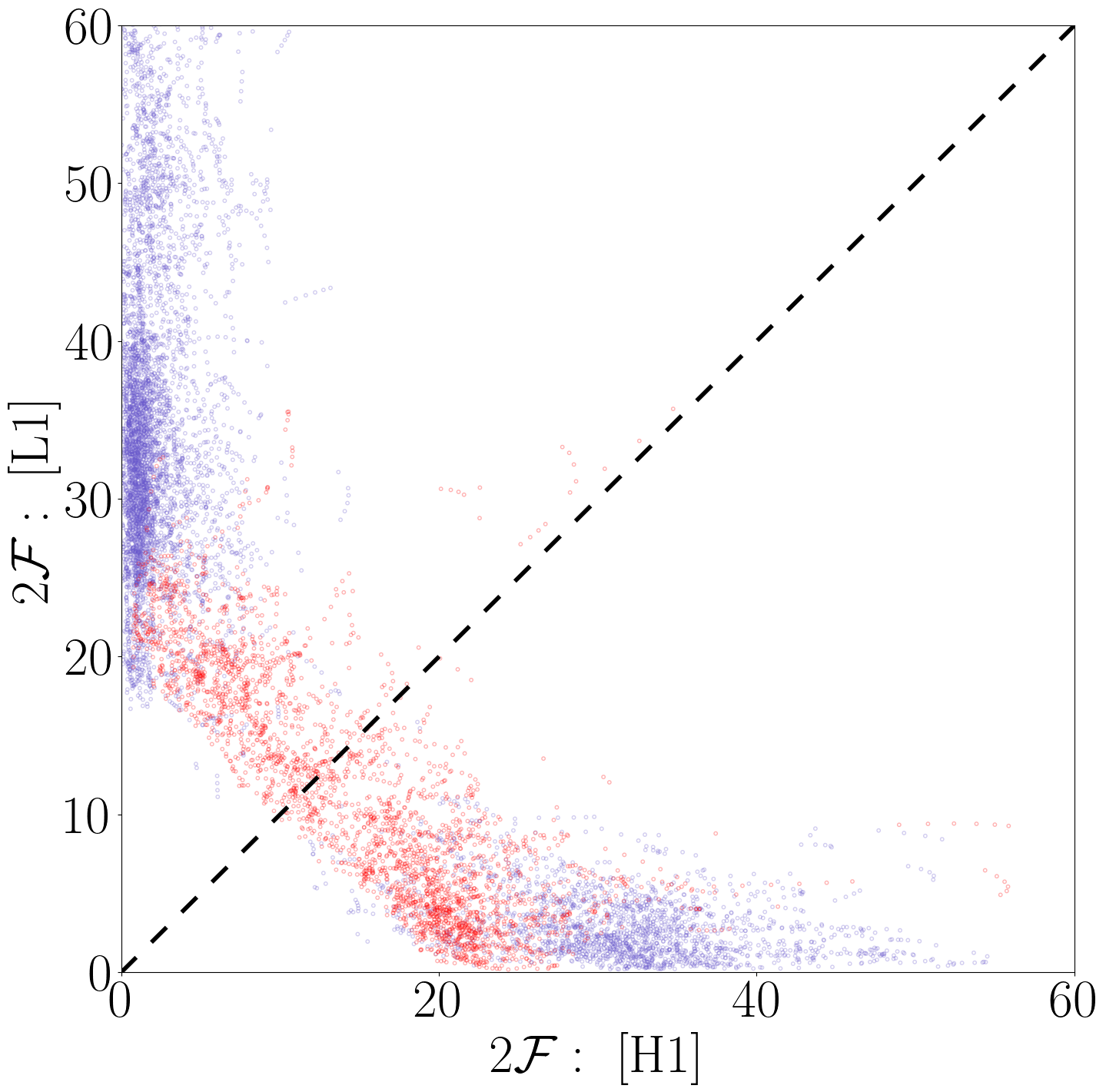}}
    \subfloat[1000--1200~Hz]{\includegraphics[width=0.33\textwidth]{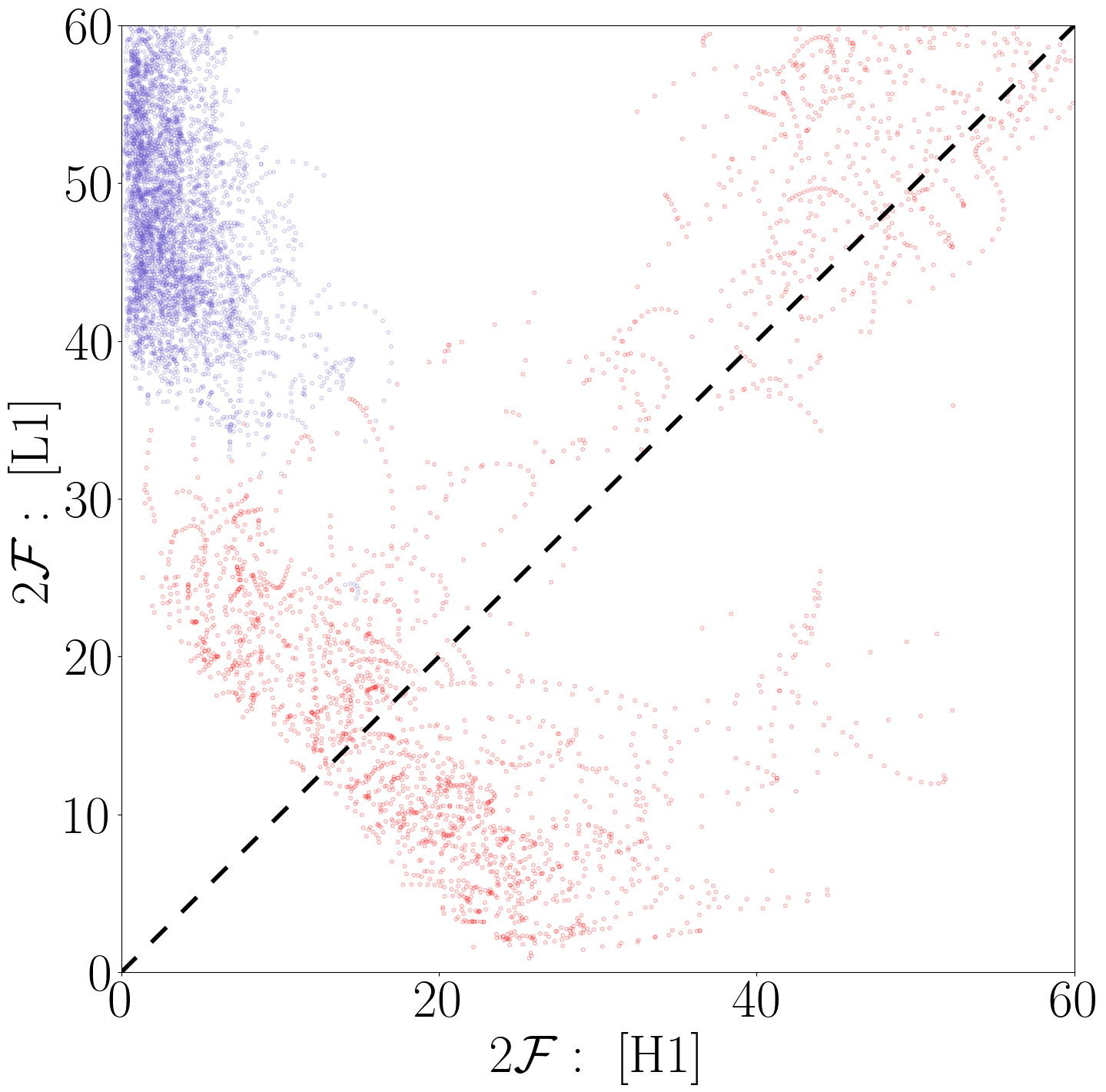}}
    \subfloat[1700--1750~Hz]{\includegraphics[width=0.33\textwidth]{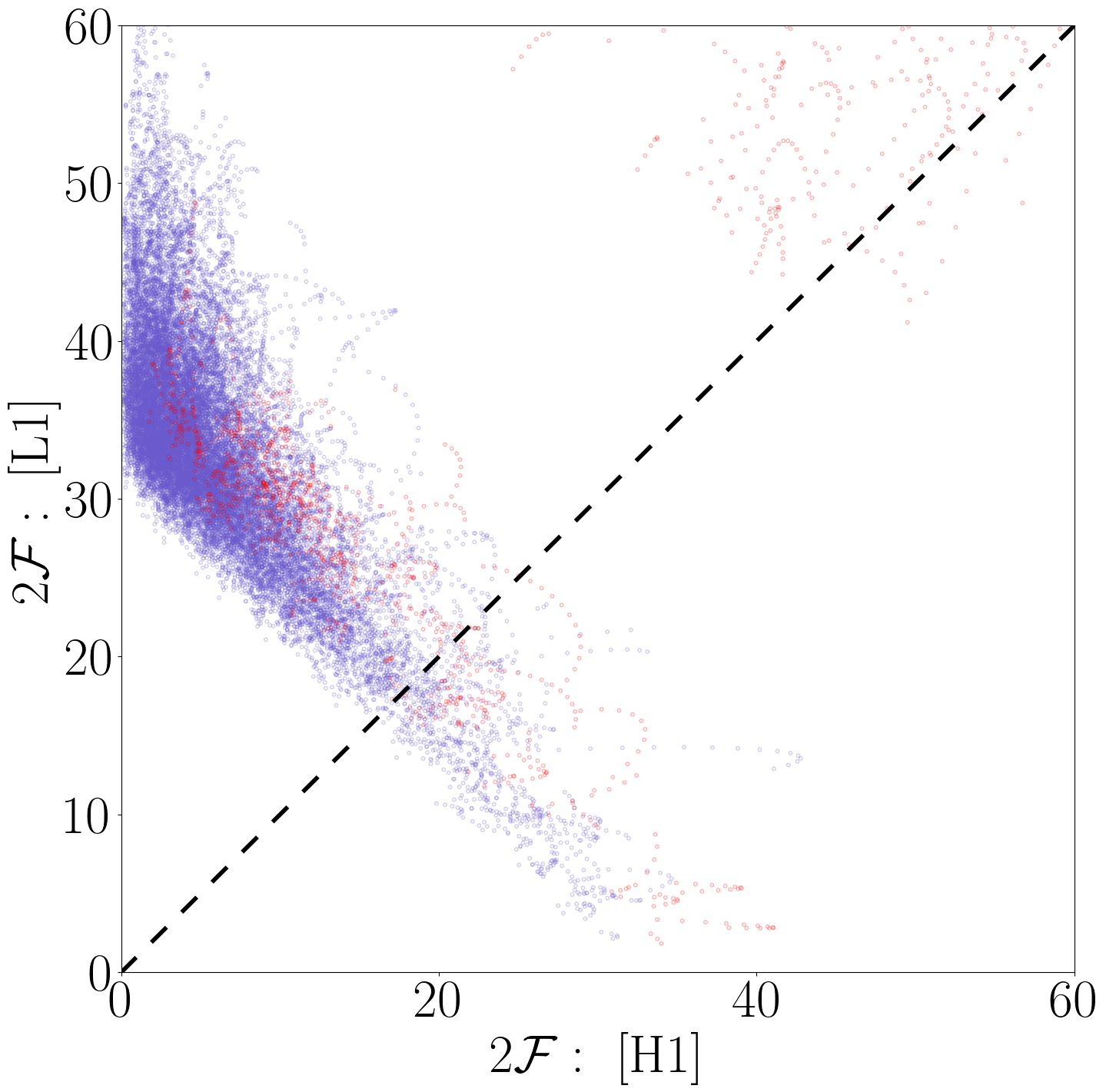}}
    \caption{Single detector $2\cF$s plotted against one another for the GW170817 search. The blue dots are the single detector $\cF$-statistics calculated during the search. These single detector $2\cF$s are calculated from the same templates which give the multi-detector $2\cF$s inFig.~\ref{fig:Fstat_Line_Vetos}. The red dots are $\cF$-statistics calculated from carrying out searches on simulated data for smaller template banks with frequency bands encompassed by that of each plot. The simulated data has signals injected with a strength corresponding to the peak $h\urss^{50\%}$ upper limit presented in Sec.~\ref{sec:Sensitivity}, $h_{0} = 1.21 \times 10^{-23}$.}
    \label{fig:GW170817_2F_vs_2F_Figures}
\end{figure*}

\begin{figure*}
    \subfloat[200--300~Hz]{\includegraphics[width=0.33\textwidth]{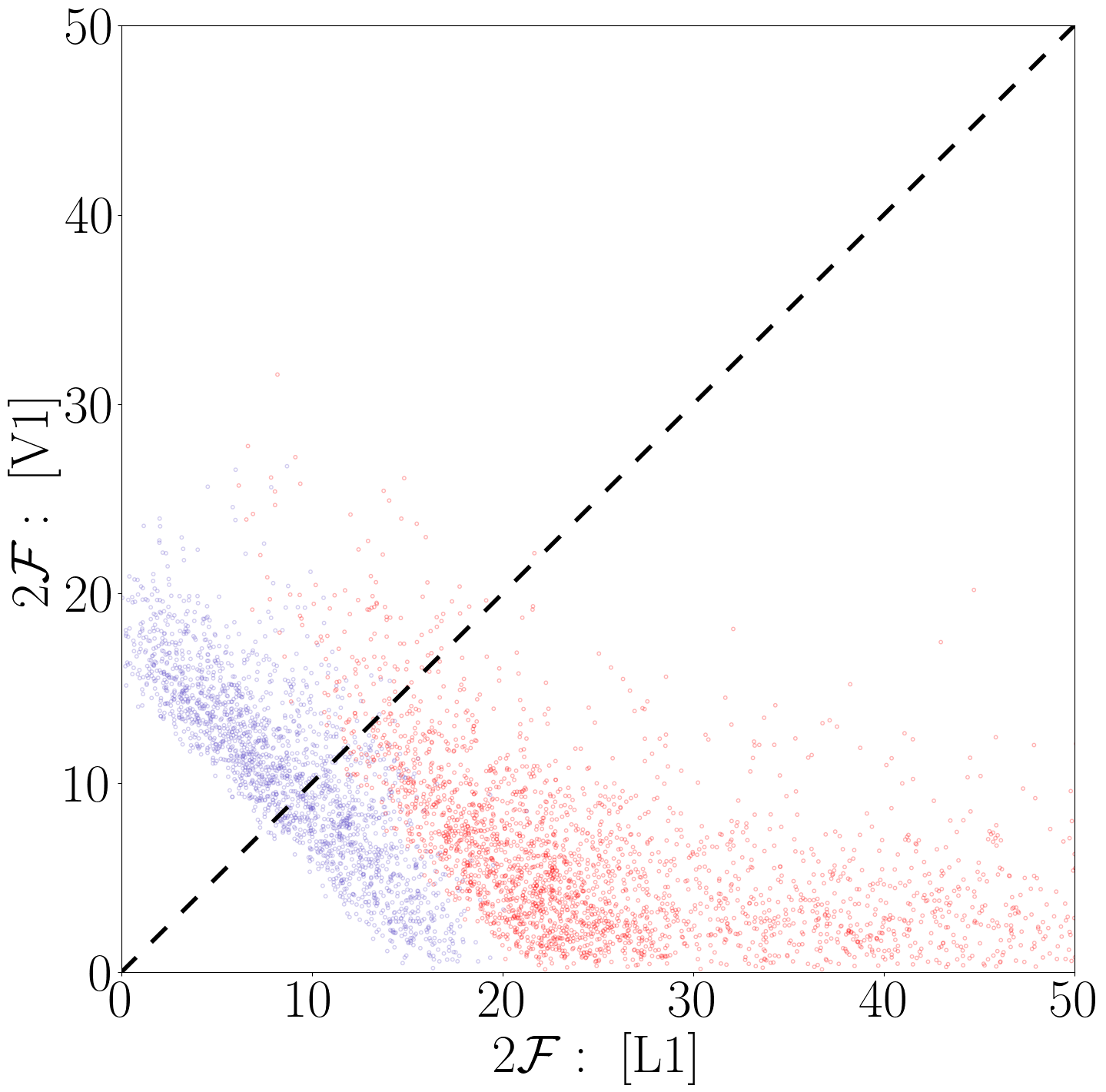}\label{fig:GW190425_200_300_Hz}}
    \subfloat[400--500~Hz]{\includegraphics[width=0.33\textwidth]{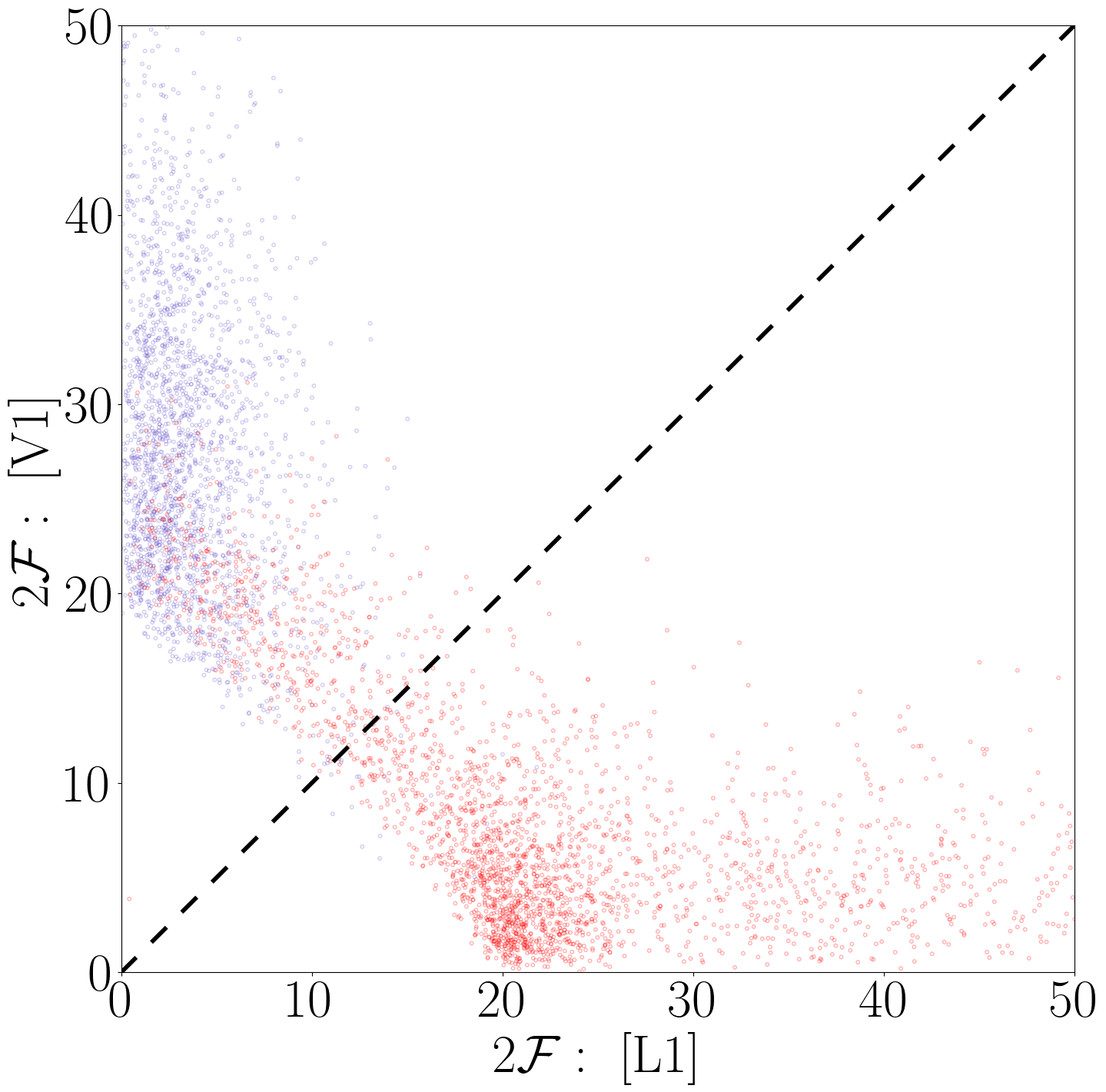}}
    \subfloat[1700--1800~Hz]{\includegraphics[width=0.33\textwidth]{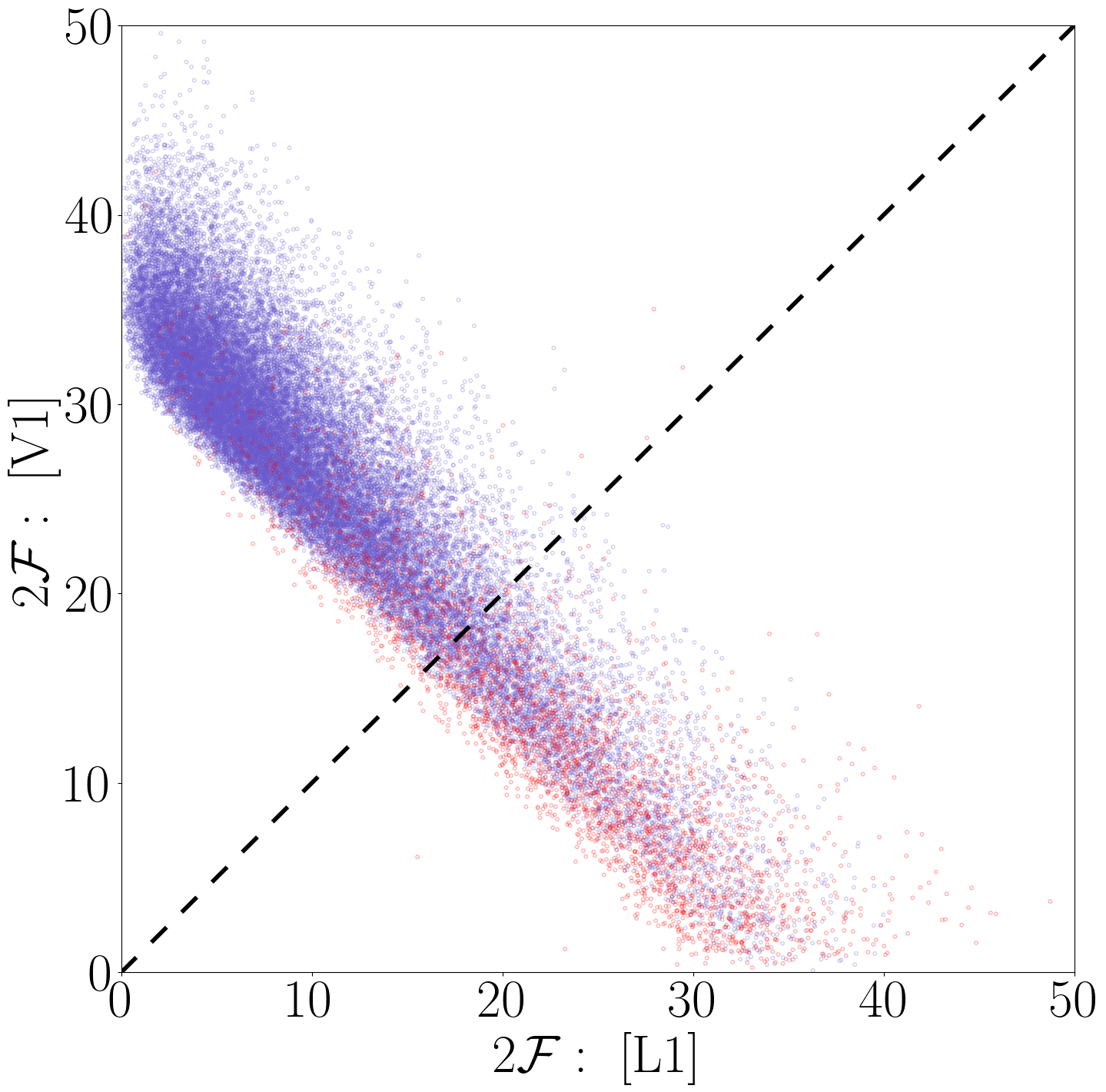}\label{fig:GW190425_1700_1800_Hz}}
    \caption{Single detector $2\cF$s plotted against one another for the GW190425 search. The blue dots are the $\cF$-statistics computed from the search and are the same as those presented in Fig.~\ref{fig:Fstat_Line_Vetos}. The red dots are $\cF$-statistics calculated from carrying out searches on simulated data for smaller template banks with frequency bands encompassed by that of each plot. The simulated data has signals injected with a strength corresponding to the peak $h\urss^{50\%}$ upper limit presented in Sec.~\ref{sec:Sensitivity}, $h_{0} = 7.4 \times 10^{-24}$.}
    \label{fig:GW190425_2F_vs_2F_Figures}
\end{figure*}

Second, we observe that, in the presence of weak lines, it is possible that the multi-detector $2\cF$ may still be greater than the contributing single detector $2\cF$s, and thereby a weak line may pass the detector $2\cF$ veto.
This behaviour is demonstrated in Figs.~\ref{fig:GW170817_2F_vs_2F_Figures} and~\ref{fig:GW190425_2F_vs_2F_Figures}, where the single detector $\cF$-statistics of the largest $2\cF$s presented in Fig.~\ref{fig:Fstat_Line_Vetos} are plotted against each other. We see that one detector typically has $\cF$-statistics which are much larger than the other. This suggests that weak lines present in that detector are contributing to the multi-detector $\cF$-statistic, while still passing the detector $2\cF$ veto. In the bottom left of each plot a cut-off in $2\cF$ values can be seen. The presented $2\cF$s are only the largest 1000 $2\cF$s from each search job; if a greater number of $2\cF$s were recorded, this cut-off would occur at lower values of the $\cF$-statistic.

We wish to compare the behaviour seen in Figs.~\ref{fig:GW170817_2F_vs_2F_Figures} and~\ref{fig:GW190425_2F_vs_2F_Figures} with that expected from an astrophysical signal. To accomplish this, we carry out searches on simulated data with an injected signal, and plot the single detector $\cF$-statistics of the top 1000 largest candidates. Each subplot shows the results from three smaller searches, with the same setup as that presented in Table~\ref{tab:SearchParams}, and different frequency ranges given in Table~\ref{tab:SimulatedSearchBands}. The simulated data contains Gaussian noise with a power spectral density (PSD) equal to the experimental data for the same source (GW170817 or GW190425) at the same frequency. Signal phase parameters are sampled at random from within the piecewise parameter space. The signal strength, $h_{0}$, is the peak strain upper limit used to determine $h\urss^{50\%}$ in Sec.~\ref{sec:Sensitivity}. This is $1.21 \times 10^{-23}~\text{Hz}^{-1/2}$ for GW170817 and $7.4 \times 10^{-24}~\text{Hz}^{-1/2}$ for GW190425.

\begin{table}
    \begin{tabularx}{\columnwidth}{ r@{.}l @{\extracolsep{\fill}} r@{\extracolsep{0pt}.}l @{\extracolsep{\fill}} r@{\extracolsep{0pt}.}l @{\extracolsep{\fill}} r@{\extracolsep{0pt}.}l }
        \hline \hline
        \multicolumn{4}{c}{GW170817} & \multicolumn{4}{c}{GW190425}  \\
        \multicolumn{2}{c}{$f\umin$} & \multicolumn{2}{c}{$f\umax$} & \multicolumn{2}{c}{$f\umin$} & \multicolumn{2}{c}{$f\umax$}  \\
        \hline
        92&0&  100&0&        192&0&  200&0 \\
        493&0&  500&0&       242&0&  250&0 \\
        999&5&  1000&0&       292&0&  300&0 \\
        1099&0&  1100&0&       392&0&  400&0 \\
        1149&0&  1150&0&       443&0&  450&0 \\
        1199&0&  1200&0&       493&0&  500&0 \\
        1699&95&  1700&0&       1699&95&  1700&0 \\
        1724&94&  1725&0&       1749&95&  1750&0 \\
        1749&95&  1750&0&       1799&95&  1800&0 \\
        \hline \hline
    \end{tabularx}
    \caption{The frequency bands of the simulated searches used for plotting the single detector $\cF$-statistics in Fig.~\ref{fig:GW170817_2F_vs_2F_Figures} for GW170817 and Fig.~\ref{fig:GW190425_2F_vs_2F_Figures} for GW190425.}
    \label{tab:SimulatedSearchBands}
\end{table}

In Fig.~\ref{fig:GW170817_2F_vs_2F_Figures} and~Fig.~\ref{fig:GW190425_2F_vs_2F_Figures}, we see that the largest single detector $2\cF$s from the simulated searches (red dots) do not overlap with the search results. For GW170817 the simulated largest $2\cF$s are generally equal in value, and do not favour one detector over another. For GW190425 the largest $2\cF$s from the simulated searches are generally louder in the L1 detector; this is to be expected because this detector is more sensitive than V1. If the candidates which passed the detector $2\cF$ veto (blue dots) were from an astrophysical signal, we would expect their distributions of single detector $2\cF$s to show similar behaviour; this is not the case, however.

In summary, the inconsistency between the single detector $2\cF$s from the GW170817 and GW190425 searches, and from the simulated searches, suggests that the experimental data contains weak instrumental lines which are the primary cause of the candidates found by the GW170817 and GW190425 searches. In the absence of a ready-made veto capable of vetoing these lines, we instead accept them as part of the background noise in the data, with the consequence that this will likely degrade the upper limits achievable by the searches (see Sec.~\ref{sec:Sensitivity}).

\subsection{Distromax}

We now look to quantify the statistical significance of the candidates found by the searches. To do this, we use the Distromax method~\cite{Tenorio2022, distromax}. Distromax gives the likelihood that the largest value of a detection statistic -- in our case, the $\cF$-statistic -- found by a search is statistically significant. This is achieved by estimating the expected distribution of the largest value of the detection statistic, assuming that no signal is present, and comparing it to the largest value found by the search. If the largest value is consistent with the distribution, we may conclude that no statistically significant candidate was found.

Distromax operates by taking in a set of detection statistics $2\cF$. This set of $2\cF$ is then randomly split into $N$ batches, each containing the same number of $2\cF$ values; this is to avoid potential correlations between $2\cF$ values in nearby regions of parameter space. Let the maxima of each of these batches be labelled $\xi^{*}_{i}$, where $i \in [1, N]$. Let us define $\xi^{*} = \max_{i} \xi^{*}_{i}$, the loudest overall $\xi^{*}$. The probability density function for $\xi^{*}$ is given by
\begin{align}
    p(\xi^{*}) &= N P(\xi^{*})\left[\int_{0}^{\xi^{*}}d\xi P(\xi)\right]^{N - 1}, \label{eq:PDFXi}
\end{align}
where $P(\xi)$ is the probability distribution function for the largest value of the detection statistic in each batch. In the case of the $\cF$-statistic, which follows a $\chi^{2}$ distribution with 4 degrees of freedom, $P(\xi)$ converges to a Gumbel distribution:
\begin{align}
    P(\xi) &= \frac{1}{\sigma}\exp\left[-\frac{\xi - \mu}{\sigma} - \exp\left(-\frac{\xi - \mu}{\sigma}\right) \right]. \label{eq:Gumbel}
\end{align}
The $\mu$ and $\sigma$ of Eq.~\eqref{eq:Gumbel} are the location and scale of the Gumbel distribution. Their values are determined by constructing a histogram of the largest $2\cF$ from each batch, and then fitting $P(\xi; \mu, \sigma)$ to the histogram. The fitted $P(\xi; \mu, \sigma)$ is then substituted into Eq.~\eqref{eq:PDFXi} to give the probability of the largest $2\cF$ from the entire search. A property of the Gumbel distribution (see~\cite{Tenorio2022} for details) is that, given $P(\xi)$ is a Gumbel distribution, $p(\xi^{*})$ is also a Gumbel distribution, with location $\mu_{*}$ and scale $\sigma_{*}$ given by
\begin{align}
    \mu_{*}  &= \mu + \sigma \ln{N} \\
    \sigma_{*} &= \sigma.
\end{align}
This change in variables effectively propagates the Gumbel distribution $P(\xi)$ along the $2\cF$ axis, and for this reason we refer to it as the ``propagated distribution''.

We use the distribution $p(\xi^{*})$ produced by Distromax to assess the statistical significance of candidates from the searches. If a $2\cF$ falls within a region of $p(\xi^{*})$ which is considered likely, then this candidate is consistent with noise and can therefore be rejected as a potential gravitational wave signal. Alternatively, if a $2\cF$ is found with a value far larger than what is predicted by $p(\xi^{*})$, then that would be considered evidence for a gravitational wave detection.

Conventionally, we would supply Distromax with the set of largest $2\cF$s from each search job. The search setup used in this work, however, only required 655 jobs (Table~\ref{tab:SearchJobSpecifics}), and that number of largest $2\cF$s is too few samples to produce reliable results with Distromax~(as discussed in~\cite{Tenorio2022}). To increase the number of samples, we use all of the recorded $2\cF$ values from each search job (after applying the line veto in Sec.~\ref{sec:line_vetoing}), instead of just the largest. Using the template density information from Fig.~\ref{fig:Temp_Density_First}, we re-partition the entire search frequency band into sub-bands containing approximately the same number of $2\cF$ values. The largest $2\cF$ from each sub-band was then selected and supplied to Distromax. In this way, the number of $2\cF$ samples was increased from 655 to 56,192 for GW190425, and 20,596 for GW170817.

For both GW170817 and GW190425 we use a total of $N = 400$ batches. To use Distromax we use the example Python notebooks provided in~\cite{distromax}. There is a trade off in choosing the appropriate value for $N$. A small $N$ results in fewer maxima $\xi^{*}_{i}$, however each contain a larger number of samples. These batches then are well modelled by a Gumbel distribution due to the large number of samples they contain, however with fewer $\xi^{*}_{i}$ the fitting parameters of Eq.~\eqref{eq:Gumbel} have a higher variance. On the other hand, a large value of $N$ provides a larger number of maxima $\xi^{*}_{i}$ which are not well modelled by Gumbel distributions but the fitting parameters of Eq.~\eqref{eq:Gumbel} are more precise. For a more detailed discussion of choosing the batch number $N$, see~\cite{Tenorio2022}. We have used a value for $N$ which has produced histograms which can be well modelled as Gumbel distributions without impacting the variance of the fitting parameters of~\cite{Tenorio2022} significantly.

\begin{figure*}
    \includegraphics[width=\textwidth]{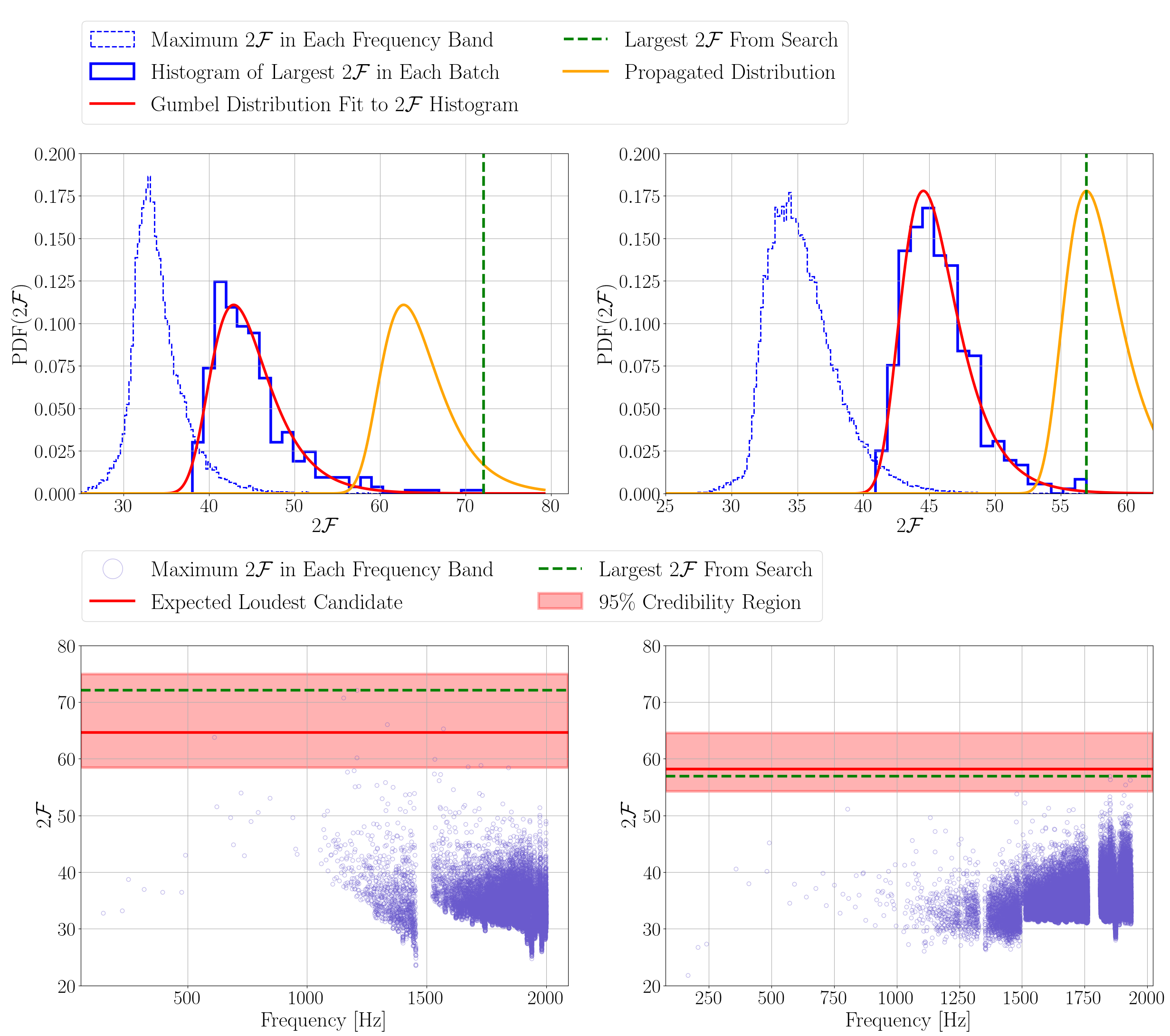}
    \caption{The probability distribution plots produced by Distromax. The left column of figures are for the GW170817 search, and the right column of figures are for the GW190425 search. The top row presents the Gumbel distributions fitted to the largest $2\cF$ histograms. The dashed blue line in the top row is a histogram of the maximum $2\cF$s from each frequency band used by Distromax. These frequency bands are those bands which have been re-partitioned to have approximately equal number of $2\cF$ values. The red Gumbel distribution is fitted to the largest $2\cF$ histograms and the yellow propagated distribution gives us the likelihood of the largest $2\cF$ from our search. The bottom row presents the 95\% confidence interval of the propagated distributions. The dashed green lines in all plots represent the largest $2\cF$s found from each search. They can be seen to sit well within the expected probability distribution for largest $2\cF$s from a search, especially for the case of GW190425 where the largest $2\cF$ coincides with peak of the distribution. In the bottom row it can be seen that these values fall within the 95\% confidence interval of the propagated distributions. This suggests that the loudest $2\cF$s found from each search are consistent with noise.}
    \label{fig:Distromax_Figures}
\end{figure*}

Figure~\ref{fig:Distromax_Figures} plots the largest $2\cF$s from the GW170817 and GW190425 search results presented in Fig.~\ref{fig:Fstat_Line_Vetos}, alongside their respective probability distributions calculated by Distromax.
For both searches, we see that the largest $2\cF$ found falls within the 95\% credible region predicted by Distromax. We conclude that the GW170817 and GW190425 search results are consistent with noise, and therefore no gravitational wave signal has been detected.

In the lower panel of Fig.~\ref{fig:Distromax_Figures}, all $\cF$-statistics used as samples by Distromax are plotted as a function of template frequency. We see that, by partitioning all $\cF$-statistics into frequency bands of similar template bank sizes, that our sample $2\cF$s are dominated by templates at higher frequencies. This is expected, as the density of templates per unit frequency increases with frequency (Fig.~\ref{fig:Temp_Density}).

\section{Sensitivity} \label{sec:Sensitivity}

\begin{table}
    \begin{tabularx}{\columnwidth}{ r@{.}l @{\extracolsep{\fill}} r@{\extracolsep{0pt}.}l @{\extracolsep{\fill}} r@{\extracolsep{0pt}.}l @{\extracolsep{\fill}} r@{\extracolsep{0pt}.}l }
        \hline \hline
        \multicolumn{4}{c}{GW170817} & \multicolumn{4}{c}{GW190425}  \\
        \multicolumn{2}{c}{$f\umin$} & \multicolumn{2}{c}{$f\umax$} & \multicolumn{2}{c}{$f\umin$} & \multicolumn{2}{c}{$f\umax$}  \\
        \hline
        92&0&  100&0&        102&0&  110&0 \\
        192&0&  200&0&       188&0&  196&0 \\
        292&0&  300&0&       267&0&  275&0 \\
        392&0&  400&0&       392&0&  400&0 \\
        490&0&  497&0&       520&0&  527&0 \\
        592&0&  597&0&       593&0&  598&0 \\
        692&0&  695&0&       697&0&  700&0 \\
        798&0&  800&0&       798&0&  800&0 \\
        879&0&  880&0&       910&0&  911&0 \\
        1019&5&  1020&0&   1028&5&  1029&0 \\
        1098&0&  1099&0&     1099&0&  1100&0 \\
        1201&0&  1202&0&     1199&0&  1200&0 \\
        1299&9&  1300&0&   1299&9&  1300&0 \\
        1399&9&  1400&0&   1399&9&  1400&0 \\
        1524&9&  1525&0&   1504&9&  1505&0 \\
        1599&95&  1600&0&  1599&95&  1600&0 \\
        1699&95&  1700&0&  1699&95&  1700&0 \\
        1799&95&  1800&0&  1799&95&  1800&0 \\
        1899&95&  1900&0&  1899&95&  1900&0 \\
        1999&95&  2000&0&  \multicolumn{4}{c}{ -- } \\
        \hline \hline
    \end{tabularx}
    \caption{Frequency ranges over which the upper limit estimates for both searches were calculated. No estimate near 2000~Hz was achieved for GW190425 as all results $>1935$~Hz were vetoed due to noise contamination.}
    \label{tab:SensitivityEstimateBands}
\end{table}

Having found no significant candidates, we now estimate the sensitivity upper limit of the search. We use Monte Carlo simulations to compute the \emph{detection probability}: the probability that a signal of strength $h_{0}$ can be recovered. We inject signals of a fixed $h_{0}$ into the same experimental data as used for the searches. The signals evolve in frequency according to the piecewise model of Eq.~\eqref{eq:PWModel}, with frequencies randomly selected from within the ranges listed in Table~\ref{tab:SensitivityEstimateBands}, and other parameters randomly selected from the parameter space given in Table~\ref{tab:SearchParams}. For each injected signal, we perform a search over the same frequency range from Table~\ref{tab:SensitivityEstimateBands}, with other parameters the same as for the searches (see Sec.~\ref{sec:Setup}).
To determine whether a search has recovered an injected signal, we use a threshold $2\cF^{*}$. Searches which have a template with a $2\cF$ above $2\cF^{*}$ are considered to have recovered the injection.
The percentage of searches where the signal was recovered becomes the detection probability for the given $h_{0}$. 

This process was then repeated for different sample $h_{0}$ values. The values of $h_{0}$ sampled were the same as those given in Table~II of~\cite{Grace2023}. As this set of $h_{0}$ is a discrete set, it is unlikely that any of these will coincide exactly with the value which gives the 50\% detection probability. To then calculate the 50\% detection probability, the first $h_{0}$'s with detection probabilities above and below the 50\% probability were linearly interpolated between to estimate this quantity.

The threshold $2\cF^{*}$ is calculated by carrying out 400 searches on simulated data with no injected signal.
This data is randomly generated for each search, and contains Gaussian noise with a PSD equal to that of the experimental data for the same source (GW170817 or GW190425) and at the same frequency band where the upper limit is being estimated. In this way, we generate independent realisations of the noise level of the experimental data.
For each search, the largest $2\cF$ is recorded; the $\cF$-statistic which occurs at the $99^{\rm th}$ percentile of the recorded $\cF$-statistics is then the threshold statistic $2\cF^{*}$. This corresponds to a 1\% \emph{false alarm probability}: the probability that a search will falsely claim to have recovered a signal.

Upper limits are ultimately quoted in terms of $h\urss^{50\%}$, the root sum squared strain amplitude which gives a 50\% detection probability~\cite{Abbott2019g}. For a time domain signal, $h\urss$ is given as
\begin{align}
    h\urss^{2} &= 2\int_{t\ustart}^{t\ufinish}\left(\left|\tilde{h}_{+}(t)\right|^{2} + \left|\tilde{h}_{\times}(t)\right|^{2}\right)dt,
\end{align}
where $t\ustart$ is given for each source in Table~\ref{tab:SearchSourceParams}, and $t\ufinish = t\ustart + 1800$~s. The functions $\tilde{h}_{+}$ and $\tilde{h}_{\times}$ are
\begin{align}
    \tilde{h}_{+} &= \frac{1}{2} A(t) \left(1 + \cos\iota\right)\cos\psi , \\
    \tilde{h}_{\times} &= A(t) \cos\iota\sin\psi , \\
    A(t) &= h_{0}\left(\frac{f\upw(t)}{f_{0}}\right)^{2}.
\end{align}
Here, $A(t)$ is the amplitude of the gravitational wave signal, and $f_{0}$ is the maximum frequency of the band in which we are estimating $h\urss^{50\%}$. As the amplitude of the gravitational wave is proportional to $f^{2}$, by Eq.\eqref{eq:strain} we set the strength of the injected signals to change in time according to $A(t)$. The parameters for $f\upw(t)$ are chosen randomly from inside the parameter space as given by Equations~\eqref{eq:f00_Condition}--\eqref{eq:fi2_Condition}.

\begin{figure*}
    \subfloat[]{\includegraphics[width=0.87\textwidth]{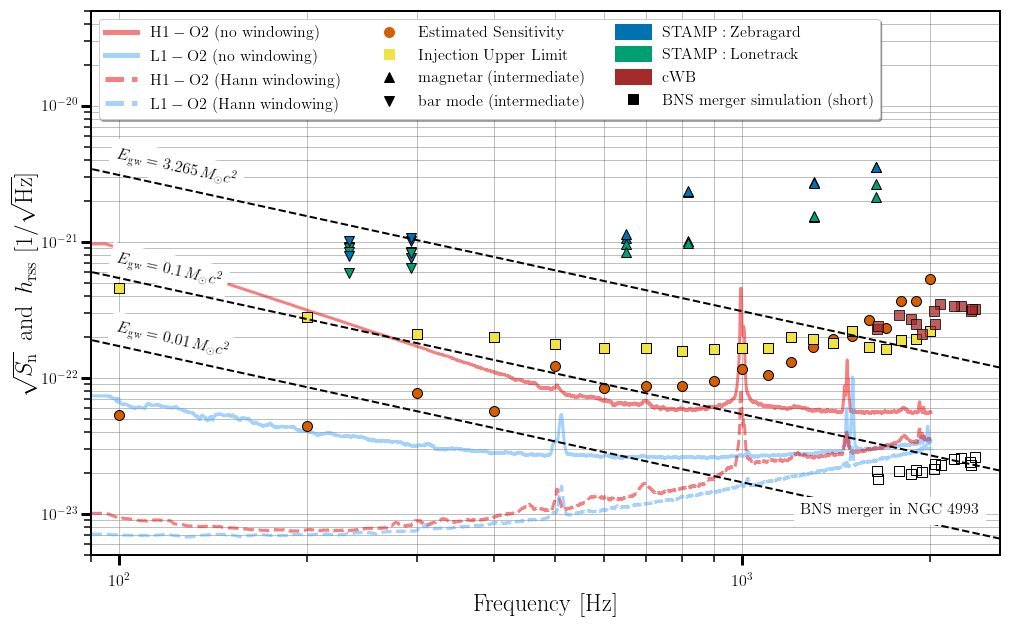}\label{fig:GW170817_sensitivity}} \\
    \subfloat[]{\includegraphics[width=0.87\textwidth]{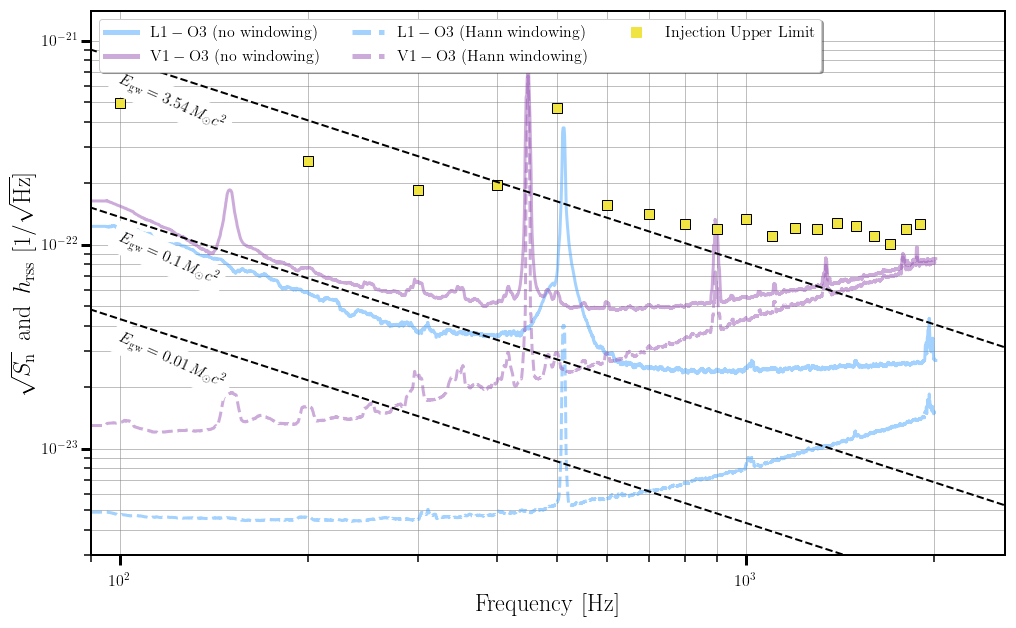}\label{fig:GW190425_sensitivity}}
    \caption{\protect\subref{fig:GW170817_sensitivity} Estimated upper limits of the piecewise search for GW170817, edited from~\cite{Abbott2017b}. \protect\subref{fig:GW190425_sensitivity} Estimated upper limits of the piecewise search for GW190425. Lines of constant $E\uGW$ indicate the gravitational wave energy required for emission at the given $h\urss^{50\%}$ and frequency. Note that $h\urss^{50\%}$ is dependent upon the distance to the source, and as such the lines of constant $E\uGW$ differ between \protect\subref{fig:GW170817_sensitivity} and \protect\subref{fig:GW190425_sensitivity}. The estimated emission strength of post-merger BNS remnants in the galaxy NGC--4993 are shown for \protect\subref{fig:GW170817_sensitivity} as open squares. The noise amplitude spectral density ($\sqrt{S_{n}}$) for the detectors used for each search are also shown. The estimated sensitivity of the piecewise method, taken from~\cite{Grace2023}, is shown in orange for comparison. Dashed lines are PSDs computed with Hann windowing; solid lines are PSDs computed with no windowing.}
    \label{fig:Sensitivity_Plots}
\end{figure*}

Figure~\ref{fig:Sensitivity_Plots} plots the upper limits of the GW170817 and GW190425 searches.
Best upper limits for both searches occur at 1700~Hz at $1.64 \times 10^{-22}~\text{Hz}^{-1/2}$ and $10^{-22}~\text{Hz}^{-1/2}$ for GW170817 and GW190425 respectively. Search upper limits for GW170817 and GW190425 are comparable at low frequencies, however the search for GW190425 shows better performance at frequencies $>700$~Hz. The improved upper limits of the GW190425 search at these frequencies is due to the lower noise levels in L1 data for this search. Figure~\ref{fig:all_sensitivities} in the Appendix plots data points from both Figs.~\ref{fig:GW170817_sensitivity} and~\ref{fig:GW190425_sensitivity} simultaneously for comparison.

At frequencies below 1400~Hz, the upper limit of the GW170817 search is worse than estimated in~\cite{Grace2023}. This is likely due to the different noise levels used in achieving the sensitivity estimates in~\cite{Grace2023} compared to the data used for the search carried out in this work. The difference between these noise levels arises from the different windowing functions being used to compute the PSDs. Windowing reduces the effects of spectral leakage, concentrating signal power into fewer bins as desired. Windowing also has the beneficial side-effect of pre-filtering the data to suppress the effects of noise artifacts, such that only specific bins are contaminated with lines, as seen in Fig.~\ref{fig:GW170817_sensitivity}. Figure~\ref{fig:Sensitivity_Plots} shows two different PSDs calculated from the search SFTs, with and without windowing. The PSDs calculated from SFTs with Hann windowing are lower than those with no windowing, likely due to the presence of weak instrumental line artifacts in the data, as illustrated in Figs.~\ref{fig:GW170817_2F_vs_2F_Figures} and~\ref{fig:GW190425_2F_vs_2F_Figures}.
The SFTs used in the searches are not windowed, and as such the upper limit estimates in Fig.~\ref{fig:GW170817_sensitivity} follow the noise level of the non-windowed PSDs. In contrast, the sensitivity estimates of~\cite{Grace2023} follow the noise levels of the windowed PSDs in Fig.~\ref{fig:GW170817_sensitivity}. The Hann windowed PSDs are characteristic of the amplitude spectral density for each detector in their respective observing runs and, for GW170817, agree with those given in~\cite{Grace2023}.

If unwindowed SFTs are more susceptible to instrumental artifacts, and thus make our upper limit estimates less sensitive, we might consider why we do not then run our search on windowed SFTs. Indeed, if we were using a power-based detection statistic~\cite[e.g.][]{MendLand2005-StcHgSrSNStt}, we would want SFTs which had been windowed. This is because power-based statistics typically assume that signal power is concentrated in a single SFT bin. The $\cF$-statistic, in contrast, is a matched filter, and as such compares the input data against a model of the expected signal~\cite{Jaranowski1998}. Both data and signal model must be consistent. If we applied windowing to the data, we would necessarily have to apply the same windowing to the expected signal. This would require a different detection statistic -- i.e.\ a windowed $\cF$-statistic -- but such a statistic has yet to be implemented. Formally, therefore, the $\cF$-statistic requires unwindowed data as its input. Consequentially we are unable to benefit from the suppression of noise artifacts windowing provides for the datasets analysed here. Future work could incorporate windowing into the $\cF$-statistic signal model, or devise a more effective veto for candidates dominated by noise in a single detector (cf. Figs.~\ref{fig:GW170817_2F_vs_2F_Figures}, \ref{fig:GW190425_2F_vs_2F_Figures}). Consequently, the different use of windowing functions between the estimated sensitivities of~\cite{Grace2023} and the upper limits given in Fig.~\ref{fig:Sensitivity_Plots} makes their comparison difficult. As the study carried out in~\cite{Grace2023} used the $\cF$-statistic, the PSDs used for injection studies should have been calculated on SFTs without windowing. This however is not the case but it does provide a good demonstration of the affects that windowed SFTs have on PSDs and upper limit estimates.

Initial upper limit estimates were carried out on frequency bands at 100~Hz intervals from 100--2000~Hz. If an instrumental line is present within any of these 100~Hz intervals, upper limit estimates cannot be achieved in these regions. This is because searches carried out on intervals with lines will have templates which achieve large $2\cF$ values. The threshold statistic, $2\cF^{*}$, is calculated on simulated data which does not contain any lines. The threshold statistic will therefore be smaller in value than the $2\cF$s found from searches on data with lines present; regardless of the strength of the injected signal $h_{0}$ used when calculating the detection probability. This means that, for each search run on experimental data, the detection probability will always contain a $2\cF$ above $2\cF^{*}$, leading to a detection probability of 100\% for every $h_{0}$. It follows that there is no strain value $h_{0}$ corresponding with a 50\% detection probability to calculate $h\urss^{50\%}$. In these cases, a frequency band free from instrumental lines close to the contaminated band was selected. Upper limit estimates for GW190425 at or near 2000~Hz could not be achieved due to noise artifacts vetoing of all candidates above 1935~Hz. The final choices of frequency bands used are listed in Table~\ref{tab:SensitivityEstimateBands}.

The energy emitted from a source radiating isotropic gravitational waves is given by~\cite{Sutton2013}
\begin{align}
    E\uGW^{\text{iso}} &= \frac{\pi^{2}c^{3}}{G}D^{2}\bar{f}^{2}h\urss^{2}, \label{eq:Radiated_Energy}
\end{align}
where $D$ is the distance to the source and $\bar{f}$ is the central frequency of the band on which we are estimating $E\uGW$. Figure~\ref{fig:Sensitivity_Plots} plots Eq.~\eqref{eq:Radiated_Energy}, rearranged to give $h\urss$ for $E\uGW^{\text{iso}} = 0.1~\Msolar c^{2}$  and $0.01~\Msolar c^{2}$ for both GW170817 and GW190425 at their respective distances. We also plot the line of the most optimistic estimate of maximum energy released after both merger events. For GW170817 this is $E\uGW = 3.2654~\Msolar c^{2}$. This value is chosen as the remaining energy left in the GW170817 system after coalescence, by subtracting the minimal estimate of the energy emitted from the system during its inspiral phase from the upper bound of the 90\% confidence interval of the system's total mass~\cite{Abbott2017b, Abbott2017}. For GW190425 the most optimistic maximum energy released is taken as the 90\% upper limit mass of the remnant left after merger; this is equivalent to the remaining energy left in the system after merger. This is estimated as $3.54~\Msolar c^{2}$ in~\cite{Sedaghat2022}.

The GW170817 search would be sensitive to post-merger remnant signals radiating a total energy of $0.1~\Msolar c^{2}$ at frequencies below 300~Hz. At frequencies $>1500$~Hz, however, the GW170817 search is only sensitive to signals radiating a total energy in the unphysical region of $>3.265~\Msolar c^{2}$. The search for GW190425 is less sensitive; at frequencies $>400$~Hz, it is only sensitive to signals radiating a total energy of $>3.54~\Msolar c^{2}$, and does not approach $0.1~\Msolar c^{2}$ at any frequencies. This is expected due to the greater distance of GW190425 compared to GW170817.

We compare our upper limits against those achieved by another search which uses three theoretical models for a post-merger neutron star~\cite{Abbott2017b}: magnetars spinning down according to the GTE~\cite{LaskyMagnetarModel}; secular bar-mode instabilities~\cite{Lai1994}; and the post-merger component of simulated BNS merger waveforms (Table I of~\cite{Abbott2017b}). For the GW170817 search, the upper limits of the piecewise model compared to those achieved by the first two of these models shows significant improvement across the explored frequency band, and by an order of magnitude at frequencies $>1300$~Hz. At the frequencies of the cWB search in~\cite{Abbott2017b}, the piecewise model shows comparable upper limits with a small improvement in the 1600--2000~Hz band.

Notwithstanding the upper limit improvement demonstrated by the piecewise model, it is unlikely that a gravitational wave signal from a post-merger remnant is detectable at current detector sensitivities. For GW170817, upper limit estimates in the frequency ranges $>1500$~Hz are almost an order of magnitude above theoretical estimates of remnant signal strengths. At frequencies below 300~Hz, the search for GW170817 shows upper limits in mass energy appropriate for potential signal detection, but remnant neutron stars are unlikely to be spinning at these frequencies~\cite{Hotokezaka2013, Bauswein2012, Takami2014, Bernuzzi2015}. The search is not sensitive enough to detect a post-merger remnant signal from a BNS merger as distant as GW190425.

\section{Conclusion} \label{sec:Conclusion}

We have presented the results of a long-transient gravitational wave search for the remnants of two binary neutron star coalescence events; GW170817 and GW190425. The searches presented here carried out a coherent $\cF$-statistic search on 1800~s of data using a piecewise model, and used identical parameter spaces and setups. Both searches were partitioned into 655 jobs, each recording the top 1000 largest candidates. Additional search jobs were run to partition wide frequency bands containing lines. Line vetoing was used to remove candidates associated with known noise sources; a detector $2\cF$ veto could not be applied, however, due to the presence of weak single-detector lines which could not be visually identified. The Distromax method was used to determine the statistical significance of the remaining candidates. All candidates were found to be consistent with noise, and no gravitational wave signal has been detected.

Estimated upper limits of the searches are quoted in terms of the $h\urss^{50\%}$, the root sum squared strain at 50\% detection probability. For GW170817 these upper limits are compared against those of previous searches~\cite{Abbott2017b}; the piecewise model has improved upper limits at frequencies below $1600$~Hz; above $1600$~Hz it is comparable to the results of the cWB search. Peak upper limits occurred at 1700~Hz of $1.64 \times 10^{-22}~\text{Hz}^{-1/2}$ for GW170817, and $10^{-22}~\text{Hz}^{-1/2}$ for GW190425. This is the first search for a post-merger remnant of GW190425.

The piecewise model was developed to more accurately follow the frequency evolution of young neutron stars than is possible with traditional continuous wave signal models. In this work we have applied it to search for the remnants of BNS mergers. The piecewise model may also be appropriate for other gravitational wave sources, notably young neutron stars in supernova remnants. One such source is SN1987A, the youngest known supernova remnant, which has been the target of previous searches~\cite{PhysRevD.94.082004, Owen2022, Owen2024}. Recent electromagnetic observations support the existence of a neutron star as the compact object at the centre of SN1987A~\cite{Fransson2024}.  It is likely to be spinning at lower frequencies than BNS remnants, but at a spin-down rate greater than traditional continuous wave source targets. The computational cost of the piecewise model is significantly reduced at lower frequencies, which would allow for longer spans of data to be used in order to increase its sensitivity.

\begin{acknowledgements}
    The authors would like to thank Rodrigio Tenorio for helpful discussions on the Distromax method.
    We also thank Julian Carlin, David Keitel, Andrew Miller, and Ling Sun for helpful comments on the manuscript.
    This research was supported by the Australian Research Council under the ARC Centre of Excellence for Gravitational Wave Discovery, grant number CE170100004. B.G.\ would like to acknowledge the funding from the Australian Government Research Training Program (AGRTP) Scholarship for their research. This work was performed on the OzSTAR national facility at Swinburne University of Technology. The OzSTAR program receives funding in part from the Astronomy National Collaborative Research Infrastructure Strategy (NCRIS) allocation provided by the Australian Government, and from the Victorian Higher Education State Investment Fund (VHESIF) provided by the Victorian Government.
    This research has made use of data or software obtained from the Gravitational Wave Open Science Center (gwosc.org), a service of the LIGO Scientific Collaboration, the Virgo Collaboration, and KAGRA. This material is based upon work supported by NSF's LIGO Laboratory which is a major facility fully funded by the National Science Foundation, as well as the Science and Technology Facilities Council (STFC) of the United Kingdom, the Max-Planck-Society (MPS), and the State of Niedersachsen/Germany for support of the construction of Advanced LIGO and construction and operation of the GEO600 detector. Additional support for Advanced LIGO was provided by the Australian Research Council. Virgo is funded, through the European Gravitational Observatory (EGO), by the French Centre National de Recherche Scientifique (CNRS), the Italian Istituto Nazionale di Fisica Nucleare (INFN) and the Dutch Nikhef, with contributions by institutions from Belgium, Germany, Greece, Hungary, Ireland, Japan, Monaco, Poland, Portugal, Spain. KAGRA is supported by the Ministry of Education, Culture, Sports, Science and Technology (MEXT), Japan Society for the Promotion of Science (JSPS) in Japan; National Research Foundation (NRF) and Ministry of Science and ICT (MSIT) in Korea; Academia Sinica (AS) and National Science and Technology Council (NSTC) in Taiwan.
    This manuscript has document number LIGO-P2400076.
\end{acknowledgements}

\appendix

\section*{Appendix}

Tables~\ref{tab:GW170817Lines} and~\ref{tab:GW190425Lines} present the lines which are visually identified for vetoing in Sec.~\ref{sec:line_vetoing}. Many of the visually identified lines could not be associated with known instrumental lines. Some lines may be unidentifiable due to the different length SFTs and windowing functions used for detector characterisation. For example, \cite{Covas2018} uses SFTs with a time base of $T_{\text{SFT}} = 1800$~s and a Tukey windowing function, whereas we have used 10~s SFTs with no windowing. As a result, such studies may not see many of the lines we observe in this work.

\begin{table}
    \begin{tabularx}{\columnwidth}[t]{r@{}l@{\extracolsep{3ex}}lX}
        \hline \hline
        \multicolumn{2}{l}{Line~(Hz)} & Detector & Origin \\
        \hline
        500&.2  & H1 & Violin mode 1st harmonic ITMX Mode 2 \\
        996&    & H1 & Violin mode 2nd harmonic ITMX \\
        1006&   & H1 & Violin mode 2nd harmonic ETMX\\
        1009&   & H1 & Violin mode 2nd harmonic ETMY\\
        1456&.3 & H1 & Violin mode 3rd harmonic \\
        1462&.3 & H1 & Violin mode 3rd harmonic \\
        1464&   & H1 & Violin mode 3rd harmonic\\
        1468&   & H1 & Violin mode 3rd harmonic \\
        1475&   & H1 & Violin mode 3rd harmonic \\
        1480&   & H1 & Unidentified line\\
        1483&.7 & H1 & Violin mode 3rd harmonic \\
        1517&   & H1 & Unidentified line\\
        1922&--1959   & H1, L1 & Violin modes 4th harmonic \\
        \hline
        306&.2 & L1 & Beam splitter violin mode 1st harmonic region \\
        315&.1 & L1 & Beam splitter violin mode 1st harmonic region \\
        499&.6 & L1 & Violin resonance \\
        512&   & L1 & Violin resonances \\
        1083&   & L1 & Calibration line\\
        1457&.7 & L1 & Violin resonance \\
        1470&.9 & L1 & Violin resonance \\
        1486&   & L1 & Unidentified line\\
        1491&   & L1   & Violin resonance\\
        1496&.1 & L1 & Violin resonances \\
        1505&.7 & L1 & Violin resonances \\
        1962&--1990   & L1 & Violin modes \\
        \hline \hline
    \end{tabularx}
    \caption{Lines used for vetoing for the GW170817 search~\cite{Covas2018, Abbott2017O1AllSky}. The detector for each line was identified by inspecting which single detector $2\cF$ from Fig.\ref{fig:Fstat_Line_Vetos} was the most significant contributor.}
    \label{tab:GW170817Lines}
\end{table}

\begin{table*}
    \begin{tabularx}{\columnwidth}[t]{r@{}l@{\extracolsep{3ex}}lX}
        \hline \hline
        \multicolumn{2}{l}{Line~(Hz)} & Detector & Origin \\
        \hline
        300&    & L1 & Power mains \\
        306&.2  & L1 & Beam splitter violin mode 1st harmonic region\\
        307&.5  & L1 & Beam splitter violin mode 1st harmonic region\\
        315&    & L1 & Beam splitter violin mode 1st harmonic region\\
        509&.2  & L1 & Beam splitter violin mode 1st harmonic region\\
        612&.5  & L1 & Beam splitter violin mode 2nd harmonic region\\
        615&    & L1 & Beam splitter violin mode 2nd harmonic region\\
        630&.1  & L1 & Beam splitter violin mode 2nd harmonic region\\
        1004&    & L1 & Calibration line mixing \\
        1012&    & L1 & Violin mode 2nd harmonic; notch filter mismatch\\
        1018&    & L1 & Violin mode 2nd harmonic; elevated noise from 1012.9--1019.75~Hz\\
        1024&.6  & L1 & Violin mode 2nd harmonic; elevated noise from 1023.25-1026~Hz\\
        1489&.3  & L1 & Violin mode 3rd harmonic \\
        1492&    & L1 & Violin mode 3rd harmonic; notch filter mismatch \\
        1498&    & L1 & Violin mode 3rd harmonic; notch filter mismatch \\
        1510&.9  & L1 & Violin mode 3rd harmonic; notch filter mismatch \\
        1936&.2  & L1 & Violin mode 4th harmonic \\
        1945&    & L1 & Violin mode 4th harmonic \\
        1952&.5  & L1 & Violin mode 4th harmonic \\
        1958&    & L1 & Violin mode 4th harmonic \\
        1961&.7  & L1 & Violin mode 4th harmonic\\
        1966&    & L1 & Violin mode 4th harmonic \\
        1977&    & L1 & Violin mode 4th harmonic \\
        1985&    & L1 & Violin mode 4th harmonic \\
        1988&    & L1 & Violin mode 4th harmonic \\
        \hline
    \end{tabularx}
    \begin{tabularx}{\columnwidth}[t]{r@{}l@{\extracolsep{3ex}}lX}
        \hline
        \multicolumn{2}{l}{Line~(Hz)} & Detector & Origin \\
        \hline
        100&    & V1 & Mains 2nd harmonic \\
        150&    & V1 & Mains sideband \\
        200&    & V1 & Environmental origin \\
        250&    & V1 & Mains harmonic \\
        279&    & V1 & Beam splitter violin mode \\
        291&    & V1 & Unidentified line \\
        295&    & V1 & Unidentified line \\
        300&    & V1 & Mains harmonic \\
        350&    & V1 & Mains harmonic \\
        356&.5  & V1 & Calibration line\\
        392&    & V1 & Turbo pump driver \\
        397&    & V1 & Environmental origin \\
        420&.3  & V1 & Scattered light \\
        444&.6  & V1 & Violin mode \\
        446&    & V1 & Violin mode \\
        450&.1  & V1 & Violin mode \\
        452&.4  & V1 & Violin mode \\
        455&.1  & V1 & Violin mode \\
        500&    & V1 & Mains harmonic\\
        550&    & V1 & Mains harmonic \\
        600&    & V1 & Mains harmonic \\
        884&    & V1 & Second order violin mode \\
        893&.1  & V1 & Second order violin mode \\
        905&    & V1 & Second order violin mode \\
        950&    & V1 & Mains harmonic \\
        1050&    & V1 & Mains harmonic \\
        1111&.5  & V1 & Calibration line \\
        1256&.6  & V1 & Mechanical mode \\
        1327&--1349 & V1 & Third order violin modes \\
        1358&    & V1 & Third order violin mode \\
        1762&--1811 & V1 & Fourth order violin modes \\
        1875&.7  & V1 & Drum mode \\
        \hline \hline
    \end{tabularx}
    \caption{Lines used for vetoing for the GW190425 search~\cite{Covas2018, Abbott2021, Davis2021}. The detector for each line was identified by inspecting which single detector $2\cF$ from Fig.\ref{fig:Fstat_Line_Vetos} was the most significant contributor.}
    \label{tab:GW190425Lines}
\end{table*}

\begin{figure*}
    \centering
    \includegraphics[width=\textwidth]{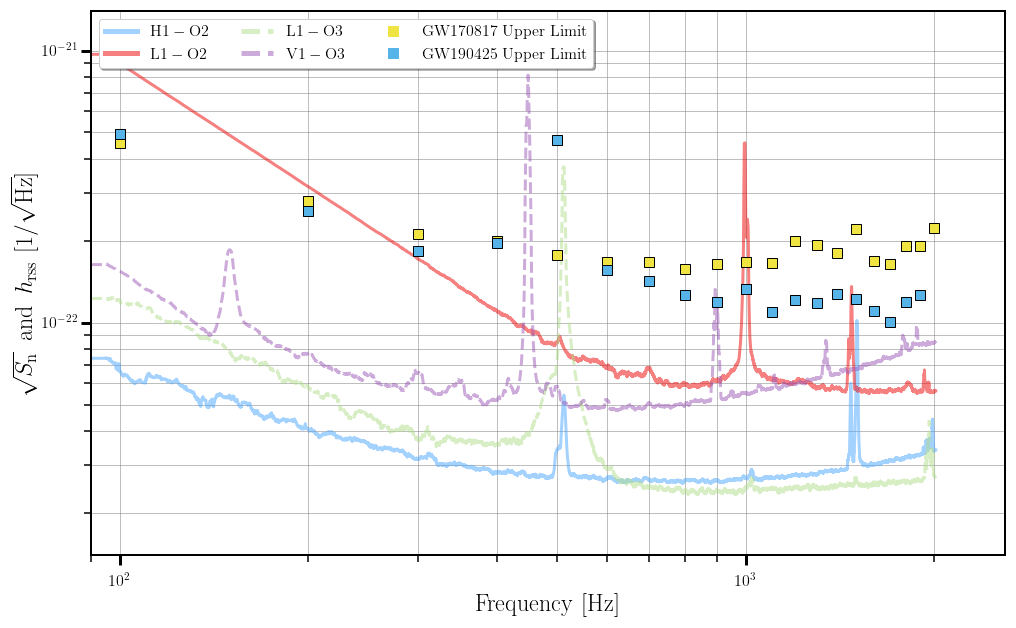}
    \caption{Combined upper limit results of the plots presented in Fig.~\ref{fig:Sensitivity_Plots}. Dashed noise lines represent those computed from the data used in the GW190425 search; solid noise lines are those computed from the data used in the GW170817 search. The PSDs shown are those from Fig.~\ref{fig:Sensitivity_Plots} with no windowing.}
    \label{fig:all_sensitivities}
\end{figure*}

Figure~\ref{fig:all_sensitivities} combines the results from Figs.~\ref{fig:GW170817_sensitivity} and~\ref{fig:GW190425_sensitivity} for ease of comparison between upper limits and noise PSDs of the GW170817 and GW190425 searches.

\clearpage


\begin{thebibliography}{94}%
\makeatletter
\providecommand \@ifxundefined [1]{%
 \@ifx{#1\undefined}
}%
\providecommand \@ifnum [1]{%
 \ifnum #1\expandafter \@firstoftwo
 \else \expandafter \@secondoftwo
 \fi
}%
\providecommand \@ifx [1]{%
 \ifx #1\expandafter \@firstoftwo
 \else \expandafter \@secondoftwo
 \fi
}%
\providecommand \natexlab [1]{#1}%
\providecommand \enquote  [1]{``#1''}%
\providecommand \bibnamefont  [1]{#1}%
\providecommand \bibfnamefont [1]{#1}%
\providecommand \citenamefont [1]{#1}%
\providecommand \href@noop [0]{\@secondoftwo}%
\providecommand \href [0]{\begingroup \@sanitize@url \@href}%
\providecommand \@href[1]{\@@startlink{#1}\@@href}%
\providecommand \@@href[1]{\endgroup#1\@@endlink}%
\providecommand \@sanitize@url [0]{\catcode `\\12\catcode `\$12\catcode
  `\&12\catcode `\#12\catcode `\^12\catcode `\_12\catcode `\%12\relax}%
\providecommand \@@startlink[1]{}%
\providecommand \@@endlink[0]{}%
\providecommand \url  [0]{\begingroup\@sanitize@url \@url }%
\providecommand \@url [1]{\endgroup\@href {#1}{\urlprefix }}%
\providecommand \urlprefix  [0]{URL }%
\providecommand \Eprint [0]{\href }%
\providecommand \doibase [0]{https://doi.org/}%
\providecommand \selectlanguage [0]{\@gobble}%
\providecommand \bibinfo  [0]{\@secondoftwo}%
\providecommand \bibfield  [0]{\@secondoftwo}%
\providecommand \translation [1]{[#1]}%
\providecommand \BibitemOpen [0]{}%
\providecommand \bibitemStop [0]{}%
\providecommand \bibitemNoStop [0]{.\EOS\space}%
\providecommand \EOS [0]{\spacefactor3000\relax}%
\providecommand \BibitemShut  [1]{\csname bibitem#1\endcsname}%
\let\auto@bib@innerbib\@empty
\bibitem [{\citenamefont {R.}\ \emph {et~al.}(2023)\citenamefont {R.} \emph
  {et~al.}}]{Collaboration2021}%
  \BibitemOpen
  \bibfield  {author} {\bibinfo {author} {\bibfnamefont {A.}~\bibnamefont {R.}}
  \emph {et~al.} (\bibinfo {collaboration} {LIGO Scientific Collaboration,
  Virgo Collaboration, and KAGRA Collaboration}),\ }\bibfield  {title}
  {\bibinfo {title} {Gwtc-3: Compact binary coalescences observed by ligo and
  virgo during the second part of the third observing run},\ }\href
  {https://doi.org/10.1103/PhysRevX.13.041039} {\bibfield  {journal} {\bibinfo
  {journal} {Phys. Rev. X}\ }\textbf {\bibinfo {volume} {13}},\ \bibinfo
  {pages} {041039} (\bibinfo {year} {2023})}\BibitemShut {NoStop}%
\bibitem [{\citenamefont {Abbott}\ \emph {et~al.}(2018)\citenamefont {Abbott}
  \emph {et~al.}}]{Abbott2018d}%
  \BibitemOpen
  \bibfield  {author} {\bibinfo {author} {\bibfnamefont {B.~P.}\ \bibnamefont
  {Abbott}} \emph {et~al.} (\bibinfo {collaboration} {LIGO Scientific
  Collaboration and Virgo Collaboration}),\ }\bibfield  {title} {\bibinfo
  {title} {{GW170817: Measurements of Neutron Star Radii and Equation of
  State}},\ }\href {https://doi.org/10.1103/PhysRevLett.121.161101} {\bibfield
  {journal} {\bibinfo  {journal} {Phys. Rev. Lett.}\ }\textbf {\bibinfo
  {volume} {121}},\ \bibinfo {pages} {161101} (\bibinfo {year}
  {2018})}\BibitemShut {NoStop}%
\bibitem [{\citenamefont {Bhat}\ and\ \citenamefont
  {Bandyopadhyay}(2019)}]{Bhat2019}%
  \BibitemOpen
  \bibfield  {author} {\bibinfo {author} {\bibfnamefont {S.~A.}\ \bibnamefont
  {Bhat}}\ and\ \bibinfo {author} {\bibfnamefont {D.}~\bibnamefont
  {Bandyopadhyay}},\ }\bibfield  {title} {\bibinfo {title} {{Neutron star
  equation of state and GW170817}},\ }\href
  {https://doi.org/10.1088/1361-6471/aaef45} {\bibfield  {journal} {\bibinfo
  {journal} {J. Phys.}\ }\textbf {\bibinfo {volume} {46}},\ \bibinfo {pages}
  {014003} (\bibinfo {year} {2019})}\BibitemShut {NoStop}%
\bibitem [{\citenamefont {Abbott}\ \emph
  {et~al.}(2017{\natexlab{a}})\citenamefont {Abbott} \emph
  {et~al.}}]{Abbott2017a}%
  \BibitemOpen
  \bibfield  {author} {\bibinfo {author} {\bibfnamefont {B.~P.}\ \bibnamefont
  {Abbott}} \emph {et~al.} (\bibinfo {collaboration} {LIGO Scientific
  Collaboration and Virgo Collaboration}),\ }\bibfield  {title} {\bibinfo
  {title} {{Multi-messenger Observations of a Binary Neutron Star Merger}},\
  }\href {https://doi.org/10.3847/2041-8213/aa91c9} {\bibfield  {journal}
  {\bibinfo  {journal} {Astrophys. J. Lett.}\ }\textbf {\bibinfo {volume}
  {848}},\ \bibinfo {pages} {L12} (\bibinfo {year}
  {2017}{\natexlab{a}})}\BibitemShut {NoStop}%
\bibitem [{\citenamefont {Tanvir}\ \emph {et~al.}(2017)\citenamefont {Tanvir}
  \emph {et~al.}}]{Tanvir2017}%
  \BibitemOpen
  \bibfield  {author} {\bibinfo {author} {\bibfnamefont {N.~R.}\ \bibnamefont
  {Tanvir}} \emph {et~al.},\ }\bibfield  {title} {\bibinfo {title} {{The
  Emergence of a Lanthanide-rich Kilonova Following the Merger of Two Neutron
  Stars}},\ }\href {https://doi.org/10.3847/2041-8213/aa90b6} {\bibfield
  {journal} {\bibinfo  {journal} {Astrophys. J. Lett.}\ }\textbf {\bibinfo
  {volume} {848}},\ \bibinfo {pages} {L27} (\bibinfo {year}
  {2017})}\BibitemShut {NoStop}%
\bibitem [{\citenamefont {Wu}\ \emph {et~al.}(2019)\citenamefont {Wu},
  \citenamefont {Barnes}, \citenamefont {Mart{\'{i}}nez-Pinedo},\ and\
  \citenamefont {Metzger}}]{Wu2019}%
  \BibitemOpen
  \bibfield  {author} {\bibinfo {author} {\bibfnamefont {M.~R.}\ \bibnamefont
  {Wu}}, \bibinfo {author} {\bibfnamefont {J.}~\bibnamefont {Barnes}}, \bibinfo
  {author} {\bibfnamefont {G.}~\bibnamefont {Mart{\'{i}}nez-Pinedo}},\ and\
  \bibinfo {author} {\bibfnamefont {B.~D.}\ \bibnamefont {Metzger}},\
  }\bibfield  {title} {\bibinfo {title} {{Fingerprints of Heavy-Element
  Nucleosynthesis in the Late-Time Lightcurves of Kilonovae}},\ }\href
  {https://doi.org/10.1103/PhysRevLett.122.062701} {\bibfield  {journal}
  {\bibinfo  {journal} {Phys. Rev. Lett.}\ }\textbf {\bibinfo {volume} {122}},\
  \bibinfo {pages} {062701} (\bibinfo {year} {2019})}\BibitemShut {NoStop}%
\bibitem [{\citenamefont {Ravi}\ and\ \citenamefont {Lasky}(2014)}]{Ravi2014}%
  \BibitemOpen
  \bibfield  {author} {\bibinfo {author} {\bibfnamefont {V.}~\bibnamefont
  {Ravi}}\ and\ \bibinfo {author} {\bibfnamefont {P.~D.}\ \bibnamefont
  {Lasky}},\ }\bibfield  {title} {\bibinfo {title} {{The birth of black holes:
  neutron star collapse times, gamma-ray bursts and fast radio bursts}},\
  }\href {https://doi.org/10.1093/mnras/stu720} {\bibfield  {journal} {\bibinfo
   {journal} {Mon. Not. R. Astron. Soc.}\ }\textbf {\bibinfo {volume} {441}},\
  \bibinfo {pages} {2433} (\bibinfo {year} {2014})}\BibitemShut {NoStop}%
\bibitem [{\citenamefont {Abbott}\ \emph
  {et~al.}(2017{\natexlab{b}})\citenamefont {Abbott} \emph
  {et~al.}}]{Abbott2017}%
  \BibitemOpen
  \bibfield  {author} {\bibinfo {author} {\bibfnamefont {R.}~\bibnamefont
  {Abbott}} \emph {et~al.} (\bibinfo {collaboration} {LIGO Scientific
  Collaboration and Virgo Collaboration}),\ }\bibfield  {title} {\bibinfo
  {title} {{GW170817: Observation of Gravitational Waves from a Binary Neutron
  Star Inspiral}},\ }\href {https://doi.org/10.1103/PhysRevLett.119.161101}
  {\bibfield  {journal} {\bibinfo  {journal} {Phys. Rev. Lett.}\ }\textbf
  {\bibinfo {volume} {119}},\ \bibinfo {pages} {161101} (\bibinfo {year}
  {2017}{\natexlab{b}})}\BibitemShut {NoStop}%
\bibitem [{\citenamefont {van Putten}\ and\ \citenamefont
  {Valle}(2019)}]{VanPutten2019}%
  \BibitemOpen
  \bibfield  {author} {\bibinfo {author} {\bibfnamefont {M.~H.}\ \bibnamefont
  {van Putten}}\ and\ \bibinfo {author} {\bibfnamefont {M.~D.}\ \bibnamefont
  {Valle}},\ }\bibfield  {title} {\bibinfo {title} {{Observational evidence for
  extended emission to GW170817}},\ }\href
  {https://doi.org/10.1093/mnrasl/sly166} {\bibfield  {journal} {\bibinfo
  {journal} {Mon. Not. R. Astron. Soc. Lett.}\ }\textbf {\bibinfo {volume}
  {482}},\ \bibinfo {pages} {L46} (\bibinfo {year} {2019})}\BibitemShut
  {NoStop}%
\bibitem [{\citenamefont {Abchouyeh}\ \emph {et~al.}(2023)\citenamefont
  {Abchouyeh}, \citenamefont {van Putten},\ and\ \citenamefont
  {Amati}}]{AghaeiAbchouyeh2023}%
  \BibitemOpen
  \bibfield  {author} {\bibinfo {author} {\bibfnamefont {M.~A.}\ \bibnamefont
  {Abchouyeh}}, \bibinfo {author} {\bibfnamefont {M.~H. P.~M.}\ \bibnamefont
  {van Putten}},\ and\ \bibinfo {author} {\bibfnamefont {L.}~\bibnamefont
  {Amati}},\ }\bibfield  {title} {\bibinfo {title} {Observational prospects of
  double neutron star mergers and their multimessenger afterglows: Ligo
  discovery power, event rates, and diversity},\ }\href
  {https://doi.org/10.3847/1538-4357/acd114} {\bibfield  {journal} {\bibinfo
  {journal} {Astrophys. J.}\ }\textbf {\bibinfo {volume} {952}},\ \bibinfo
  {pages} {157} (\bibinfo {year} {2023})}\BibitemShut {NoStop}%
\bibitem [{\citenamefont {Abbott}\ \emph {et~al.}(2020)\citenamefont {Abbott}
  \emph {et~al.}}]{Abbott190425}%
  \BibitemOpen
  \bibfield  {author} {\bibinfo {author} {\bibfnamefont {B.~P.}\ \bibnamefont
  {Abbott}} \emph {et~al.} (\bibinfo {collaboration} {LIGO Scientific
  Collaboration and Virgo Collaboration}),\ }\bibfield  {title} {\bibinfo
  {title} {{GW190425: Observation of a Compact Binary Coalescence with Total
  Mass $\sim$ 3.4 $M_{\odot}$}},\ }\href
  {https://doi.org/10.3847/2041-8213/ab75f5} {\bibfield  {journal} {\bibinfo
  {journal} {Astrophys. J. Lett.}\ }\textbf {\bibinfo {volume} {892}},\
  \bibinfo {pages} {L3} (\bibinfo {year} {2020})}\BibitemShut {NoStop}%
\bibitem [{\citenamefont {Abbott}\ \emph
  {et~al.}(2017{\natexlab{c}})\citenamefont {Abbott} \emph
  {et~al.}}]{Abbott2017b}%
  \BibitemOpen
  \bibfield  {author} {\bibinfo {author} {\bibfnamefont {B.~P.}\ \bibnamefont
  {Abbott}} \emph {et~al.} (\bibinfo {collaboration} {LIGO Scientific
  Collaboration and Virgo Collaboration}),\ }\bibfield  {title} {\bibinfo
  {title} {{Search for Post-merger Gravitational Waves from the Remnant of the
  Binary Neutron Star Merger GW170817}},\ }\href
  {https://doi.org/10.3847/2041-8213/aa9a35} {\bibfield  {journal} {\bibinfo
  {journal} {Astrophys. J. Lett.}\ }\textbf {\bibinfo {volume} {851}},\
  \bibinfo {pages} {L16} (\bibinfo {year} {2017}{\natexlab{c}})}\BibitemShut
  {NoStop}%
\bibitem [{\citenamefont {Abbott}\ \emph
  {et~al.}(2019{\natexlab{a}})\citenamefont {Abbott} \emph
  {et~al.}}]{TheLIGOScientificCollaboration2019}%
  \BibitemOpen
  \bibfield  {author} {\bibinfo {author} {\bibfnamefont {B.~P.}\ \bibnamefont
  {Abbott}} \emph {et~al.} (\bibinfo {collaboration} {LIGO Scientific
  Collaboration and Virgo Collaboration}),\ }\bibfield  {title} {\bibinfo
  {title} {{Search for gravitational waves from a long-lived remnant of the
  binary neutron star merger GW170817}},\ }\href
  {https://doi.org/10.3847/1538-4357/ab0f3d} {\bibfield  {journal} {\bibinfo
  {journal} {Astrophys. J.}\ }\textbf {\bibinfo {volume} {875}},\ \bibinfo
  {pages} {160} (\bibinfo {year} {2019}{\natexlab{a}})}\BibitemShut {NoStop}%
\bibitem [{\citenamefont {Abbott}\ \emph
  {et~al.}(2019{\natexlab{b}})\citenamefont {Abbott} \emph
  {et~al.}}]{Abbott2019}%
  \BibitemOpen
  \bibfield  {author} {\bibinfo {author} {\bibfnamefont {B.~P.}\ \bibnamefont
  {Abbott}} \emph {et~al.} (\bibinfo {collaboration} {{LIGO Scientific
  Collaboration, Virgo Collaboration}}),\ }\bibfield  {title} {\bibinfo {title}
  {Properties of the binary neutron star merger gw170817},\ }\href
  {https://doi.org/10.1103/PhysRevX.9.011001} {\bibfield  {journal} {\bibinfo
  {journal} {Phys. Rev. X}\ }\textbf {\bibinfo {volume} {9}},\ \bibinfo {pages}
  {011001} (\bibinfo {year} {2019}{\natexlab{b}})}\BibitemShut {NoStop}%
\bibitem [{\citenamefont {Miller}\ \emph {et~al.}(2019)\citenamefont {Miller}
  \emph {et~al.}}]{Miller:2019jtp}%
  \BibitemOpen
  \bibfield  {author} {\bibinfo {author} {\bibfnamefont {A.~L.}\ \bibnamefont
  {Miller}} \emph {et~al.},\ }\bibfield  {title} {\bibinfo {title} {How
  effective is machine learning to detect long transient gravitational waves
  from neutron stars in a real search?},\ }\href
  {https://doi.org/10.1103/PhysRevD.100.062005} {\bibfield  {journal} {\bibinfo
   {journal} {Phys. Rev. D}\ }\textbf {\bibinfo {volume} {100}},\ \bibinfo
  {pages} {062005} (\bibinfo {year} {2019})}\BibitemShut {NoStop}%
\bibitem [{\citenamefont {van Putten}\ \emph {et~al.}(2019)\citenamefont {van
  Putten}, \citenamefont {Valle},\ and\ \citenamefont
  {Levinson}}]{VanPutten2019b}%
  \BibitemOpen
  \bibfield  {author} {\bibinfo {author} {\bibfnamefont {M.~H. P.~M.}\
  \bibnamefont {van Putten}}, \bibinfo {author} {\bibfnamefont {M.~D.}\
  \bibnamefont {Valle}},\ and\ \bibinfo {author} {\bibfnamefont
  {A.}~\bibnamefont {Levinson}},\ }\bibfield  {title} {\bibinfo {title}
  {Multi-messenger extended emission from the compact remnant in gw170817},\
  }\href {https://doi.org/10.3847/2041-8213/ab18a2} {\bibfield  {journal}
  {\bibinfo  {journal} {Astrophys. J. Lett.}\ }\textbf {\bibinfo {volume}
  {876}},\ \bibinfo {pages} {L2} (\bibinfo {year} {2019})}\BibitemShut
  {NoStop}%
\bibitem [{\citenamefont {{Oliver}}\ \emph {et~al.}(2019)\citenamefont
  {{Oliver}}, \citenamefont {{Keitel}}, \citenamefont {{Miller}}, \citenamefont
  {{Estelles}},\ and\ \citenamefont
  {{Sintes}}}]{OlivEtAl2019-MtcSEBSNDtGrvExEmG}%
  \BibitemOpen
  \bibfield  {author} {\bibinfo {author} {\bibfnamefont {M.}~\bibnamefont
  {{Oliver}}}, \bibinfo {author} {\bibfnamefont {D.}~\bibnamefont {{Keitel}}},
  \bibinfo {author} {\bibfnamefont {A.~L.}\ \bibnamefont {{Miller}}}, \bibinfo
  {author} {\bibfnamefont {H.}~\bibnamefont {{Estelles}}},\ and\ \bibinfo
  {author} {\bibfnamefont {A.~M.}\ \bibnamefont {{Sintes}}},\ }\bibfield
  {title} {\bibinfo {title} {Matched-filter study and energy budget suggest no
  detectable gravitational-wave `extended emission' from {GW}170817},\ }\href
  {https://doi.org/10.1093/mnras/stz439} {\bibfield  {journal} {\bibinfo
  {journal} {Mon. Not. R. Astron. Soc.}\ }\textbf {\bibinfo {volume} {485}},\
  \bibinfo {pages} {843} (\bibinfo {year} {2019})}\BibitemShut {NoStop}%
\bibitem [{\citenamefont {Dhurandhar}\ \emph {et~al.}(2008)\citenamefont
  {Dhurandhar}, \citenamefont {Krishnan}, \citenamefont {Mukhopadhyay},\ and\
  \citenamefont {Whelan}}]{PhysRevD.77.082001}%
  \BibitemOpen
  \bibfield  {author} {\bibinfo {author} {\bibfnamefont {S.}~\bibnamefont
  {Dhurandhar}}, \bibinfo {author} {\bibfnamefont {B.}~\bibnamefont
  {Krishnan}}, \bibinfo {author} {\bibfnamefont {H.}~\bibnamefont
  {Mukhopadhyay}},\ and\ \bibinfo {author} {\bibfnamefont {J.~T.}\ \bibnamefont
  {Whelan}},\ }\bibfield  {title} {\bibinfo {title} {Cross-correlation search
  for periodic gravitational waves},\ }\href
  {https://doi.org/10.1103/PhysRevD.77.082001} {\bibfield  {journal} {\bibinfo
  {journal} {Phys. Rev. D}\ }\textbf {\bibinfo {volume} {77}},\ \bibinfo
  {pages} {082001} (\bibinfo {year} {2008})}\BibitemShut {NoStop}%
\bibitem [{\citenamefont {Chung}\ \emph {et~al.}(2011)\citenamefont {Chung},
  \citenamefont {Melatos}, \citenamefont {Krishnan},\ and\ \citenamefont
  {Whelan}}]{10.1111/j.1365-2966.2011.18585.x}%
  \BibitemOpen
  \bibfield  {author} {\bibinfo {author} {\bibfnamefont {C.~T.~Y.}\
  \bibnamefont {Chung}}, \bibinfo {author} {\bibfnamefont {A.}~\bibnamefont
  {Melatos}}, \bibinfo {author} {\bibfnamefont {B.}~\bibnamefont {Krishnan}},\
  and\ \bibinfo {author} {\bibfnamefont {J.~T.}\ \bibnamefont {Whelan}},\
  }\bibfield  {title} {\bibinfo {title} {{Designing a cross-correlation search
  for continuous-wave gravitational radiation from a neutron star in the
  supernova remnant SNR 1987A*}},\ }\href
  {https://doi.org/10.1111/j.1365-2966.2011.18585.x} {\bibfield  {journal}
  {\bibinfo  {journal} {Monthly Notices of the Royal Astronomical Society}\
  }\textbf {\bibinfo {volume} {414}},\ \bibinfo {pages} {2650} (\bibinfo {year}
  {2011})}\BibitemShut {NoStop}%
\bibitem [{\citenamefont {Coyne}\ \emph {et~al.}(2016)\citenamefont {Coyne},
  \citenamefont {Corsi},\ and\ \citenamefont {Owen}}]{PhysRevD.93.104059}%
  \BibitemOpen
  \bibfield  {author} {\bibinfo {author} {\bibfnamefont {R.}~\bibnamefont
  {Coyne}}, \bibinfo {author} {\bibfnamefont {A.}~\bibnamefont {Corsi}},\ and\
  \bibinfo {author} {\bibfnamefont {B.~J.}\ \bibnamefont {Owen}},\ }\bibfield
  {title} {\bibinfo {title} {Cross-correlation method for intermediate-duration
  gravitational wave searches associated with gamma-ray bursts},\ }\href
  {https://doi.org/10.1103/PhysRevD.93.104059} {\bibfield  {journal} {\bibinfo
  {journal} {Phys. Rev. D}\ }\textbf {\bibinfo {volume} {93}},\ \bibinfo
  {pages} {104059} (\bibinfo {year} {2016})}\BibitemShut {NoStop}%
\bibitem [{\citenamefont {Sowell}\ \emph {et~al.}(2019)\citenamefont {Sowell},
  \citenamefont {Corsi},\ and\ \citenamefont {Coyne}}]{PhysRevD.100.124041}%
  \BibitemOpen
  \bibfield  {author} {\bibinfo {author} {\bibfnamefont {E.}~\bibnamefont
  {Sowell}}, \bibinfo {author} {\bibfnamefont {A.}~\bibnamefont {Corsi}},\ and\
  \bibinfo {author} {\bibfnamefont {R.}~\bibnamefont {Coyne}},\ }\bibfield
  {title} {\bibinfo {title} {Multiwaveform cross-correlation search method for
  intermediate-duration gravitational waves from gamma-ray bursts},\ }\href
  {https://doi.org/10.1103/PhysRevD.100.124041} {\bibfield  {journal} {\bibinfo
   {journal} {Phys. Rev. D}\ }\textbf {\bibinfo {volume} {100}},\ \bibinfo
  {pages} {124041} (\bibinfo {year} {2019})}\BibitemShut {NoStop}%
\bibitem [{\citenamefont {Cornish}\ and\ \citenamefont
  {Romano}(2013)}]{PhysRevD.87.122003}%
  \BibitemOpen
  \bibfield  {author} {\bibinfo {author} {\bibfnamefont {N.~J.}\ \bibnamefont
  {Cornish}}\ and\ \bibinfo {author} {\bibfnamefont {J.~D.}\ \bibnamefont
  {Romano}},\ }\bibfield  {title} {\bibinfo {title} {Towards a unified
  treatment of gravitational-wave data analysis},\ }\href
  {https://doi.org/10.1103/PhysRevD.87.122003} {\bibfield  {journal} {\bibinfo
  {journal} {Phys. Rev. D}\ }\textbf {\bibinfo {volume} {87}},\ \bibinfo
  {pages} {122003} (\bibinfo {year} {2013})}\BibitemShut {NoStop}%
\bibitem [{\citenamefont {Grace}\ \emph {et~al.}(2023)\citenamefont {Grace},
  \citenamefont {Wette}, \citenamefont {Scott},\ and\ \citenamefont
  {Sun}}]{Grace2023}%
  \BibitemOpen
  \bibfield  {author} {\bibinfo {author} {\bibfnamefont {B.}~\bibnamefont
  {Grace}}, \bibinfo {author} {\bibfnamefont {K.}~\bibnamefont {Wette}},
  \bibinfo {author} {\bibfnamefont {S.~M.}\ \bibnamefont {Scott}},\ and\
  \bibinfo {author} {\bibfnamefont {L.}~\bibnamefont {Sun}},\ }\bibfield
  {title} {\bibinfo {title} {{Piecewise frequency model for searches for
  long-transient gravitational waves from young neutron stars}},\ }\href
  {https://doi.org/10.1103/PhysRevD.108.123045} {\bibfield  {journal} {\bibinfo
   {journal} {Phys. Rev. D}\ }\textbf {\bibinfo {volume} {108}},\ \bibinfo
  {pages} {123045} (\bibinfo {year} {2023})}\BibitemShut {NoStop}%
\bibitem [{\citenamefont {Jaranowski}\ \emph {et~al.}(1998)\citenamefont
  {Jaranowski}, \citenamefont {Kr{\'{o}}lak},\ and\ \citenamefont
  {Schutz}}]{Jaranowski1998}%
  \BibitemOpen
  \bibfield  {author} {\bibinfo {author} {\bibfnamefont {P.}~\bibnamefont
  {Jaranowski}}, \bibinfo {author} {\bibfnamefont {A.}~\bibnamefont
  {Kr{\'{o}}lak}},\ and\ \bibinfo {author} {\bibfnamefont {B.F.}~\bibnamefont
  {Schutz}},\ }\bibfield  {title} {\bibinfo {title} {{Data analysis of
  gravitational-wave signals from spinning neutron stars: The signal and its
  detection}},\ }\href {https://doi.org/10.1103/PhysRevD.58.063001} {\bibfield
  {journal} {\bibinfo  {journal} {Phys. Rev. D}\ }\textbf {\bibinfo {volume}
  {58}},\ \bibinfo {pages} {063001} (\bibinfo {year} {1998})}\BibitemShut
  {NoStop}%
\bibitem [{\citenamefont {Prix}(2007)}]{Prix2007}%
  \BibitemOpen
  \bibfield  {author} {\bibinfo {author} {\bibfnamefont {R.}~\bibnamefont
  {Prix}},\ }\bibfield  {title} {\bibinfo {title} {{Search for continuous
  gravitational waves: Metric of the multi-detector $\mathcal{F}$-statistic}},\
  }\bibfield  {journal} {\bibinfo  {journal} {Phys. Rev. D}\ }\textbf {\bibinfo
  {volume} {75}},\ \bibinfo {pages} {023004}, \ \href {https://doi.org/10.1103/PhysRevD.75.023004}
  {10.1103/PhysRevD.75.023004} (\bibinfo {year} {2007})\BibitemShut {NoStop}%
\bibitem [{\citenamefont {Prix}\ and\ \citenamefont {Krishnan}(2009)}]{Prix}%
  \BibitemOpen
  \bibfield  {author} {\bibinfo {author} {\bibfnamefont {R.}~\bibnamefont
  {Prix}}\ and\ \bibinfo {author} {\bibfnamefont {B.}~\bibnamefont
  {Krishnan}},\ }\bibfield  {title} {\bibinfo {title} {{Targeted search for
  continuous gravitational waves: Bayesian versus maximum-likelihood
  statistics}},\ }\href {https://doi.org/10.1088/0264-9381/26/20/204013}
  {\bibfield  {journal} {\bibinfo  {journal} {Class. Quantum Gravity}\ }\textbf
  {\bibinfo {volume} {26}},\ \bibinfo {pages} {204013} (\bibinfo {year}
  {2009})}\BibitemShut {NoStop}%
\bibitem [{\citenamefont {{Owen}}(1996)}]{Owen1996-STmGrvWInsBnCTmS}%
  \BibitemOpen
  \bibfield  {author} {\bibinfo {author} {\bibfnamefont {B.~J.}\ \bibnamefont
  {{Owen}}},\ }\bibfield  {title} {\bibinfo {title} {Search templates for
  gravitational waves from inspiraling binaries: {C}hoice of template
  spacing},\ }\href {https://doi.org/10.1103/PhysRevD.53.6749} {\bibfield
  {journal} {\bibinfo  {journal} {Phys. Rev. D}\ }\textbf {\bibinfo {volume}
  {53}},\ \bibinfo {pages} {6749} (\bibinfo {year} {1996})}\BibitemShut
  {NoStop}%
\bibitem [{\citenamefont {Astone}\ \emph {et~al.}(2002)\citenamefont {Astone},
  \citenamefont {Borkowski}, \citenamefont {Jaranowski},\ and\ \citenamefont
  {Kr{\'{o}}lak}}]{Astone2002}%
  \BibitemOpen
  \bibfield  {author} {\bibinfo {author} {\bibfnamefont {P.}~\bibnamefont
  {Astone}}, \bibinfo {author} {\bibfnamefont {K.~M.}\ \bibnamefont
  {Borkowski}}, \bibinfo {author} {\bibfnamefont {P.}~\bibnamefont
  {Jaranowski}},\ and\ \bibinfo {author} {\bibfnamefont {A.}~\bibnamefont
  {Kr{\'{o}}lak}},\ }\bibfield  {title} {\bibinfo {title} {{Data analysis of
  gravitational-wave signals from spinning neutron stars. IV. An all-sky
  search}},\ }\href {https://doi.org/10.1103/PhysRevD.65.042003} {\bibfield
  {journal} {\bibinfo  {journal} {Phys. Rev. D}\ }\textbf {\bibinfo {volume}
  {65}},\ \bibinfo {pages} {042003} (\bibinfo {year} {2002})}\BibitemShut
  {NoStop}%
\bibitem [{\citenamefont {Brady}\ \emph {et~al.}(1998)\citenamefont {Brady},
  \citenamefont {Creighton}, \citenamefont {Cutler},\ and\ \citenamefont
  {Schutz}}]{Brady1998}%
  \BibitemOpen
  \bibfield  {author} {\bibinfo {author} {\bibfnamefont {P.~R.}\ \bibnamefont
  {Brady}}, \bibinfo {author} {\bibfnamefont {T.}~\bibnamefont {Creighton}},
  \bibinfo {author} {\bibfnamefont {C.}~\bibnamefont {Cutler}},\ and\ \bibinfo
  {author} {\bibfnamefont {B.~F.}\ \bibnamefont {Schutz}},\ }\bibfield  {title}
  {\bibinfo {title} {{Searching for periodic sources with LIGO}},\ }\href
  {https://doi.org/10.1103/PhysRevD.57.2101} {\bibfield  {journal} {\bibinfo
  {journal} {Phys. Rev. D}\ }\textbf {\bibinfo {volume} {57}},\ \bibinfo
  {pages} {2101} (\bibinfo {year} {1998})}\BibitemShut {NoStop}%
\bibitem [{\citenamefont {Abbott}\ \emph
  {et~al.}(2022{\natexlab{a}})\citenamefont {Abbott} \emph
  {et~al.}}]{Abbott2022a}%
  \BibitemOpen
  \bibfield  {author} {\bibinfo {author} {\bibfnamefont {R.}~\bibnamefont
  {Abbott}} \emph {et~al.} (\bibinfo {collaboration} {LIGO Scientific
  Collaboration and Virgo Collaboration}),\ }\bibfield  {title} {\bibinfo
  {title} {{Search of the early O3 LIGO data for continuous gravitational waves
  from the Cassiopeia A and Vela Jr. supernova remnants}},\ }\href
  {https://doi.org/10.1103/PhysRevD.105.082005} {\bibfield  {journal} {\bibinfo
   {journal} {Phys. Rev. D}\ }\textbf {\bibinfo {volume} {105}},\ \bibinfo
  {pages} {082005} (\bibinfo {year} {2022}{\natexlab{a}})}\BibitemShut
  {NoStop}%
\bibitem [{\citenamefont {Owen}\ \emph {et~al.}(2022)\citenamefont {Owen},
  \citenamefont {Lindblom},\ and\ \citenamefont {Pinheiro}}]{Owen2022}%
  \BibitemOpen
  \bibfield  {author} {\bibinfo {author} {\bibfnamefont {B.~J.}\ \bibnamefont
  {Owen}}, \bibinfo {author} {\bibfnamefont {L.}~\bibnamefont {Lindblom}},\
  and\ \bibinfo {author} {\bibfnamefont {L.~S.}\ \bibnamefont {Pinheiro}},\
  }\bibfield  {title} {\bibinfo {title} {{First Constraining Upper Limits on
  Gravitational-wave Emission from NS 1987A in SNR 1987A}},\ }\href
  {https://doi.org/10.3847/2041-8213/ac84dc} {\bibfield  {journal} {\bibinfo
  {journal} {Astrophys. J. Lett.}\ }\textbf {\bibinfo {volume} {935}},\
  \bibinfo {pages} {L7} (\bibinfo {year} {2022})}\BibitemShut {NoStop}%
\bibitem [{\citenamefont {Abbott}\ \emph
  {et~al.}(2022{\natexlab{b}})\citenamefont {Abbott} \emph
  {et~al.}}]{Abbott2022b}%
  \BibitemOpen
  \bibfield  {author} {\bibinfo {author} {\bibfnamefont {R.}~\bibnamefont
  {Abbott}} \emph {et~al.} (\bibinfo {collaboration} {LIGO Scientific
  Collaboration, Virgo Collaboration and KAGRA Collaboration}),\ }\bibfield
  {title} {\bibinfo {title} {{All-sky search for continuous gravitational waves
  from isolated neutron stars using Advanced LIGO and Advanced Virgo O3
  data}},\ }\href {https://doi.org/10.1103/PhysRevD.106.102008} {\bibfield
  {journal} {\bibinfo  {journal} {Phys. Rev. D}\ }\textbf {\bibinfo {volume}
  {106}},\ \bibinfo {pages} {102008} (\bibinfo {year}
  {2022}{\natexlab{b}})}\BibitemShut {NoStop}%
\bibitem [{\citenamefont {Wette}(2014)}]{Wette2014}%
  \BibitemOpen
  \bibfield  {author} {\bibinfo {author} {\bibfnamefont {K.}~\bibnamefont
  {Wette}},\ }\bibfield  {title} {\bibinfo {title} {{Lattice template placement
  for coherent all-sky searches for gravitational-wave pulsars}},\ }\href
  {https://doi.org/10.1103/PhysRevD.90.122010} {\bibfield  {journal} {\bibinfo
  {journal} {Phys. Rev. D}\ }\textbf {\bibinfo {volume} {90}},\ \bibinfo
  {pages} {122010} (\bibinfo {year} {2014})}\BibitemShut {NoStop}%
\bibitem [{\citenamefont {{Prix}}(2007)}]{Prix2007-TmpSrGrvWEfLCFPrS}%
  \BibitemOpen
  \bibfield  {author} {\bibinfo {author} {\bibfnamefont {R.}~\bibnamefont
  {{Prix}}},\ }\bibfield  {title} {\bibinfo {title} {Template-based searches
  for gravitational waves: efficient lattice covering of flat parameter
  spaces},\ }\href {https://doi.org/10.1088/0264-9381/24/19/S11} {\bibfield
  {journal} {\bibinfo  {journal} {Class. Quantum Gravity}\ }\textbf {\bibinfo
  {volume} {24}},\ \bibinfo {pages} {S481} (\bibinfo {year}
  {2007})}\BibitemShut {NoStop}%
\bibitem [{\citenamefont {Cutler}\ and\ \citenamefont
  {Schutz}(2005)}]{Cutler2005}%
  \BibitemOpen
  \bibfield  {author} {\bibinfo {author} {\bibfnamefont {C.}~\bibnamefont
  {Cutler}}\ and\ \bibinfo {author} {\bibfnamefont {B.~F.}\ \bibnamefont
  {Schutz}},\ }\bibfield  {title} {\bibinfo {title} {{Generalized
  $\mathcal{F}$-statistic: Multiple detectors and multiple gravitational wave
  pulsars}},\ }\href {https://doi.org/10.1103/PhysRevD.72.063006} {\bibfield
  {journal} {\bibinfo  {journal} {Phys. Rev. D}\ }\textbf {\bibinfo {volume}
  {72}},\ \bibinfo {pages} {063006} (\bibinfo {year} {2005})}\BibitemShut
  {NoStop}%
\bibitem [{\citenamefont {Romani}\ \emph {et~al.}(2022)\citenamefont {Romani},
  \citenamefont {Kandel}, \citenamefont {Filippenko}, \citenamefont {Brink},\
  and\ \citenamefont {Zheng}}]{Romani2022}%
  \BibitemOpen
  \bibfield  {author} {\bibinfo {author} {\bibfnamefont {R.~W.}\ \bibnamefont
  {Romani}}, \bibinfo {author} {\bibfnamefont {D.}~\bibnamefont {Kandel}},
  \bibinfo {author} {\bibfnamefont {A.~V.}\ \bibnamefont {Filippenko}},
  \bibinfo {author} {\bibfnamefont {T.~G.}\ \bibnamefont {Brink}},\ and\
  \bibinfo {author} {\bibfnamefont {W.}~\bibnamefont {Zheng}},\ }\bibfield
  {title} {\bibinfo {title} {{PSR J0952-0607: The fastest and heaviest known
  galactic neutron star}},\ }\href {https://doi.org/10.3847/2041-8213/ac8007}
  {\bibfield  {journal} {\bibinfo  {journal} {Astrophys. J. Lett.}\ }\textbf
  {\bibinfo {volume} {934}},\ \bibinfo {pages} {L17} (\bibinfo {year}
  {2022})}\BibitemShut {NoStop}%
\bibitem [{\citenamefont {Antoniadis}\ \emph {et~al.}(2013)\citenamefont
  {Antoniadis} \emph {et~al.}}]{Antoniadis2013}%
  \BibitemOpen
  \bibfield  {author} {\bibinfo {author} {\bibfnamefont {J.}~\bibnamefont
  {Antoniadis}} \emph {et~al.},\ }\bibfield  {title} {\bibinfo {title} {{A
  massive pulsar in a compact relativistic binary}},\ }\href
  {https://doi.org/10.1126/science.1233232} {\bibfield  {journal} {\bibinfo
  {journal} {Science}\ }\textbf {\bibinfo {volume} {340}},\ \bibinfo {pages}
  {448} (\bibinfo {year} {2013})}\BibitemShut {NoStop}%
\bibitem [{\citenamefont {Bauswein}\ and\ \citenamefont
  {Stergioulas}(2015)}]{Bauswein2015}%
  \BibitemOpen
  \bibfield  {author} {\bibinfo {author} {\bibfnamefont {A.}~\bibnamefont
  {Bauswein}}\ and\ \bibinfo {author} {\bibfnamefont {N.}~\bibnamefont
  {Stergioulas}},\ }\bibfield  {title} {\bibinfo {title} {{Unified picture of
  the post-merger dynamics and gravitational wave emission in neutron star
  mergers}},\ }\href {https://doi.org/10.1103/PhysRevD.91.124056} {\bibfield
  {journal} {\bibinfo  {journal} {Phys. Rev. D}\ }\textbf {\bibinfo {volume}
  {91}},\ \bibinfo {pages} {124056} (\bibinfo {year} {2015})}\BibitemShut
  {NoStop}%
\bibitem [{\citenamefont {Takami}\ \emph {et~al.}(2015)\citenamefont {Takami},
  \citenamefont {Rezzolla},\ and\ \citenamefont {Baiotti}}]{Takami2015}%
  \BibitemOpen
  \bibfield  {author} {\bibinfo {author} {\bibfnamefont {K.}~\bibnamefont
  {Takami}}, \bibinfo {author} {\bibfnamefont {L.}~\bibnamefont {Rezzolla}},\
  and\ \bibinfo {author} {\bibfnamefont {L.}~\bibnamefont {Baiotti}},\
  }\bibfield  {title} {\bibinfo {title} {{Spectral properties of the
  post-merger gravitational-wave signal from binary neutron stars}},\ }\href
  {https://doi.org/10.1103/PhysRevD.91.064001} {\bibfield  {journal} {\bibinfo
  {journal} {Phys. Rev. D}\ }\textbf {\bibinfo {volume} {91}},\ \bibinfo
  {pages} {064001} (\bibinfo {year} {2015})}\BibitemShut {NoStop}%
\bibitem [{\citenamefont {Ai}\ \emph {et~al.}(2020)\citenamefont {Ai},
  \citenamefont {Gao},\ and\ \citenamefont {Zhang}}]{Ai2020}%
  \BibitemOpen
  \bibfield  {author} {\bibinfo {author} {\bibfnamefont {S.}~\bibnamefont
  {Ai}}, \bibinfo {author} {\bibfnamefont {H.}~\bibnamefont {Gao}},\ and\
  \bibinfo {author} {\bibfnamefont {B.}~\bibnamefont {Zhang}},\ }\bibfield
  {title} {\bibinfo {title} {{What Constraints on the Neutron Star Maximum Mass
  Can One Pose from GW170817 Observations?}},\ }\href
  {https://doi.org/10.3847/1538-4357/ab80bd} {\bibfield  {journal} {\bibinfo
  {journal} {Astrophys. J.}\ }\textbf {\bibinfo {volume} {893}},\ \bibinfo
  {pages} {146} (\bibinfo {year} {2020})}\BibitemShut {NoStop}%
\bibitem [{\citenamefont {Baumgarte}\ \emph {et~al.}(2000)\citenamefont
  {Baumgarte}, \citenamefont {Shapiro},\ and\ \citenamefont
  {Shibata}}]{Baumgarte2000}%
  \BibitemOpen
  \bibfield  {author} {\bibinfo {author} {\bibfnamefont {T.~W.}\ \bibnamefont
  {Baumgarte}}, \bibinfo {author} {\bibfnamefont {S.~L.}\ \bibnamefont
  {Shapiro}},\ and\ \bibinfo {author} {\bibfnamefont {M.}~\bibnamefont
  {Shibata}},\ }\bibfield  {title} {\bibinfo {title} {{On the Maximum Mass of
  Differentially Rotating Neutron Stars}},\ }\href
  {https://doi.org/10.1086/312425} {\bibfield  {journal} {\bibinfo  {journal}
  {Astrophys. J.}\ }\textbf {\bibinfo {volume} {528}},\ \bibinfo {pages} {L29}
  (\bibinfo {year} {2000})}\BibitemShut {NoStop}%
\bibitem [{\citenamefont {Hotokezaka}\ \emph {et~al.}(2013)\citenamefont
  {Hotokezaka}, \citenamefont {Kiuchi}, \citenamefont {Kyutoku}, \citenamefont
  {Muranushi}, \citenamefont {Sekiguchi}, \citenamefont {Shibata},\ and\
  \citenamefont {Taniguchi}}]{Hotokezaka2013}%
  \BibitemOpen
  \bibfield  {author} {\bibinfo {author} {\bibfnamefont {K.}~\bibnamefont
  {Hotokezaka}}, \bibinfo {author} {\bibfnamefont {K.}~\bibnamefont {Kiuchi}},
  \bibinfo {author} {\bibfnamefont {K.}~\bibnamefont {Kyutoku}}, \bibinfo
  {author} {\bibfnamefont {T.}~\bibnamefont {Muranushi}}, \bibinfo {author}
  {\bibfnamefont {Y.i.}\ \bibnamefont {Sekiguchi}}, \bibinfo {author}
  {\bibfnamefont {M.}~\bibnamefont {Shibata}},\ and\ \bibinfo {author}
  {\bibfnamefont {K.}~\bibnamefont {Taniguchi}},\ }\bibfield  {title} {\bibinfo
  {title} {{Remnant massive neutron stars of binary neutron star mergers:
  Evolution process and gravitational waveform}},\ }\href
  {https://doi.org/10.1103/PhysRevD.88.044026} {\bibfield  {journal} {\bibinfo
  {journal} {Phys. Rev. D}\ }\textbf {\bibinfo {volume} {88}},\ \bibinfo
  {pages} {044026} (\bibinfo {year} {2013})}\BibitemShut {NoStop}%
\bibitem [{\citenamefont {Shapiro}(2000)}]{Shapiro2000}%
  \BibitemOpen
  \bibfield  {author} {\bibinfo {author} {\bibfnamefont {S.~L.}\ \bibnamefont
  {Shapiro}},\ }\bibfield  {title} {\bibinfo {title} {{Differential Rotation in
  Neutron Stars: Magnetic Braking and Viscous Damping}},\ }\href
  {https://doi.org/10.1086/317209} {\bibfield  {journal} {\bibinfo  {journal}
  {Astrophys. J.}\ }\textbf {\bibinfo {volume} {544}},\ \bibinfo {pages} {397}
  (\bibinfo {year} {2000})}\BibitemShut {NoStop}%
\bibitem [{\citenamefont {Suvorov}\ and\ \citenamefont
  {Glampedakis}(2022)}]{Suvorov2022}%
  \BibitemOpen
  \bibfield  {author} {\bibinfo {author} {\bibfnamefont {A.~G.}\ \bibnamefont
  {Suvorov}}\ and\ \bibinfo {author} {\bibfnamefont {K.}~\bibnamefont
  {Glampedakis}},\ }\bibfield  {title} {\bibinfo {title} {{Magnetically
  supramassive neutron stars}},\ }\href
  {https://doi.org/10.1103/PhysRevD.105.L061302} {\bibfield  {journal}
  {\bibinfo  {journal} {Phys. Rev. D}\ }\textbf {\bibinfo {volume} {105}},\
  \bibinfo {pages} {L061302} (\bibinfo {year} {2022})}\BibitemShut {NoStop}%
\bibitem [{\citenamefont {Shibata}\ \emph {et~al.}(2005)\citenamefont
  {Shibata}, \citenamefont {Taniguchi},\ and\ \citenamefont
  {Uryu}}]{Shibata2005}%
  \BibitemOpen
  \bibfield  {author} {\bibinfo {author} {\bibfnamefont {M.}~\bibnamefont
  {Shibata}}, \bibinfo {author} {\bibfnamefont {K.}~\bibnamefont {Taniguchi}},\
  and\ \bibinfo {author} {\bibfnamefont {K.}~\bibnamefont {Uryu}},\ }\bibfield
  {title} {\bibinfo {title} {{Merger of binary neutron stars with realistic
  equations of state in full general relativity}},\ }\href
  {https://doi.org/10.1103/PhysRevD.71.084021} {\bibfield  {journal} {\bibinfo
  {journal} {Phys. Rev. D}\ }\textbf {\bibinfo {volume} {71}},\ \bibinfo
  {pages} {084021} (\bibinfo {year} {2005})}\BibitemShut {NoStop}%
\bibitem [{\citenamefont {Colaiuda}\ \emph {et~al.}(2008)\citenamefont
  {Colaiuda}, \citenamefont {Ferrari}, \citenamefont {Gualtieri},\ and\
  \citenamefont {Pons}}]{Colaiuda2008}%
  \BibitemOpen
  \bibfield  {author} {\bibinfo {author} {\bibfnamefont {A.}~\bibnamefont
  {Colaiuda}}, \bibinfo {author} {\bibfnamefont {V.}~\bibnamefont {Ferrari}},
  \bibinfo {author} {\bibfnamefont {L.}~\bibnamefont {Gualtieri}},\ and\
  \bibinfo {author} {\bibfnamefont {J.~A.}\ \bibnamefont {Pons}},\ }\bibfield
  {title} {\bibinfo {title} {{Relativistic models of magnetars: Structure and
  deformations}},\ }\href {https://doi.org/10.1111/j.1365-2966.2008.12966.x}
  {\bibfield  {journal} {\bibinfo  {journal} {Mon. Not. R. Astron. Soc.}\
  }\textbf {\bibinfo {volume} {385}},\ \bibinfo {pages} {2080} (\bibinfo {year}
  {2008})}\BibitemShut {NoStop}%
\bibitem [{\citenamefont {Haskell}\ \emph {et~al.}(2007)\citenamefont
  {Haskell}, \citenamefont {Samuelsson}, \citenamefont {Glampedakis},\ and\
  \citenamefont {Andersson}}]{Haskell2007a}%
  \BibitemOpen
  \bibfield  {author} {\bibinfo {author} {\bibfnamefont {B.}~\bibnamefont
  {Haskell}}, \bibinfo {author} {\bibfnamefont {L.}~\bibnamefont {Samuelsson}},
  \bibinfo {author} {\bibfnamefont {K.}~\bibnamefont {Glampedakis}},\ and\
  \bibinfo {author} {\bibfnamefont {N.}~\bibnamefont {Andersson}},\ }\bibfield
  {title} {\bibinfo {title} {{Modelling magnetically deformed neutron stars}},\
  }\href {https://doi.org/10.1111/j.1365-2966.2008.12861.x} {\bibfield
  {journal} {\bibinfo  {journal} {Mon. Not. R. Astron. Soc.}\ }\textbf
  {\bibinfo {volume} {385}},\ \bibinfo {pages} {531} (\bibinfo {year}
  {2007})}\BibitemShut {NoStop}%
\bibitem [{\citenamefont {Ciolfi}\ \emph {et~al.}(2010)\citenamefont {Ciolfi},
  \citenamefont {Ferrari},\ and\ \citenamefont {Gualtieri}}]{Ciolfi2010}%
  \BibitemOpen
  \bibfield  {author} {\bibinfo {author} {\bibfnamefont {R.}~\bibnamefont
  {Ciolfi}}, \bibinfo {author} {\bibfnamefont {V.}~\bibnamefont {Ferrari}},\
  and\ \bibinfo {author} {\bibfnamefont {L.}~\bibnamefont {Gualtieri}},\
  }\bibfield  {title} {\bibinfo {title} {{Structure and deformations of
  strongly magnetized neutron stars with twisted-torus configurations}},\
  }\href {https://doi.org/10.1111/j.1365-2966.2010.16847.x} {\bibfield
  {journal} {\bibinfo  {journal} {Mon. Not. R. Astron. Soc.}\ }\textbf
  {\bibinfo {volume} {406}},\ \bibinfo {pages} {2540} (\bibinfo {year}
  {2010})}\BibitemShut {NoStop}%
\bibitem [{\citenamefont {Haskell}\ \emph {et~al.}(2006)\citenamefont
  {Haskell}, \citenamefont {Jones},\ and\ \citenamefont
  {Andersson}}]{Haskell2006}%
  \BibitemOpen
  \bibfield  {author} {\bibinfo {author} {\bibfnamefont {B.}~\bibnamefont
  {Haskell}}, \bibinfo {author} {\bibfnamefont {D.~I.}\ \bibnamefont {Jones}},\
  and\ \bibinfo {author} {\bibfnamefont {N.}~\bibnamefont {Andersson}},\
  }\bibfield  {title} {\bibinfo {title} {{Mountains on Neutron Stars: Accreted
  vs. Non-Accreted crusts}},\ }\href
  {https://doi.org/10.1111/j.1365-2966.2006.10998.x} {\bibfield  {journal}
  {\bibinfo  {journal} {Mon. Not. R. Astron. Soc.}\ }\textbf {\bibinfo {volume}
  {373}},\ \bibinfo {pages} {1423} (\bibinfo {year} {2006})}\BibitemShut
  {NoStop}%
\bibitem [{\citenamefont {Ushomirsky}\ \emph {et~al.}(2000)\citenamefont
  {Ushomirsky}, \citenamefont {Cutler},\ and\ \citenamefont
  {Bildsten}}]{Ushomirsky2000}%
  \BibitemOpen
  \bibfield  {author} {\bibinfo {author} {\bibfnamefont {G.}~\bibnamefont
  {Ushomirsky}}, \bibinfo {author} {\bibfnamefont {C.}~\bibnamefont {Cutler}},\
  and\ \bibinfo {author} {\bibfnamefont {L.}~\bibnamefont {Bildsten}},\
  }\bibfield  {title} {\bibinfo {title} {{Deformations of accreting neutron
  star crusts and gravitational wave emission}},\ }\href
  {https://doi.org/10.1046/j.1365-8711.2000.03938.x} {\bibfield  {journal}
  {\bibinfo  {journal} {Mon. Not. R. Astron. Soc.}\ }\textbf {\bibinfo {volume}
  {319}},\ \bibinfo {pages} {902} (\bibinfo {year} {2000})}\BibitemShut
  {NoStop}%
\bibitem [{\citenamefont {{De Lillo}}\ \emph {et~al.}(2022)\citenamefont {{De
  Lillo}}, \citenamefont {Suresh},\ and\ \citenamefont {Miller}}]{DeLillo2022}%
  \BibitemOpen
  \bibfield  {author} {\bibinfo {author} {\bibfnamefont {F.}~\bibnamefont {{De
  Lillo}}}, \bibinfo {author} {\bibfnamefont {J.}~\bibnamefont {Suresh}},\ and\
  \bibinfo {author} {\bibfnamefont {A.~L.}\ \bibnamefont {Miller}},\ }\bibfield
   {title} {\bibinfo {title} {{Stochastic gravitational-wave background
  searches and constraints on neutron-star ellipticity}},\ }\href
  {https://doi.org/10.1093/mnras/stac984} {\bibfield  {journal} {\bibinfo
  {journal} {Mon. Not. R. Astron. Soc.}\ }\textbf {\bibinfo {volume} {513}},\
  \bibinfo {pages} {1105} (\bibinfo {year} {2022})}\BibitemShut {NoStop}%
\bibitem [{\citenamefont {Gill}\ \emph {et~al.}(2019)\citenamefont {Gill},
  \citenamefont {Nathanail},\ and\ \citenamefont {Rezzolla}}]{Gill2019}%
  \BibitemOpen
  \bibfield  {author} {\bibinfo {author} {\bibfnamefont {R.}~\bibnamefont
  {Gill}}, \bibinfo {author} {\bibfnamefont {A.}~\bibnamefont {Nathanail}},\
  and\ \bibinfo {author} {\bibfnamefont {L.}~\bibnamefont {Rezzolla}},\
  }\bibfield  {title} {\bibinfo {title} {{When Did the Remnant of GW170817
  Collapse to a Black Hole?}},\ }\href
  {https://doi.org/10.3847/1538-4357/ab16da} {\bibfield  {journal} {\bibinfo
  {journal} {Astrophys. J.}\ }\textbf {\bibinfo {volume} {876}},\ \bibinfo
  {pages} {139} (\bibinfo {year} {2019})}\BibitemShut {NoStop}%
\bibitem [{\citenamefont {Thrane}\ \emph {et~al.}(2011)\citenamefont {Thrane},
  \citenamefont {Kandhasamy}, \citenamefont {Ott}, \citenamefont {Anderson},
  \citenamefont {Christensen}, \citenamefont {Coughlin}, \citenamefont
  {Dorsher}, \citenamefont {Giampanis}, \citenamefont {Mandic}, \citenamefont
  {Mytidis}, \citenamefont {Prestegard}, \citenamefont {Raffai},\ and\
  \citenamefont {Whiting}}]{Thrane2011}%
  \BibitemOpen
  \bibfield  {author} {\bibinfo {author} {\bibfnamefont {E.}~\bibnamefont
  {Thrane}}, \bibinfo {author} {\bibfnamefont {S.}~\bibnamefont {Kandhasamy}},
  \bibinfo {author} {\bibfnamefont {C.~D.}\ \bibnamefont {Ott}}, \bibinfo
  {author} {\bibfnamefont {W.~G.}\ \bibnamefont {Anderson}}, \bibinfo {author}
  {\bibfnamefont {N.~L.}\ \bibnamefont {Christensen}}, \bibinfo {author}
  {\bibfnamefont {M.~W.}\ \bibnamefont {Coughlin}}, \bibinfo {author}
  {\bibfnamefont {S.}~\bibnamefont {Dorsher}}, \bibinfo {author} {\bibfnamefont
  {S.}~\bibnamefont {Giampanis}}, \bibinfo {author} {\bibfnamefont
  {V.}~\bibnamefont {Mandic}}, \bibinfo {author} {\bibfnamefont
  {A.}~\bibnamefont {Mytidis}}, \bibinfo {author} {\bibfnamefont
  {T.}~\bibnamefont {Prestegard}}, \bibinfo {author} {\bibfnamefont
  {P.}~\bibnamefont {Raffai}},\ and\ \bibinfo {author} {\bibfnamefont
  {B.}~\bibnamefont {Whiting}},\ }\bibfield  {title} {\bibinfo {title} {{Long
  gravitational-wave transients and associated detection strategies for a
  network of terrestrial interferometers}},\ }\href
  {https://doi.org/10.1103/PhysRevD.83.083004} {\bibfield  {journal} {\bibinfo
  {journal} {Phys. Rev. D}\ }\textbf {\bibinfo {volume} {83}},\ \bibinfo
  {pages} {083004} (\bibinfo {year} {2011})}\BibitemShut {NoStop}%
\bibitem [{\citenamefont {Klimenko}\ \emph {et~al.}(2008)\citenamefont
  {Klimenko}, \citenamefont {Yakushin}, \citenamefont {Mercer},\ and\
  \citenamefont {Mitselmakher}}]{Klimenko2008}%
  \BibitemOpen
  \bibfield  {author} {\bibinfo {author} {\bibfnamefont {S.}~\bibnamefont
  {Klimenko}}, \bibinfo {author} {\bibfnamefont {I.}~\bibnamefont {Yakushin}},
  \bibinfo {author} {\bibfnamefont {A.}~\bibnamefont {Mercer}},\ and\ \bibinfo
  {author} {\bibfnamefont {G.}~\bibnamefont {Mitselmakher}},\ }\bibfield
  {title} {\bibinfo {title} {{A coherent method for detection of gravitational
  wave bursts}},\ }\href {https://doi.org/10.1088/0264-9381/25/11/114029}
  {\bibfield  {journal} {\bibinfo  {journal} {Class. Quantum Gravity}\ }\textbf
  {\bibinfo {volume} {25}},\ \bibinfo {pages} {114029} (\bibinfo {year}
  {2008})}\BibitemShut {NoStop}%
\bibitem [{\citenamefont {Viterbi}(1967)}]{Viterbi1967}%
  \BibitemOpen
  \bibfield  {author} {\bibinfo {author} {\bibfnamefont {A.~J.}\ \bibnamefont
  {Viterbi}},\ }\bibfield  {title} {\bibinfo {title} {{Error Bounds for
  Convolutional Codes and an Asymptotically Optimum Decoding Algorithm}},\
  }\href {https://doi.org/10.1109/tit.1967.1054010} {\bibfield  {journal}
  {\bibinfo  {journal} {IEEE Trans. Inf. Theory}\ }\textbf {\bibinfo {volume}
  {13}},\ \bibinfo {pages} {260} (\bibinfo {year} {1967})}\BibitemShut
  {NoStop}%
\bibitem [{\citenamefont {Suvorova}\ \emph {et~al.}(2016)\citenamefont
  {Suvorova}, \citenamefont {Sun}, \citenamefont {Melatos}, \citenamefont
  {Moran},\ and\ \citenamefont {Evans}}]{Suvorova2016}%
  \BibitemOpen
  \bibfield  {author} {\bibinfo {author} {\bibfnamefont {S.}~\bibnamefont
  {Suvorova}}, \bibinfo {author} {\bibfnamefont {L.}~\bibnamefont {Sun}},
  \bibinfo {author} {\bibfnamefont {A.}~\bibnamefont {Melatos}}, \bibinfo
  {author} {\bibfnamefont {W.}~\bibnamefont {Moran}},\ and\ \bibinfo {author}
  {\bibfnamefont {R.~J.}\ \bibnamefont {Evans}},\ }\bibfield  {title} {\bibinfo
  {title} {{Hidden Markov model tracking of continuous gravitational waves from
  a neutron star with wandering spin}},\ }\href
  {https://doi.org/10.1103/PhysRevD.93.123009} {\bibfield  {journal} {\bibinfo
  {journal} {Phys. Rev. D}\ }\textbf {\bibinfo {volume} {93}},\ \bibinfo
  {pages} {123009} (\bibinfo {year} {2016})}\BibitemShut {NoStop}%
\bibitem [{\citenamefont {Sun}\ and\ \citenamefont {Melatos}(2019)}]{Sun2019}%
  \BibitemOpen
  \bibfield  {author} {\bibinfo {author} {\bibfnamefont {L.}~\bibnamefont
  {Sun}}\ and\ \bibinfo {author} {\bibfnamefont {A.}~\bibnamefont {Melatos}},\
  }\bibfield  {title} {\bibinfo {title} {{Application of hidden Markov model
  tracking to the search for long-duration transient gravitational waves from
  the remnant of the binary neutron star merger GW170817}},\ }\href
  {https://doi.org/10.1103/PhysRevD.99.123003} {\bibfield  {journal} {\bibinfo
  {journal} {Phys. Rev. D}\ }\textbf {\bibinfo {volume} {99}},\ \bibinfo
  {pages} {123003} (\bibinfo {year} {2019})}\BibitemShut {NoStop}%
\bibitem [{\citenamefont {{Hough}}(1959)}]{Houg1959-McAnlBbChPct}%
  \BibitemOpen
  \bibfield  {author} {\bibinfo {author} {\bibfnamefont {P.~V.~C.}\
  \bibnamefont {{Hough}}},\ }\bibfield  {title} {\bibinfo {title} {Machine
  {A}nalysis of {B}ubble {C}hamber {P}ictures},\ }in\ \href@noop {} {\emph
  {\bibinfo {booktitle} {2nd {I}nternational {C}onference on {H}igh-{E}nergy
  {A}ccelerators and {I}nstrumentation}}},\ Vol.\ \bibinfo {volume} {C590914},\
  \bibinfo {editor} {edited by\ \bibinfo {editor} {\bibfnamefont
  {L.}~\bibnamefont {{Kowarski}}}}\ (\bibinfo  {publisher} {CERN},\ \bibinfo
  {address} {Geneva},\ \bibinfo {year} {1959})\ p.\ \bibinfo {pages}
  {554}\BibitemShut {NoStop}%
\bibitem [{\citenamefont {Astone}\ \emph {et~al.}(2014)\citenamefont {Astone},
  \citenamefont {Colla}, \citenamefont {D'Antonio}, \citenamefont {Frasca},\
  and\ \citenamefont {Palomba}}]{Astone:2014esa}%
  \BibitemOpen
  \bibfield  {author} {\bibinfo {author} {\bibfnamefont {P.}~\bibnamefont
  {Astone}}, \bibinfo {author} {\bibfnamefont {A.}~\bibnamefont {Colla}},
  \bibinfo {author} {\bibfnamefont {S.}~\bibnamefont {D'Antonio}}, \bibinfo
  {author} {\bibfnamefont {S.}~\bibnamefont {Frasca}},\ and\ \bibinfo {author}
  {\bibfnamefont {C.}~\bibnamefont {Palomba}},\ }\bibfield  {title} {\bibinfo
  {title} {{Method for all-sky searches of continuous gravitational wave
  signals using the frequency-Hough transform}},\ }\href
  {https://doi.org/10.1103/PhysRevD.90.042002} {\bibfield  {journal} {\bibinfo
  {journal} {Phys. Rev. D}\ }\textbf {\bibinfo {volume} {90}},\ \bibinfo
  {pages} {042002} (\bibinfo {year} {2014})}\BibitemShut {NoStop}%
\bibitem [{\citenamefont {Miller}\ \emph {et~al.}(2018)\citenamefont {Miller}
  \emph {et~al.}}]{Miller:2018rbg}%
  \BibitemOpen
  \bibfield  {author} {\bibinfo {author} {\bibfnamefont {A.}~\bibnamefont
  {Miller}} \emph {et~al.},\ }\bibfield  {title} {\bibinfo {title} {Method to
  search for long duration gravitational wave transients from isolated neutron
  stars using the generalized frequency-hough transform},\ }\href
  {https://doi.org/10.1103/PhysRevD.98.102004} {\bibfield  {journal} {\bibinfo
  {journal} {Phys. Rev. D}\ }\textbf {\bibinfo {volume} {98}},\ \bibinfo
  {pages} {102004} (\bibinfo {year} {2018})}\BibitemShut {NoStop}%
\bibitem [{\citenamefont {Oliver}\ \emph {et~al.}(2019)\citenamefont {Oliver},
  \citenamefont {Keitel},\ and\ \citenamefont {Sintes}}]{Oliver2019}%
  \BibitemOpen
  \bibfield  {author} {\bibinfo {author} {\bibfnamefont {M.}~\bibnamefont
  {Oliver}}, \bibinfo {author} {\bibfnamefont {D.}~\bibnamefont {Keitel}},\
  and\ \bibinfo {author} {\bibfnamefont {A.~M.}\ \bibnamefont {Sintes}},\
  }\bibfield  {title} {\bibinfo {title} {{Adaptive transient Hough method for
  long-duration gravitational wave transients}},\ }\href
  {https://doi.org/10.1103/PhysRevD.99.104067} {\bibfield  {journal} {\bibinfo
  {journal} {Phys. Rev. D}\ }\textbf {\bibinfo {volume} {99}},\ \bibinfo
  {pages} {104067} (\bibinfo {year} {2019})}\BibitemShut {NoStop}%
\bibitem [{\citenamefont {Helou}\ \emph {et~al.}(2015)\citenamefont {Helou},
  \citenamefont {Musco}, \citenamefont {{Miller -}}, \citenamefont
  {Berra-Montiel}, \citenamefont {Molgado}, \citenamefont
  {{Rodr{\'{i}}guez-L{\'{o}}pez -}}, \citenamefont {Shao}, \citenamefont {Li},
  \citenamefont {Cornish},\ and\ \citenamefont {Littenberg}}]{Helou2015}%
  \BibitemOpen
  \bibfield  {author} {\bibinfo {author} {\bibfnamefont {A.}~\bibnamefont
  {Helou}}, \bibinfo {author} {\bibfnamefont {I.}~\bibnamefont {Musco}},
  \bibinfo {author} {\bibfnamefont {J.~C.}\ \bibnamefont {{Miller -}}},
  \bibinfo {author} {\bibfnamefont {J.}~\bibnamefont {Berra-Montiel}}, \bibinfo
  {author} {\bibfnamefont {A.}~\bibnamefont {Molgado}}, \bibinfo {author}
  {\bibfnamefont {{\'{A}}.}~\bibnamefont {{Rodr{\'{i}}guez-L{\'{o}}pez -}}},
  \bibinfo {author} {\bibfnamefont {H.-C.}\ \bibnamefont {Shao}}, \bibinfo
  {author} {\bibfnamefont {T.}~\bibnamefont {Li}}, \bibinfo {author}
  {\bibfnamefont {N.~J.}\ \bibnamefont {Cornish}},\ and\ \bibinfo {author}
  {\bibfnamefont {T.~B.}\ \bibnamefont {Littenberg}},\ }\bibfield  {title}
  {\bibinfo {title} {{Bayeswave: Bayesian inference for gravitational wave
  bursts and instrument glitches}},\ }\href
  {https://doi.org/10.1088/0264-9381/32/13/135012} {\bibfield  {journal}
  {\bibinfo  {journal} {Class. Quantum Gravity}\ }\textbf {\bibinfo {volume}
  {32}},\ \bibinfo {pages} {135012} (\bibinfo {year} {2015})}\BibitemShut
  {NoStop}%
\bibitem [{\citenamefont {Coughlin}\ \emph {et~al.}(2019)\citenamefont
  {Coughlin} \emph {et~al.}}]{Coughlin2019}%
  \BibitemOpen
  \bibfield  {author} {\bibinfo {author} {\bibfnamefont {M.~W.}\ \bibnamefont
  {Coughlin}} \emph {et~al.},\ }\bibfield  {title} {\bibinfo {title} {{GROWTH
  on S190425z: Searching Thousands of Square Degrees to Identify an Optical or
  Infrared Counterpart to a Binary Neutron Star Merger with the Zwicky
  Transient Facility and Palomar Gattini-IR}},\ }\href
  {https://doi.org/10.3847/2041-8213/ab4ad8} {\bibfield  {journal} {\bibinfo
  {journal} {Astrophys. J. Lett.}\ }\textbf {\bibinfo {volume} {885}},\
  \bibinfo {pages} {L19} (\bibinfo {year} {2019})}\BibitemShut {NoStop}%
\bibitem [{\citenamefont {Hosseinzadeh}\ \emph {et~al.}(2019)\citenamefont
  {Hosseinzadeh} \emph {et~al.}}]{Hosseinzadeh2019}%
  \BibitemOpen
  \bibfield  {author} {\bibinfo {author} {\bibfnamefont {G.}~\bibnamefont
  {Hosseinzadeh}} \emph {et~al.},\ }\bibfield  {title} {\bibinfo {title}
  {{Follow-up of the Neutron Star Bearing Gravitational-wave Candidate Events
  S190425z and S190426c with MMT and SOAR}},\ }\href
  {https://doi.org/10.3847/2041-8213/ab271c} {\bibfield  {journal} {\bibinfo
  {journal} {Astrophys. J. Lett.}\ }\textbf {\bibinfo {volume} {880}},\
  \bibinfo {pages} {L4} (\bibinfo {year} {2019})}\BibitemShut {NoStop}%
\bibitem [{\citenamefont {Ostriker}\ and\ \citenamefont
  {Gunn}(1969)}]{Ostriker1969}%
  \BibitemOpen
  \bibfield  {author} {\bibinfo {author} {\bibfnamefont {J.~P.}\ \bibnamefont
  {Ostriker}}\ and\ \bibinfo {author} {\bibfnamefont {J.~E.}\ \bibnamefont
  {Gunn}},\ }\bibfield  {title} {\bibinfo {title} {{On the Nature of Pulsars.
  I. Theory}},\ }\href {https://doi.org/10.1086/150160} {\bibfield  {journal}
  {\bibinfo  {journal} {Astrophys. J.}\ }\textbf {\bibinfo {volume} {157}},\
  \bibinfo {pages} {1395} (\bibinfo {year} {1969})}\BibitemShut {NoStop}%
\bibitem [{\citenamefont {Aasi}\ \emph {et~al.}(2015)\citenamefont {Aasi} \emph
  {et~al.}}]{Aasi2015}%
  \BibitemOpen
  \bibfield  {author} {\bibinfo {author} {\bibfnamefont {J.}~\bibnamefont
  {Aasi}} \emph {et~al.} (\bibinfo {collaboration} {LIGO Scientific
  Collaboration}),\ }\bibfield  {title} {\bibinfo {title} {{Advanced LIGO}},\
  }\href {https://doi.org/10.1088/0264-9381/32/7/074001} {\bibfield  {journal}
  {\bibinfo  {journal} {Class. Quantum Gravity}\ }\textbf {\bibinfo {volume}
  {32}},\ \bibinfo {pages} {074001} (\bibinfo {year} {2015})}\BibitemShut
  {NoStop}%
\bibitem [{\citenamefont {Acernese}\ \emph {et~al.}(2014)\citenamefont
  {Acernese} \emph {et~al.}}]{Acernese2014}%
  \BibitemOpen
  \bibfield  {author} {\bibinfo {author} {\bibfnamefont {F.}~\bibnamefont
  {Acernese}} \emph {et~al.},\ }\bibfield  {title} {\bibinfo {title} {{Advanced
  Virgo: a second-generation interferometric gravitational wave detector}},\
  }\href {https://doi.org/10.1088/0264-9381/32/2/024001} {\bibfield  {journal}
  {\bibinfo  {journal} {Class. Quantum Gravity}\ }\textbf {\bibinfo {volume}
  {32}},\ \bibinfo {pages} {024001} (\bibinfo {year} {2014})}\BibitemShut
  {NoStop}%
\bibitem [{\citenamefont {Abbott}\ \emph {et~al.}(2023)\citenamefont {Abbott}
  \emph {et~al.}}]{Abbott2023}%
  \BibitemOpen
  \bibfield  {author} {\bibinfo {author} {\bibfnamefont {R.}~\bibnamefont
  {Abbott}} \emph {et~al.} (\bibinfo {collaboration} {LIGO Scientific
  Collaboration, the Virgo Collaboration and the KAGRA Collaboration}),\
  }\bibfield  {title} {\bibinfo {title} {{Open Data from the Third Observing
  Run of LIGO, Virgo, KAGRA, and GEO}},\ }\href
  {https://doi.org/10.3847/1538-4365/acdc9f} {\bibfield  {journal} {\bibinfo
  {journal} {Astrophys. J. Suppl. Ser.}\ }\textbf {\bibinfo {volume} {267}},\
  \bibinfo {pages} {29} (\bibinfo {year} {2023})}\BibitemShut {NoStop}%
\bibitem [{\citenamefont {{Cahillane}}\ \emph {et~al.}(2017)\citenamefont
  {{Cahillane}}, \citenamefont {{Betzwieser}}, \citenamefont {{Brown}},
  \citenamefont {{Goetz}}, \citenamefont {{Hall}}, \citenamefont {{Izumi}},
  \citenamefont {{Kandhasamy}}, \citenamefont {{Karki}}, \citenamefont
  {{Kissel}}, \citenamefont {{Mendell}}, \citenamefont {{Savage}},
  \citenamefont {{Tuyenbayev}}, \citenamefont {{Urban}}, \citenamefont
  {{Viets}}, \citenamefont {{Wade}},\ and\ \citenamefont
  {{Weinstein}}}]{CahiEtAl2017-ClbUncAdLFScObR}%
  \BibitemOpen
  \bibfield  {author} {\bibinfo {author} {\bibfnamefont {C.}~\bibnamefont
  {{Cahillane}}}, \bibinfo {author} {\bibfnamefont {J.}~\bibnamefont
  {{Betzwieser}}}, \bibinfo {author} {\bibfnamefont {D.~A.}\ \bibnamefont
  {{Brown}}}, \bibinfo {author} {\bibfnamefont {E.}~\bibnamefont {{Goetz}}},
  \bibinfo {author} {\bibfnamefont {E.~D.}\ \bibnamefont {{Hall}}}, \bibinfo
  {author} {\bibfnamefont {K.}~\bibnamefont {{Izumi}}}, \bibinfo {author}
  {\bibfnamefont {S.}~\bibnamefont {{Kandhasamy}}}, \bibinfo {author}
  {\bibfnamefont {S.}~\bibnamefont {{Karki}}}, \bibinfo {author} {\bibfnamefont
  {J.~S.}\ \bibnamefont {{Kissel}}}, \bibinfo {author} {\bibfnamefont
  {G.}~\bibnamefont {{Mendell}}}, \bibinfo {author} {\bibfnamefont {R.~L.}\
  \bibnamefont {{Savage}}}, \bibinfo {author} {\bibfnamefont {D.}~\bibnamefont
  {{Tuyenbayev}}}, \bibinfo {author} {\bibfnamefont {A.}~\bibnamefont
  {{Urban}}}, \bibinfo {author} {\bibfnamefont {A.}~\bibnamefont {{Viets}}},
  \bibinfo {author} {\bibfnamefont {M.}~\bibnamefont {{Wade}}},\ and\ \bibinfo
  {author} {\bibfnamefont {A.~J.}\ \bibnamefont {{Weinstein}}},\ }\bibfield
  {title} {\bibinfo {title} {Calibration uncertainty for {A}dvanced {LIGO}'s
  first and second observing runs},\ }\href
  {https://doi.org/10.1103/PhysRevD.96.102001} {\bibfield  {journal} {\bibinfo
  {journal} {Physical Review D}\ }\textbf {\bibinfo {volume} {96}},\ \bibinfo
  {pages} {102001} (\bibinfo {year} {2017})}\BibitemShut {NoStop}%
\bibitem [{\citenamefont {{Sun}}\ \emph {et~al.}(2020)\citenamefont {{Sun}},
  \citenamefont {{Goetz}}, \citenamefont {{Kissel}}, \citenamefont
  {{Betzwieser}}, \citenamefont {{Karki}}, \citenamefont {{Viets}},
  \citenamefont {{Wade}}, \citenamefont {{Bhattacharjee}}, \citenamefont
  {{Bossilkov}}, \citenamefont {{Covas}}, \citenamefont {{Datrier}},
  \citenamefont {{Gray}}, \citenamefont {{Kandhasamy}}, \citenamefont
  {{Lecoeuche}}, \citenamefont {{Mendell}}, \citenamefont {{Mistry}},
  \citenamefont {{Payne}}, \citenamefont {{Savage}}, \citenamefont
  {{Weinstein}}, \citenamefont {{Aston}}, \citenamefont {{Buikema}},
  \citenamefont {{Cahillane}}, \citenamefont {{Driggers}}, \citenamefont
  {{Dwyer}}, \citenamefont {{Kumar}},\ and\ \citenamefont
  {{Urban}}}]{SunEtAl2020-ChrSyErAdLClb}%
  \BibitemOpen
  \bibfield  {author} {\bibinfo {author} {\bibfnamefont {L.}~\bibnamefont
  {{Sun}}}, \bibinfo {author} {\bibfnamefont {E.}~\bibnamefont {{Goetz}}},
  \bibinfo {author} {\bibfnamefont {J.~S.}\ \bibnamefont {{Kissel}}}, \bibinfo
  {author} {\bibfnamefont {J.}~\bibnamefont {{Betzwieser}}}, \bibinfo {author}
  {\bibfnamefont {S.}~\bibnamefont {{Karki}}}, \bibinfo {author} {\bibfnamefont
  {A.}~\bibnamefont {{Viets}}}, \bibinfo {author} {\bibfnamefont
  {M.}~\bibnamefont {{Wade}}}, \bibinfo {author} {\bibfnamefont
  {D.}~\bibnamefont {{Bhattacharjee}}}, \bibinfo {author} {\bibfnamefont
  {V.}~\bibnamefont {{Bossilkov}}}, \bibinfo {author} {\bibfnamefont {P.~B.}\
  \bibnamefont {{Covas}}}, \bibinfo {author} {\bibfnamefont {L.~E.~H.}\
  \bibnamefont {{Datrier}}}, \bibinfo {author} {\bibfnamefont {R.}~\bibnamefont
  {{Gray}}}, \bibinfo {author} {\bibfnamefont {S.}~\bibnamefont
  {{Kandhasamy}}}, \bibinfo {author} {\bibfnamefont {Y.~K.}\ \bibnamefont
  {{Lecoeuche}}}, \bibinfo {author} {\bibfnamefont {G.}~\bibnamefont
  {{Mendell}}}, \bibinfo {author} {\bibfnamefont {T.}~\bibnamefont {{Mistry}}},
  \bibinfo {author} {\bibfnamefont {E.}~\bibnamefont {{Payne}}}, \bibinfo
  {author} {\bibfnamefont {R.~L.}\ \bibnamefont {{Savage}}}, \bibinfo {author}
  {\bibfnamefont {A.~J.}\ \bibnamefont {{Weinstein}}}, \bibinfo {author}
  {\bibfnamefont {S.}~\bibnamefont {{Aston}}}, \bibinfo {author} {\bibfnamefont
  {A.}~\bibnamefont {{Buikema}}}, \bibinfo {author} {\bibfnamefont
  {C.}~\bibnamefont {{Cahillane}}}, \bibinfo {author} {\bibfnamefont {J.~C.}\
  \bibnamefont {{Driggers}}}, \bibinfo {author} {\bibfnamefont {S.~E.}\
  \bibnamefont {{Dwyer}}}, \bibinfo {author} {\bibfnamefont {R.}~\bibnamefont
  {{Kumar}}},\ and\ \bibinfo {author} {\bibfnamefont {A.}~\bibnamefont
  {{Urban}}},\ }\bibfield  {title} {\bibinfo {title} {Characterization of
  systematic error in {A}dvanced {LIGO} calibration},\ }\href
  {https://doi.org/10.1088/1361-6382/abb14e} {\bibfield  {journal} {\bibinfo
  {journal} {Classical and Quantum Gravity}\ }\textbf {\bibinfo {volume}
  {37}},\ \bibinfo {pages} {225008} (\bibinfo {year} {2020})}\BibitemShut
  {NoStop}%
\bibitem [{\citenamefont {Davis}\ \emph {et~al.}(2021)\citenamefont {Davis}
  \emph {et~al.}}]{Davis2021}%
  \BibitemOpen
  \bibfield  {author} {\bibinfo {author} {\bibfnamefont {D.}~\bibnamefont
  {Davis}} \emph {et~al.},\ }\bibfield  {title} {\bibinfo {title} {{LIGO
  detector characterization in the second and third observing runs}},\ }\href
  {https://doi.org/10.1088/1361-6382/abfd85} {\bibfield  {journal} {\bibinfo
  {journal} {Class. Quantum Gravity}\ }\textbf {\bibinfo {volume} {38}},\
  \bibinfo {pages} {135014} (\bibinfo {year} {2021})}\BibitemShut {NoStop}%
\bibitem [{\citenamefont {{LIGO Scientific Collaboration, the Virgo
  Collaboration and the KAGRA Collaboration}}(2017)}]{GlitchSub}%
  \BibitemOpen
  \bibfield  {author} {\bibinfo {author} {\bibnamefont {{LIGO Scientific
  Collaboration, the Virgo Collaboration and the KAGRA Collaboration}}},\
  }\href {https://dcc.ligo.org/LIGO-T1700406-v3/public} {\emph {\bibinfo
  {title} {{BayesWave Glitch Subtraction for GW170817}}}},\ \bibinfo {type}
  {Tech. Rep.}\ \bibinfo {number} {T1700406-v3}\ (\bibinfo  {institution}
  {LIGO},\ \bibinfo {year} {2017})\BibitemShut {NoStop}%
\bibitem [{\citenamefont {{LIGO Scientific Collaboration}}\ \emph
  {et~al.}(2018)\citenamefont {{LIGO Scientific Collaboration}}, \citenamefont
  {{Virgo Collaboration}},\ and\ \citenamefont {{KAGRA
  Collaboration}}}]{lalsuite}%
  \BibitemOpen
  \bibfield  {author} {\bibinfo {author} {\bibnamefont {{LIGO Scientific
  Collaboration}}}, \bibinfo {author} {\bibnamefont {{Virgo Collaboration}}},\
  and\ \bibinfo {author} {\bibnamefont {{KAGRA Collaboration}}},\ }\href
  {https://doi.org/10.7935/GT1W-FZ16} {\bibinfo {title} {{LVK} {A}lgorithm
  {L}ibrary - {LALS}uite}},\ \bibinfo {howpublished} {Free software (GPL)}
  (\bibinfo {year} {2018})\BibitemShut {NoStop}%
\bibitem [{\citenamefont {Covas}\ \emph {et~al.}(2018)\citenamefont {Covas}
  \emph {et~al.}}]{Covas2018}%
  \BibitemOpen
  \bibfield  {author} {\bibinfo {author} {\bibfnamefont {P.~B.}\ \bibnamefont
  {Covas}} \emph {et~al.},\ }\bibfield  {title} {\bibinfo {title}
  {{Identification and mitigation of narrow spectral artifacts that degrade
  searches for persistent gravitational waves in the first two observing runs
  of Advanced LIGO}},\ }\href {https://doi.org/10.1103/PhysRevD.97.082002}
  {\bibfield  {journal} {\bibinfo  {journal} {Phys. Rev. D}\ }\textbf {\bibinfo
  {volume} {97}},\ \bibinfo {pages} {082002} (\bibinfo {year}
  {2018})}\BibitemShut {NoStop}%
\bibitem [{\citenamefont {Goetz}\ \emph {et~al.}(2021)\citenamefont {Goetz}
  \emph {et~al.}}]{GoetzO3Lines}%
  \BibitemOpen
  \bibfield  {author} {\bibinfo {author} {\bibfnamefont {E.}~\bibnamefont
  {Goetz}} \emph {et~al.},\ }\href
  {https://https://dcc.ligo.org/T2100200/public} {\emph {\bibinfo {title} {{O3
  lines and combs in found in self-gated C01 data}}}},\ \bibinfo {type} {Tech.
  Rep.}\ \bibinfo {number} {T2100200-v2}\ (\bibinfo  {institution} {LIGO},\
  \bibinfo {year} {2021})\BibitemShut {NoStop}%
\bibitem [{\citenamefont {{Aasi}}\ \emph
  {et~al.}(2013{\natexlab{a}})\citenamefont {{Aasi}} \emph
  {et~al.}}]{LIGOVirg2013-EnsAlSrPrGrvWLSD}%
  \BibitemOpen
  \bibfield  {author} {\bibinfo {author} {\bibfnamefont {J.}~\bibnamefont
  {{Aasi}}} \emph {et~al.} (\bibinfo {collaboration} {{LIGO Scientific
  Collaboration} and {Virgo Collaboration}}),\ }\bibfield  {title} {\bibinfo
  {title} {Einstein@{H}ome all-sky search for periodic gravitational waves in
  {LIGO} {S}5 data},\ }\href {https://doi.org/10.1103/PhysRevD.87.042001}
  {\bibfield  {journal} {\bibinfo  {journal} {Phys. Rev. D}\ }\textbf {\bibinfo
  {volume} {87}},\ \bibinfo {pages} {042001} (\bibinfo {year}
  {2013}{\natexlab{a}})}\BibitemShut {NoStop}%
\bibitem [{\citenamefont {{Aasi}}\ \emph
  {et~al.}(2013{\natexlab{b}})\citenamefont {{Aasi}} \emph
  {et~al.}}]{LIGOVirg2013-DrSrCntGrvWGlC}%
  \BibitemOpen
  \bibfield  {author} {\bibinfo {author} {\bibfnamefont {J.}~\bibnamefont
  {{Aasi}}} \emph {et~al.} (\bibinfo {collaboration} {{LIGO Scientific
  Collaboration} and {Virgo Collaboration}}),\ }\bibfield  {title} {\bibinfo
  {title} {Directed search for continuous gravitational waves from the
  {G}alactic center},\ }\href {https://doi.org/10.1103/PhysRevD.88.102002}
  {\bibfield  {journal} {\bibinfo  {journal} {Phys. Rev. D}\ }\textbf {\bibinfo
  {volume} {88}},\ \bibinfo {pages} {102002} (\bibinfo {year}
  {2013}{\natexlab{b}})}\BibitemShut {NoStop}%
\bibitem [{\citenamefont {Leaci}(2015)}]{Leaci2015}%
  \BibitemOpen
  \bibfield  {author} {\bibinfo {author} {\bibfnamefont {P.}~\bibnamefont
  {Leaci}},\ }\bibfield  {title} {\bibinfo {title} {{Methods to filter out
  spurious disturbances in continuous-wave searches from gravitational-wave
  detectors}},\ }\href {https://doi.org/10.1088/0031-8949/90/12/125001}
  {\bibfield  {journal} {\bibinfo  {journal} {Phys. Scr.}\ }\textbf {\bibinfo
  {volume} {90}},\ \bibinfo {pages} {125001} (\bibinfo {year}
  {2015})}\BibitemShut {NoStop}%
\bibitem [{\citenamefont {Tenorio}\ \emph {et~al.}(2022)\citenamefont
  {Tenorio}, \citenamefont {Modafferi}, \citenamefont {Keitel},\ and\
  \citenamefont {Sintes}}]{Tenorio2022}%
  \BibitemOpen
  \bibfield  {author} {\bibinfo {author} {\bibfnamefont {R.}~\bibnamefont
  {Tenorio}}, \bibinfo {author} {\bibfnamefont {L.~M.}\ \bibnamefont
  {Modafferi}}, \bibinfo {author} {\bibfnamefont {D.}~\bibnamefont {Keitel}},\
  and\ \bibinfo {author} {\bibfnamefont {A.~M.}\ \bibnamefont {Sintes}},\
  }\bibfield  {title} {\bibinfo {title} {{Empirically estimating the
  distribution of the loudest candidate from a gravitational-wave search}},\
  }\href {https://doi.org/10.1103/PhysRevD.105.044029} {\bibfield  {journal}
  {\bibinfo  {journal} {Phys. Rev. D}\ }\textbf {\bibinfo {volume} {105}},\
  \bibinfo {pages} {044029} (\bibinfo {year} {2022})}\BibitemShut {NoStop}%
\bibitem [{\citenamefont {Tenorio}\ \emph {et~al.}(2021)\citenamefont
  {Tenorio}, \citenamefont {Modafferi}, \citenamefont {Keitel},\ and\
  \citenamefont {Sintes}}]{distromax}%
  \BibitemOpen
  \bibfield  {author} {\bibinfo {author} {\bibfnamefont {R.}~\bibnamefont
  {Tenorio}}, \bibinfo {author} {\bibfnamefont {L.~M.}\ \bibnamefont
  {Modafferi}}, \bibinfo {author} {\bibfnamefont {D.}~\bibnamefont {Keitel}},\
  and\ \bibinfo {author} {\bibfnamefont {A.~M.}\ \bibnamefont {Sintes}},\
  }\href {https://doi.org/10.5281/zenodo.5763765} {\bibinfo {title} {distromax:
  Empirically estimating the distribution of the loudest candidate from a
  gravitational-wave search}} (\bibinfo {year} {2021})\BibitemShut {NoStop}%
\bibitem [{\citenamefont {Abbott}\ \emph
  {et~al.}(2019{\natexlab{c}})\citenamefont {Abbott} \emph
  {et~al.}}]{Abbott2019g}%
  \BibitemOpen
  \bibfield  {author} {\bibinfo {author} {\bibfnamefont {B.~P.}\ \bibnamefont
  {Abbott}} \emph {et~al.} (\bibinfo {collaboration} {LIGO Scientific
  Collaboration and Virgo Collaboration}),\ }\bibfield  {title} {\bibinfo
  {title} {{All-sky search for long-duration gravitational-wave transients in
  the second Advanced LIGO observing run}},\ }\href
  {https://doi.org/10.1103/PhysRevD.99.104033} {\bibfield  {journal} {\bibinfo
  {journal} {Phys. Rev. D}\ }\textbf {\bibinfo {volume} {99}},\ \bibinfo
  {pages} {104033} (\bibinfo {year} {2019}{\natexlab{c}})}\BibitemShut
  {NoStop}%
\bibitem [{\citenamefont {{Mendell}}\ and\ \citenamefont
  {{Landry}}(2005)}]{MendLand2005-StcHgSrSNStt}%
  \BibitemOpen
  \bibfield  {author} {\bibinfo {author} {\bibfnamefont {G.}~\bibnamefont
  {{Mendell}}}\ and\ \bibinfo {author} {\bibfnamefont {M.}~\bibnamefont
  {{Landry}}},\ }\href {https://dcc.ligo.org/LIGO-T050003-x0/public} {\emph
  {\bibinfo {title} {Stack{S}lide and {H}ough {S}earch {SNR} and
  {S}tatistics}}},\ \bibinfo {type} {Tech. Rep.}\ \bibinfo {number}
  {T050003-x0}\ (\bibinfo  {institution} {LIGO},\ \bibinfo {year}
  {2005})\BibitemShut {NoStop}%
\bibitem [{\citenamefont {Sutton}(2013)}]{Sutton2013}%
  \BibitemOpen
  \bibfield  {author} {\bibinfo {author} {\bibfnamefont {P.~J.}\ \bibnamefont
  {Sutton}},\ }\href {https://arxiv.org/abs/1304.0210} {\bibinfo {title} {{A
  Rule of Thumb for the Detectability of Gravitational-Wave Bursts}}},\
  \bibinfo {howpublished} {arXiv:1304.0210} (\bibinfo {year}
  {2013})\BibitemShut {NoStop}%
\bibitem [{\citenamefont {Sedaghat}\ \emph {et~al.}(2022)\citenamefont
  {Sedaghat}, \citenamefont {Zebarjad}, \citenamefont {Bordbar}, \citenamefont
  {{Eslam Panah}},\ and\ \citenamefont {Moradi}}]{Sedaghat2022}%
  \BibitemOpen
  \bibfield  {author} {\bibinfo {author} {\bibfnamefont {J.}~\bibnamefont
  {Sedaghat}}, \bibinfo {author} {\bibfnamefont {S.~M.}\ \bibnamefont
  {Zebarjad}}, \bibinfo {author} {\bibfnamefont {G.~H.}\ \bibnamefont
  {Bordbar}}, \bibinfo {author} {\bibfnamefont {B.}~\bibnamefont {{Eslam
  Panah}}},\ and\ \bibinfo {author} {\bibfnamefont {R.}~\bibnamefont
  {Moradi}},\ }\bibfield  {title} {\bibinfo {title} {{Is the remnant of
  GW190425 a strange quark star?}},\ }\href
  {https://doi.org/10.1016/J.PHYSLETB.2022.137388} {\bibfield  {journal}
  {\bibinfo  {journal} {Phys. Lett. B}\ }\textbf {\bibinfo {volume} {833}},\
  \bibinfo {pages} {137388} (\bibinfo {year} {2022})}\BibitemShut {NoStop}%
\bibitem [{\citenamefont {Lasky}\ \emph {et~al.}(2017)\citenamefont {Lasky},
  \citenamefont {Sarin},\ and\ \citenamefont {Sammut}}]{LaskyMagnetarModel}%
  \BibitemOpen
  \bibfield  {author} {\bibinfo {author} {\bibfnamefont {P.}~\bibnamefont
  {Lasky}}, \bibinfo {author} {\bibfnamefont {N.}~\bibnamefont {Sarin}},\ and\
  \bibinfo {author} {\bibfnamefont {L.}~\bibnamefont {Sammut}},\ }\href
  {https://dcc.ligo.org/LIGO-T1700408/public} {\emph {\bibinfo {title}
  {{Long-duration waveform models for millisecond magnetars born in neutron
  star mergers}}}},\ \bibinfo {type} {Tech. Rep.}\ \bibinfo {number}
  {T1700408-v2}\ (\bibinfo  {institution} {LIGO},\ \bibinfo {year}
  {2017})\BibitemShut {NoStop}%
\bibitem [{\citenamefont {Lai}\ and\ \citenamefont {Shapiro}(1994)}]{Lai1994}%
  \BibitemOpen
  \bibfield  {author} {\bibinfo {author} {\bibfnamefont {D.}~\bibnamefont
  {Lai}}\ and\ \bibinfo {author} {\bibfnamefont {S.}~\bibnamefont {Shapiro}},\
  }\bibfield  {title} {\bibinfo {title} {{Gravitational Radiation from Rapidly
  Rotating Nascent Neutron Stars}},\ }\href {https://doi.org/10.1086/175438}
  {\bibfield  {journal} {\bibinfo  {journal} {Astrophys. J.}\ }\textbf
  {\bibinfo {volume} {442}},\ \bibinfo {pages} {259} (\bibinfo {year}
  {1994})}\BibitemShut {NoStop}%
\bibitem [{\citenamefont {Bauswein}\ and\ \citenamefont
  {Janka}(2012)}]{Bauswein2012}%
  \BibitemOpen
  \bibfield  {author} {\bibinfo {author} {\bibfnamefont {A.}~\bibnamefont
  {Bauswein}}\ and\ \bibinfo {author} {\bibfnamefont {H.~T.}\ \bibnamefont
  {Janka}},\ }\bibfield  {title} {\bibinfo {title} {{Measuring neutron-star
  properties via gravitational waves from neutron-star mergers}},\ }\href
  {https://doi.org/10.1103/PhysRevLett.108.011101} {\bibfield  {journal}
  {\bibinfo  {journal} {Phys. Rev. Lett.}\ }\textbf {\bibinfo {volume} {108}},\
  \bibinfo {pages} {011101} (\bibinfo {year} {2012})}\BibitemShut {NoStop}%
\bibitem [{\citenamefont {Takami}\ \emph {et~al.}(2014)\citenamefont {Takami},
  \citenamefont {Rezzolla},\ and\ \citenamefont {Baiotti}}]{Takami2014}%
  \BibitemOpen
  \bibfield  {author} {\bibinfo {author} {\bibfnamefont {K.}~\bibnamefont
  {Takami}}, \bibinfo {author} {\bibfnamefont {L.}~\bibnamefont {Rezzolla}},\
  and\ \bibinfo {author} {\bibfnamefont {L.}~\bibnamefont {Baiotti}},\
  }\bibfield  {title} {\bibinfo {title} {{Constraining the equation of state of
  neutron stars from binary mergers}},\ }\href
  {https://doi.org/10.1103/PhysRevLett.113.091104} {\bibfield  {journal}
  {\bibinfo  {journal} {Phys. Rev. Lett.}\ }\textbf {\bibinfo {volume} {113}},\
  \bibinfo {pages} {091104} (\bibinfo {year} {2014})}\BibitemShut {NoStop}%
\bibitem [{\citenamefont {Bernuzzi}\ \emph {et~al.}(2015)\citenamefont
  {Bernuzzi}, \citenamefont {Dietrich},\ and\ \citenamefont
  {Nagar}}]{Bernuzzi2015}%
  \BibitemOpen
  \bibfield  {author} {\bibinfo {author} {\bibfnamefont {S.}~\bibnamefont
  {Bernuzzi}}, \bibinfo {author} {\bibfnamefont {T.}~\bibnamefont {Dietrich}},\
  and\ \bibinfo {author} {\bibfnamefont {A.}~\bibnamefont {Nagar}},\ }\bibfield
   {title} {\bibinfo {title} {{Modeling the Complete Gravitational Wave
  Spectrum of Neutron Star Mergers}},\ }\href
  {https://doi.org/10.1103/PhysRevLett.115.091101} {\bibfield  {journal}
  {\bibinfo  {journal} {Phys. Rev. Lett.}\ }\textbf {\bibinfo {volume} {115}},\
  \bibinfo {pages} {091101} (\bibinfo {year} {2015})}\BibitemShut {NoStop}%
\bibitem [{\citenamefont {Sun}\ \emph {et~al.}(2016)\citenamefont {Sun},
  \citenamefont {Melatos}, \citenamefont {Lasky}, \citenamefont {Chung},\ and\
  \citenamefont {Darman}}]{PhysRevD.94.082004}%
  \BibitemOpen
  \bibfield  {author} {\bibinfo {author} {\bibfnamefont {L.}~\bibnamefont
  {Sun}}, \bibinfo {author} {\bibfnamefont {A.}~\bibnamefont {Melatos}},
  \bibinfo {author} {\bibfnamefont {P.~D.}\ \bibnamefont {Lasky}}, \bibinfo
  {author} {\bibfnamefont {C.~T.~Y.}\ \bibnamefont {Chung}},\ and\ \bibinfo
  {author} {\bibfnamefont {N.~S.}\ \bibnamefont {Darman}},\ }\bibfield  {title}
  {\bibinfo {title} {Cross-correlation search for continuous gravitational
  waves from a compact object in snr 1987a in ligo science run 5},\ }\href
  {https://doi.org/10.1103/PhysRevD.94.082004} {\bibfield  {journal} {\bibinfo
  {journal} {Phys. Rev. D}\ }\textbf {\bibinfo {volume} {94}},\ \bibinfo
  {pages} {082004} (\bibinfo {year} {2016})}\BibitemShut {NoStop}%
\bibitem [{\citenamefont {Owen}\ \emph {et~al.}(2024)\citenamefont {Owen},
  \citenamefont {Lindblom}, \citenamefont {Pinheiro},\ and\ \citenamefont
  {Rajbhandari}}]{Owen2024}%
  \BibitemOpen
  \bibfield  {author} {\bibinfo {author} {\bibfnamefont {B.~J.}\ \bibnamefont
  {Owen}}, \bibinfo {author} {\bibfnamefont {L.}~\bibnamefont {Lindblom}},
  \bibinfo {author} {\bibfnamefont {L.~S.}\ \bibnamefont {Pinheiro}},\ and\
  \bibinfo {author} {\bibfnamefont {B.}~\bibnamefont {Rajbhandari}},\
  }\bibfield  {title} {\bibinfo {title} {{Improved Upper Limits on
  Gravitational-wave Emission from NS 1987A in SNR 1987A}},\ }\href
  {https://doi.org/10.3847/2041-8213/AD2263} {\bibfield  {journal} {\bibinfo
  {journal} {Astrophys. J. Lett.}\ }\textbf {\bibinfo {volume} {962}},\
  \bibinfo {pages} {L23} (\bibinfo {year} {2024})}\BibitemShut {NoStop}%
\bibitem [{\citenamefont {Fransson}\ \emph {et~al.}(2024)\citenamefont
  {Fransson} \emph {et~al.}}]{Fransson2024}%
  \BibitemOpen
  \bibfield  {author} {\bibinfo {author} {\bibfnamefont {C.}~\bibnamefont
  {Fransson}} \emph {et~al.},\ }\bibfield  {title} {\bibinfo {title} {{Emission
  lines due to ionizing radiation from a compact object in the remnant of
  Supernova 1987A}},\ }\href {https://doi.org/10.1126/SCIENCE.ADJ5796}
  {\bibfield  {journal} {\bibinfo  {journal} {Science}\ }\textbf {\bibinfo
  {volume} {383}},\ \bibinfo {pages} {898} (\bibinfo {year}
  {2024})}\BibitemShut {NoStop}%
\bibitem [{\citenamefont {Abbott}\ \emph
  {et~al.}(2017{\natexlab{d}})\citenamefont {Abbott} \emph
  {et~al.}}]{Abbott2017O1AllSky}%
  \BibitemOpen
  \bibfield  {author} {\bibinfo {author} {\bibfnamefont {B.~P.}\ \bibnamefont
  {Abbott}} \emph {et~al.} (\bibinfo {collaboration} {LIGO Scientific
  Collaboration and Virgo Collaboration}),\ }\bibfield  {title} {\bibinfo
  {title} {{All-sky search for periodic gravitational waves in the O1 LIGO
  data}},\ }\href {https://doi.org/10.1103/PhysRevD.96.062002} {\bibfield
  {journal} {\bibinfo  {journal} {Phys. Rev. D}\ }\textbf {\bibinfo {volume}
  {96}},\ \bibinfo {pages} {062002} (\bibinfo {year}
  {2017}{\natexlab{d}})}\BibitemShut {NoStop}%
\bibitem [{\citenamefont {Abbott}\ \emph {et~al.}(2021)\citenamefont {Abbott}
  \emph {et~al.}}]{Abbott2021}%
  \BibitemOpen
  \bibfield  {author} {\bibinfo {author} {\bibfnamefont {R.}~\bibnamefont
  {Abbott}} \emph {et~al.} (\bibinfo {collaboration} {LIGO Scientific
  Collaboration and Virgo Collaboration}),\ }\bibfield  {title} {\bibinfo
  {title} {{All-sky search in early O3 LIGO data for continuous
  gravitational-wave signals from unknown neutron stars in binary systems}},\
  }\href {https://doi.org/10.1103/PhysRevD.103.064017} {\bibfield  {journal}
  {\bibinfo  {journal} {Phys. Rev. D}\ }\textbf {\bibinfo {volume} {103}},\
  \bibinfo {pages} {064017} (\bibinfo {year} {2021})}\BibitemShut {NoStop}%
\end{thebibliography}
\end{document}